%% file: main.tex
\documentclass[preprint,12pt]{elsarticle}
\input{notation}

\journal{Computer Methods in Applied Mechanics and Engineering}
\usepackage{threeparttable}
\usepackage{graphicx}
\usepackage{amsmath}
\usepackage{placeins}
\begin{document}
\begin{frontmatter}
\title{AI-Q: Quadrature by learned reconstruction for the material point method}

\author[inst1]{Yupeng Jiang \corref{cor1}}
\ead{yupeng.jiang@ibnm.uni-hannover.de}
\author[inst2,inst3]{Minchen Li \corref{cor1}}
\ead{minchernl@gmail.com}

\cortext[cor1]{Corresponding author.}

% --------------------------------------------------
% Affiliation
% --------------------------------------------------
\affiliation[inst1]{
    organization={IBNM, Leibniz Universit\"at Hannover},
    city={Hannover},
    postcode={30167},
    country={Germany}
}
\affiliation[inst2]{
    organization={Computer Science Department, Carnegie Mellon University},
    city={Pittsburgh},
    postcode={15213},
    country={United States}
}

\affiliation[inst3]{
    organization={Genesis AI},
    city={San Carlos, CA},
    postcode={94070},
    country={United States}
}

% --------------------------------------------------
% Abstract
% This file already contains:
% \begin{abstract} ... \end{abstract}
% --------------------------------------------------
\input{abstract_ai_q}

% --------------------------------------------------
% Keywords
% --------------------------------------------------

\end{frontmatter}
\setcounter{section}{0}
\input{introduction}
\setcounter{section}{1}
\input{methodology}

\setcounter{section}{2}
\input{single_cell_tests}

\setcounter{section}{3}
\input{numerical_results}

\setcounter{section}{4}
\input{discussion_ai_q}
\setcounter{section}{5}
\input{conclusions_ai_q}
\input{acknowledgement}
\appendix
\input{theory_appendix}
\bibliographystyle{elsarticle-num}
\bibliography{references}
\end{document}

\maketitle

This document provides the basis definitions, quadrature construction,
geometric descriptors, and sampling parameters used in the accompanying
article. Notation follows the main text unless stated otherwise.

% Paste the appendix bodies below their new section headings.
% Omit their old \section commands, \appendix commands, and counter resets.
% Remove manual \tag commands so supplementary numbers update automatically.

\section{Background basis functions and derivatives}
\label{sec:supp-basis}

For a one-dimensional grid node at $x_i$, let $r=(x-x_i)/h$ be the dimensionless nodal distance. The centered degree-$k$ cardinal kernel is denoted by $B_k(r)$. The kernels used in the structured formulation are
\begin{equation}
 B_1(r)=\begin{cases}1-\abs r,&\abs r<1,\\0,&\abs r\ge1,\end{cases}
 \qquad
 B_2(r)=\begin{cases}
 \frac34-r^2,&\abs r<\frac12,\\
 \frac12(\frac32-\abs r)^2,&\frac12\le\abs r<\frac32,\\
 0,&\abs r\ge\frac32,
 \end{cases}
 \label{eq:basis12}
\end{equation}
\begin{equation}
 B_3(r)=\begin{cases}
 \frac23-r^2+\frac12\abs r^3,&\abs r<1,\\
 \frac16(2-\abs r)^3,&1\le\abs r<2,\\
 0,&\abs r\ge2.
 \end{cases}
 \label{eq:basis3}
\end{equation}
They can equivalently be evaluated by the standard B-spline recurrence~\cite{deBoor1972}. For a grid node $\vct x_i=(x_{i,1},\ldots,x_{i,d})^{\T}$, the tensor-product basis and its physical derivative are
\begin{equation}
 \begin{aligned}
 N_i(\vct x)&=\prod_{a=1}^dB_k\!\left(\frac{x_a-x_{i,a}}h\right),\\
 \frac{\partial N_i}{\partial x_b}(\vct x)
 &=\frac1hB_k'\!\left(\frac{x_b-x_{i,b}}h\right)
       \prod_{a\ne b}B_k\!\left(\frac{x_a-x_{i,a}}h\right).
 \end{aligned}
 \label{eq:tensorbasis}
\end{equation}
The prime differentiates the dimensionless kernel argument. The complete basis satisfies partition of unity and linear precision, whereas a truncated support may not. Boundary-modified bases require their own compatible evaluation and are not identified with these interior cardinal expressions.

With mechanical nodes at integer multiples of $h$, the degree-one and degree-three polynomial spans lie between consecutive integer grid coordinates. A span $[j_0h,(j_0+1)h)$ has active nodes $j_0,j_0+1$ for degree one and $j_0-1,j_0,j_0+1,j_0+2$ for degree three. Degree-two knots lie at half-integer coordinates; its span $[(j_0+1/2)h,(j_0+3/2)h)$ has active nodes $j_0,j_0+1,j_0+2$. Taking tensor products gives $4,9,16$ active nodes in two dimensions and $8,27,64$ in three dimensions. These are stencil counts, not particle counts or Gauss-point counts.

\section{Simplex quadrature and polynomial exactness}
\label{sec:supp-quadrature}

The reference-simplex rule in Eq.~\eqref{eq:simplexrule} can be constructed from ordinary Gauss--Legendre rules on the unit interval through a collapsed-coordinate map. Let $\vct u=(u_1,\ldots,u_d)^{\T}\in[0,1]^d$ and define
\begin{equation}
 \eta_a=u_a\prod_{b=1}^{a-1}(1-u_b),\qquad a=1,\ldots,d,
 \qquad
 J_{\mathrm D}(\vct u)=\prod_{b=1}^{d-1}(1-u_b)^{d-b}.
 \label{eq:duffymap}
\end{equation}
An empty product equals one. The coordinates satisfy $\eta_a\ge0$ and $\sum_a\eta_a\le1$; $J_{\mathrm D}$ is the absolute Jacobian determinant. In two dimensions this gives $(\eta_1,\eta_2)=(u_1,(1-u_1)u_2)$ with Jacobian $1-u_1$. In three dimensions the Jacobian is $(1-u_1)^2(1-u_2)$. Thus the dimensional extension changes the quadrature measure as well as the number of coordinates.

Let $u_a^{(j)}$ and $\omega_a^{(j)}$, $a=1,\ldots,g_{\simp}$, be the one-dimensional abscissae and weights for coordinate $j$ on $[0,1]$. A multi-index $q=(a_1,\ldots,a_d)$ defines $\vct u_q$ and the reference-simplex rule
\begin{equation}
 Q_{\simp}=g_{\simp}^{\,d},
 \qquad
 \vct\eta_q=\vct\eta(\vct u_q),
 \qquad
 \widehat\omega_q=J_{\mathrm D}(\vct u_q)
                     \prod_{j=1}^d\omega_{a_j}^{(j)}.
 \label{eq:duffyrule}
\end{equation}
The weights are positive. Taking $g_{\simp}\ge\lceil d/2\rceil$ integrates the Jacobian exactly, so their sum is the reference volume $1/d!$. The two-dimensional G2 and G3 rules consequently use four and nine points per triangle. This mapped tensor-product naming convention is not the same as naming a symmetric simplex rule by its polynomial degree.

Let $r_\sigma$ be the total degree of a polynomial stress component. On a full tensor-product span, a degree-$k$ basis derivative has coordinatewise degrees $k-1$ in the differentiated direction and $k$ in the other directions. A sufficient exactness condition for $d\ge2$ is therefore
\begin{equation}
 2g_{\full}-1\ge r_\sigma+k,
 \qquad\text{or}\qquad
 g_{\full}\ge\left\lceil\frac{r_\sigma+k+1}{2}\right\rceil.
 \label{eq:fullorder}
\end{equation}
For $d=1$, the numerator inside the ceiling is instead $r_\sigma+k$. This statement concerns exact stress values in the integrand, not the degree of a neural reconstruction.

On a general simplex inside the tensor-product parent, a basis derivative has total degree at most $dk-1$. An affine simplex map preserves this degree, so the stress--gradient integrand has degree at most $m=r_\sigma+dk-1$. Under Eq.~\eqref{eq:duffymap}, its degree in any collapsed coordinate is at most $m$, and the Jacobian adds at most $d-1$. Ordinary tensor-product Gauss--Legendre quadrature is therefore sufficient when
\begin{equation}
 2g_{\simp}-1\ge r_\sigma+dk+d-2,
 \qquad\text{or}\qquad
 g_{\simp}\ge\left\lceil\frac{r_\sigma+dk+d-1}{2}\right\rceil.
 \label{eq:simplexorder}
\end{equation}
This is an orientation-independent sufficient bound, not a necessary point count for every individual polynomial. In particular, a generic oblique simplex can require more points than a full Cartesian span even though both use the same background basis.

\begin{table}[htbp]
\centering\small
\caption{Sufficient one-dimensional orders for a polynomial stress times a background gradient. The full-cell rows apply in $d\ge2$; the simplex rows below specialize Eq.~\eqref{eq:simplexorder} to triangles.}
\label{tab:orders}
\begin{tabular}{@{}llccc@{}}
\toprule
Domain & Basis degree $k$ & $r_\sigma=0$ & $r_\sigma=1$ & $r_\sigma=2$\\
\midrule
Full cell & 1 & 1 & 2 & 2\\
Full cell & 2 & 2 & 2 & 3\\
Full cell & 3 & 2 & 3 & 3\\
Triangle & 1 & 2 & 2 & 3\\
Triangle & 2 & 3 & 3 & 4\\
Triangle & 3 & 4 & 4 & 5\\
\bottomrule
\end{tabular}
\end{table}

Table~\ref{tab:orders} explains why a low-order cut-cell target should not automatically be equated with the continuum integral. For trigonometric stresses there is no finite polynomial-exactness argument. Increasing the quadrature order and resampling the same analytical field assesses that discrepancy without changing the learned target or, for cut cells, the represented boundary geometry.

\section{Cut-cell geometric descriptors and normalization}
\label{sec:supp-cut-context}

At a frozen state, define $\vct w_{\mathrm g}=\sum_iN_i\vct w_i$ as in Eq.~\eqref{eq:gridfields}. Direct substitution of the reconstructed force gives
\begin{equation}
 \sum_{i\in\mathcal I_c}\vct w_i\cdot\widehat{\vct f}_{ic}^{\full}
 =-\sum_{q=1}^{Q_{\full}}W_{cq}\widehat\stress_{cq}^{\full}
                             :\nabla\vct w_{\mathrm g}(\vct x_{cq}).
 \label{eq:discretework}
\end{equation}
This is an exact identity for the discrete stress and quadrature, and it does not assume continuum integration exactness. If the complete active basis satisfies
\begin{equation}
 \sum_iN_i(\vct x)=1,
 \qquad
 \sum_i\vct x_iN_i(\vct x)=\vct x,
 \label{eq:basisprecision}
\end{equation}
then its gradients satisfy $\sum_i\nabla N_i=\vct0$ and $\sum_i\vct x_i\otimes\nabla N_i=\Id$. Taking a constant virtual velocity in Eq.~\eqref{eq:discretework} shows that the internal resultant vanishes. Taking $\vct w_i=\mat\Omega\vct x_i$, where $\mat\Omega^{\T}=-\mat\Omega$ is an arbitrary infinitesimal rotation, gives zero because the stress is symmetric. In dimension-independent form the results are
\begin{equation}
 \sum_i\widehat{\vct f}_{ic}^{\full}=\vct0,
 \qquad
 \sum_i\left(\vct x_i\otimes\widehat{\vct f}_{ic}^{\full}
      -\widehat{\vct f}_{ic}^{\full}\otimes\vct x_i\right)=\mat0.
 \label{eq:internalbalance}
\end{equation}
The second equality is the vanishing antisymmetric moment tensor, equivalent to zero torque in two and three dimensions. The same proof holds with the cut-cell child sum. These identities follow from complete basis assembly and stress symmetry, not from hard affine stress reproduction or a learned conservation penalty. Standard particle integration has the same structural identities. They hold before boundary reactions, damping, or other momentum-changing operations, and they cannot justify discarding forces at unsupported nodes.

\section{Two-dimensional full-cell sampling parameters}
\label{sec:supp-sampling}

The arbitrary-cardinality generator first constructs a near-square candidate stratification and selects exactly $P_c$ sites. Particle order is randomized. Position jitter is applied relative to the local strata, and configurations with duplicate or nearly coincident sites, nonfinite coordinates, or collapsed coverage are rejected. The following ranges specify the two-dimensional mixture of Table~\ref{tab:layouts}; they do not define an additional online filter.

\begin{table}[htbp]
\centering\small
\caption{Detailed two-dimensional full-cell sampling parameters in the specified V3 profile. Coordinates in this table are normalized to the unit square $[0,1]^2$, not to the network's centered interval.}
\label{tab:layoutdetails}
\begin{tabularx}{\textwidth}{@{}lY@{}}
\toprule
Operation & Sampling parameter\\
\midrule
Position jitter & Maximum $0.30$ of each local stratum width\\
Random-gap candidate count & $P_c+\max\{2,\lceil0.20P_c\rceil\}$ before removal\\
Affine stretch & Each diagonal entry in $[0.86,1.14]$\\
Affine shear & Upper and lower entries in $[-0.16,0.16]$ and $[-0.08,0.08]$\\
Cluster center & $[0.30,0.70]^2$\\
Cluster strength & $[0.08,0.24]$\\
Correlated sine displacement & Initial amplitude $[0.008,0.035]$ and phase $[0,2\pi)$\\
Boundary-crossing shift & Random axis and sign; magnitude $[0.08,0.22]$; modulo wrapping\\
Minimum pair-distance gate & Greater than $10^{-8}$ of the model-cell width\\
Convex-hull area-fraction gate & Greater than $10^{-4}$\\
Maximum resampling attempts & 64 per requested geometry\\
\bottomrule
\end{tabularx}
\end{table}

The generator rescales or reduces proposed displacements when needed to keep the configuration within the parent, rather than relying on substantial clipping of out-of-bounds sites. With zero volume jitter, each particle has equal representative volume and the volume sum matches the full model cell. These settings characterize the synthetic training distribution. They should not be interpreted as restrictions imposed on the physical deformation or particle masses in the MPM calculation.

\section{Additional single-cell diagnostics}
\label{sec:supp-single-cell}

\begin{figure}[htbp]
  \centering
  \includegraphics[width=\linewidth]{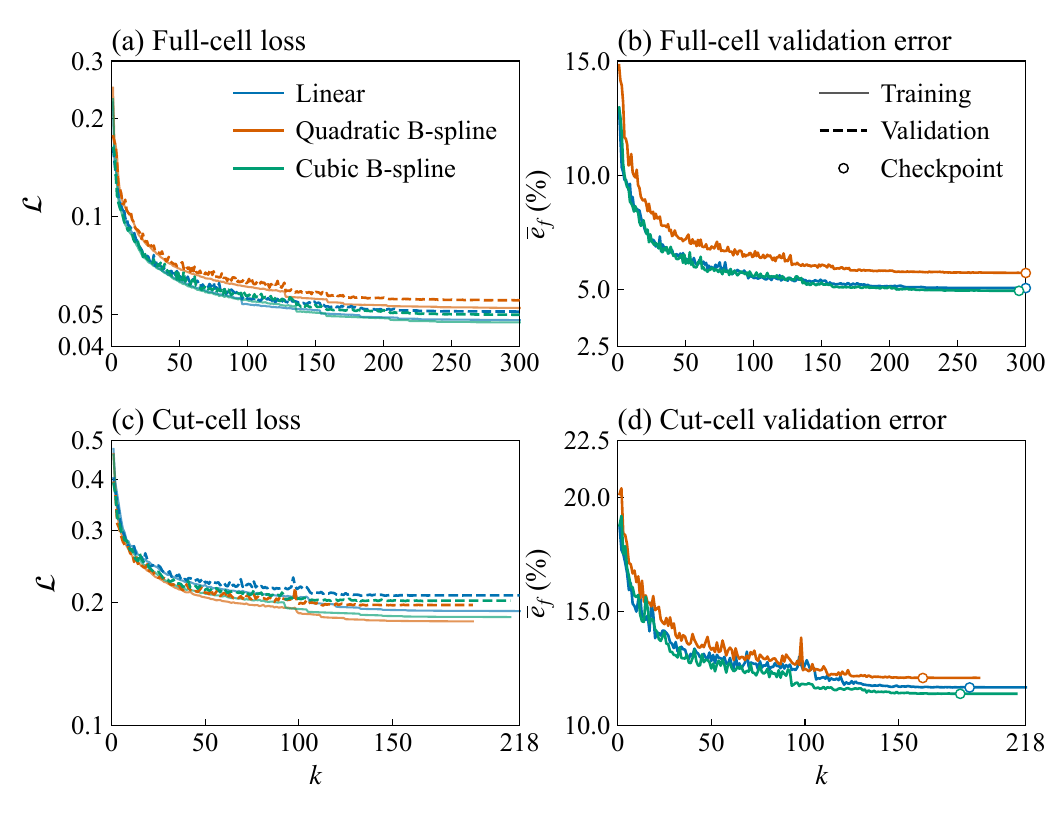}
  \caption{Two-dimensional AI-Q training and validation histories for the full- and cut-cell models. Open circles mark the selected checkpoints.}
  \label{fig:supp-training-2d}
\end{figure}

\begin{figure}[htbp]
  \centering
  \includegraphics[width=\linewidth]{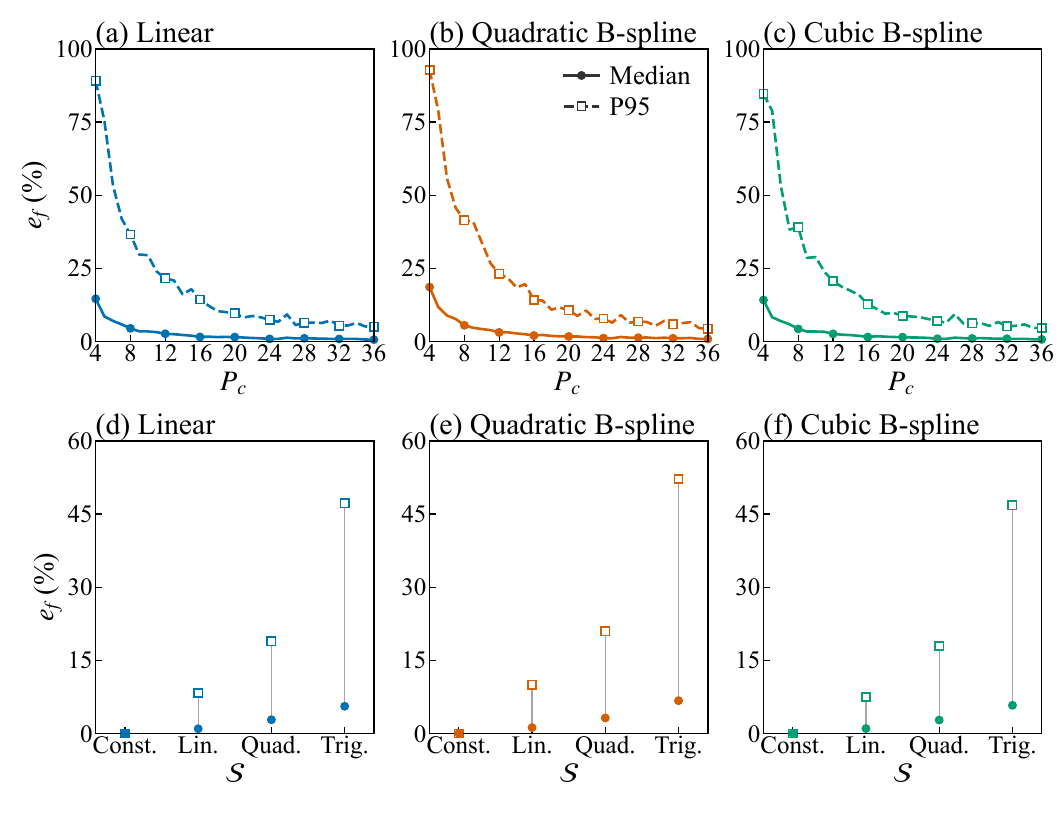}
  \caption{AI-Q full-cell internal-force error by particle count and analytical stress family for linear, quadratic B-spline, and cubic B-spline bases. Errors use the independent G10 reference.}
  \label{fig:supp-full-accuracy}
\end{figure}

\begin{figure}[htbp]
  \centering
  \includegraphics[width=\linewidth]{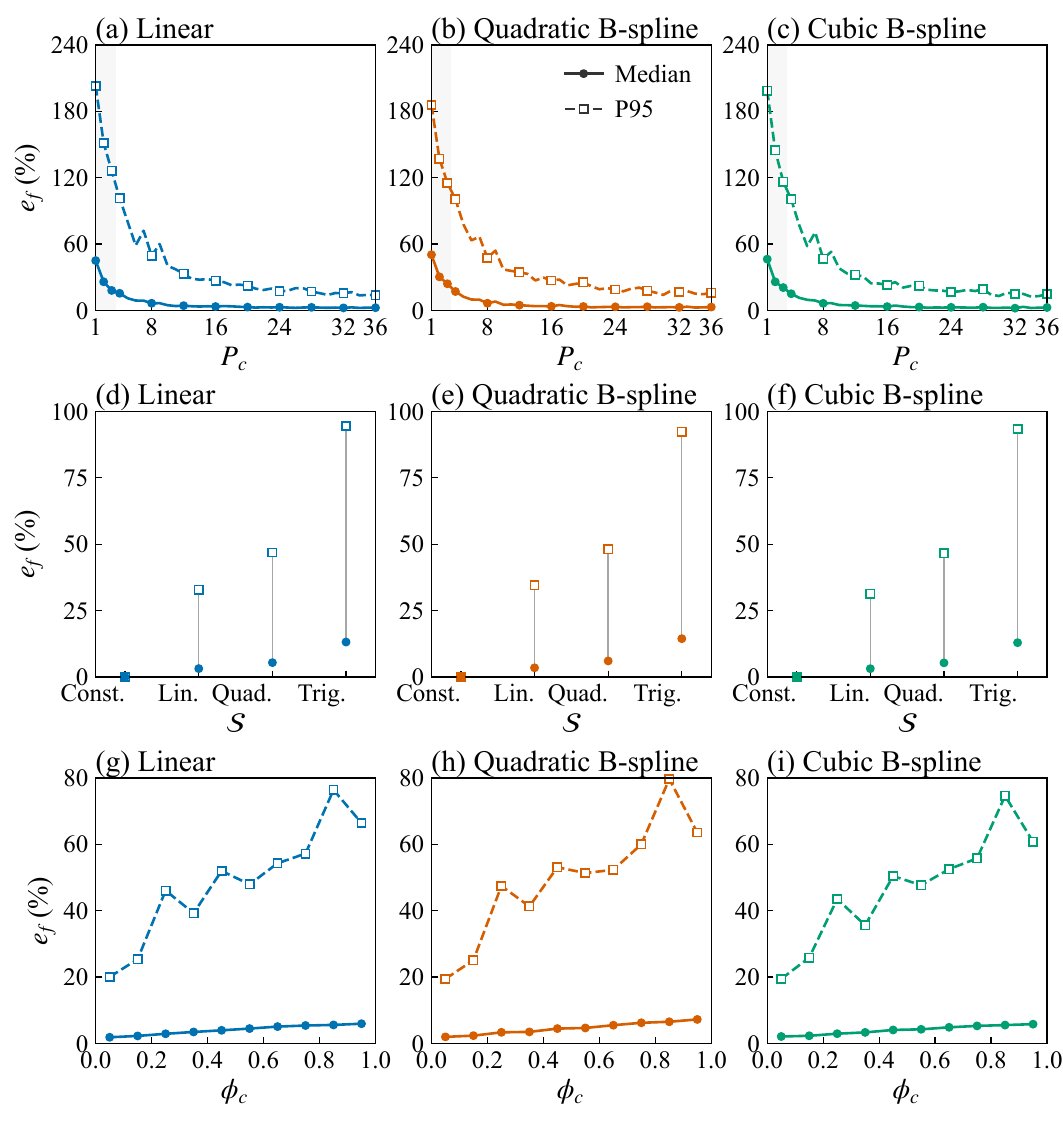}
  \caption{AI-Q cut-cell internal-force error by parent particle count, analytical stress family, and represented material fraction. Errors use the complete-parent G3 composite-triangle target.}
  \label{fig:supp-cut-accuracy}
\end{figure}

\begin{figure}[htbp]
  \centering
  \includegraphics[width=\linewidth]{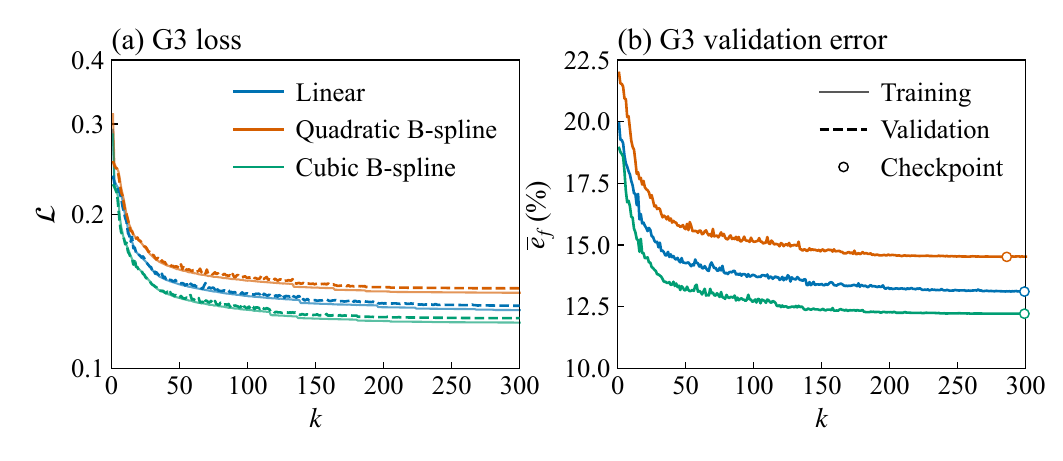}
  \caption{Three-dimensional full-cell AI-Q training and validation histories for G3 quadrature. Open circles mark the selected checkpoints.}
  \label{fig:supp-training-3d}
\end{figure}

\begin{figure}[htbp]
  \centering
  \includegraphics[width=\linewidth]{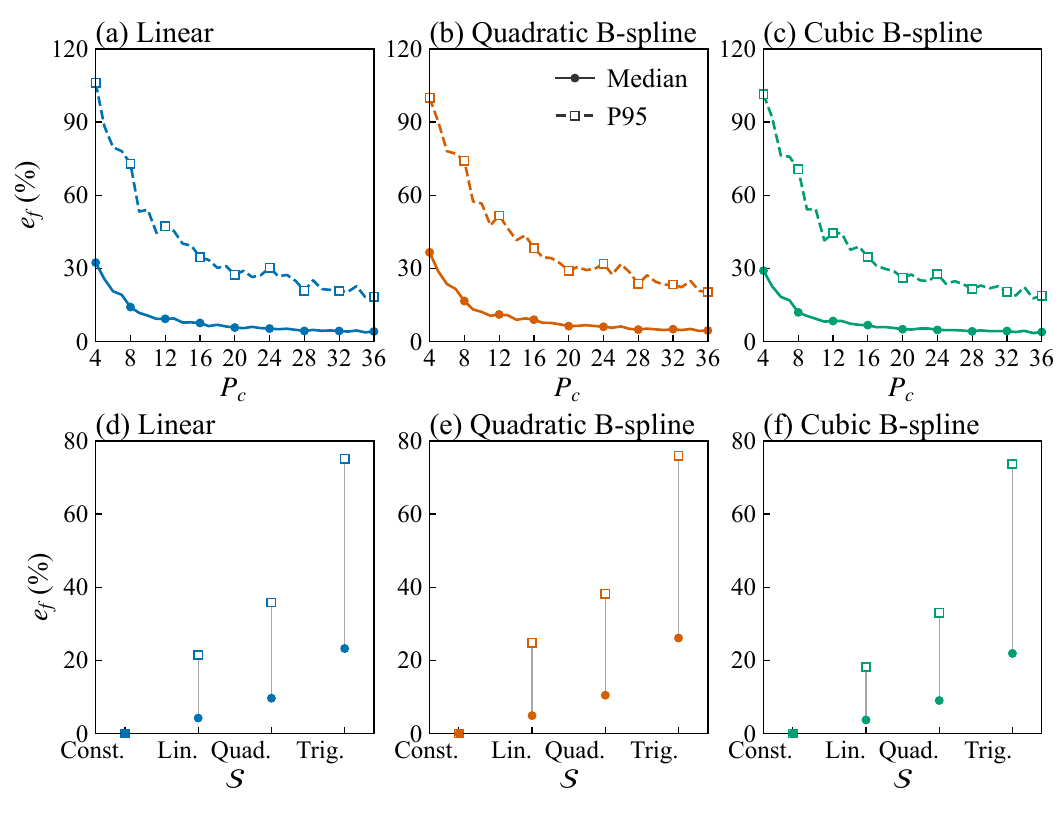}
  \caption{Three-dimensional full-cell AI-Q internal-force error by particle count and analytical stress family, evaluated against the independent higher-order reference.}
  \label{fig:supp-3d-accuracy}
\end{figure}

Figure~\ref{fig:supp-training-2d} reports the two-dimensional training histories that accompany the validation errors summarized in the main paper. Figures~\ref{fig:supp-full-accuracy} and~\ref{fig:supp-cut-accuracy} give the AI-Q-only error breakdowns by particle count and stress family; the cut-cell plot additionally resolves represented material fraction. The corresponding three-dimensional training and AI-Q-only accuracy plots are shown in Figs.~\ref{fig:supp-training-3d} and~\ref{fig:supp-3d-accuracy}. These plots are separated from the main article because the method-comparison figures there already convey the primary comparative evidence.

%% file: notation.tex
\usepackage{amsmath,amssymb,amsthm,bm}
\usepackage{stmaryrd}
\usepackage{booktabs,array,tabularx,longtable}
\usepackage{microtype}
\usepackage{needspace}
\usepackage[hidelinks]{hyperref}
\newcommand{\R}{\mathbb R}
\newcommand{\vct}[1]{\bm{#1}}
\newcommand{\mat}[1]{\bm{#1}}
\newcommand{\dd}{\,\mathrm d}
\newcommand{\T}{\mathsf T}

\newcommand{\norm}[1]{\left\lVert#1\right\rVert}
\newcommand{\abs}[1]{\left|#1\right|}
\newcommand{\stress}{\mat\sigma}
\newcommand{\MPM}{\mathrm{MPM}}

\newcommand{\full}{\mathrm F}
\newcommand{\cut}{\mathrm C}
\newcommand{\simp}{\mathrm S}
\newcommand{\Gauss}{\mathrm G}
\newcommand{\Frob}{\mathrm{Frob}}
\DeclareMathOperator{\sym}{sym}

\DeclareMathOperator{\vvec}{vec}

\numberwithin{equation}{section}
\allowdisplaybreaks[1]
\newcolumntype{Y}{>{\raggedright\arraybackslash}X}

%% file: abstract_ai_q.tex
% Abstract for the current 72-page manuscript, read on 2026-11-03.
% Quantitative reductions: Table 5, PDF p. 41; relative reductions of
% pooled mean relative force errors against standard MPM, in 2D only.
% For elsarticle, place this environment inside frontmatter.
\begin{abstract}
The dual role of material points as carriers of constitutive history and quadrature sites couples the integration accuracy of the material point method (MPM) to the evolving particle distribution. This work introduces AI-assisted quadrature (AI-Q), a framework that separates these roles through constrained learning of a local stress-reconstruction operator. The central idea is to design reconstruction for the stress moments entering the discrete momentum balance, making the assembled internal forces the principal learning objective. Exact constant reproduction and soft affine consistency constrain the approximation, while auxiliary stress supervision regularizes modes left undetermined by force matching. The network uses the particle configuration to determine the coefficients that map particle stresses to quadrature points. Constitutive updates remain on the particles, and the same reconstruction principle applies to both full and cut cells. Numerical studies demonstrate reduced mean and high-percentile internal-force errors relative to conventional particle quadrature and classical reconstruction methods over mixed stress fields and particle configurations. These improvements carry through to the stress solution in boundary-value problems, while three-dimensional tests demonstrate compatibility with large deformation. By separating material representation from force integration, AI-Q provides a general alternative to particle quadrature and a systematic route to resolving the long-standing integration problem in MPM.
\end{abstract}

%% file: introduction.tex
% Introduction for:
% Quadrature by reconstruction: Constrained learning for the material point method
% Merge introduction_references.bib into the manuscript bibliography.
% This section uses standard LaTeX commands and requires no custom math macros.

\section{Introduction}
\label{sec:introduction}

The material point method (MPM) combines a Lagrangian description of material history with a background-grid solution of the momentum equations. Material points carry mass, stress and constitutive state, while the grid supplies the spatial approximation used to determine their motion \cite{Sulsky1994,Sulsky1995, jiang2020hybrid, jiang2022hybrid}. Resetting the grid independently of the material accommodates large deformation without the mesh tangling associated with a body-fitted Lagrangian discretization. The computational mesh need not follow the evolving material boundary, and history variables remain attached to the particles. These attributes make MPM particularly attractive for problems involving large motion, changing material geometry and path-dependent behavior.

Numerical integration remains a persistent obstacle to realizing this flexibility with reliable accuracy. In conventional MPM, particles serve simultaneously as material tracers and integration points. Their motion follows the solution, whereas accurate force integration requires suitable sampling of the product of stress and background-basis gradients over the occupied material region. Deformation can therefore degrade the integration rule even when the particle stresses themselves are accurate. The resulting force errors contribute to stress oscillations and impaired convergence, with difficulties arising both from cell crossing and from irregular sampling within a cell \cite{Steffen2008,Baumgarten2023}. In implicit analysis, spatial integration also enters the equilibrium residual and tangent operator \cite{GonzalezAcosta2020, luo2026overlapping}. The common challenge is therefore to maintain a reliable integration rule as the material samples evolve relative to the background basis.

Several families of methods address this integration error from complementary directions. Generalized and convected particle-domain formulations integrate basis functions over finite particle domains, while partitioned variants split those domains at background-cell boundaries to respect the piecewise grid basis \cite{Bardenhagen2004,Sadeghirad2011,Wilson2021,SunZhou2021}. B-spline and related formulations further improve domain integration, including recent extensions to implicit coupled porous-media analysis \cite{Sun2022,Zhang2025,Ma2026Partitioned}. These methods improve the geometric integration of the particle representation, but their accuracy still depends on how well the evolving particles resolve the stress variation being integrated.

Other approaches modify the sampling or the discrete force itself. Particle repositioning can reduce volume and moment defects in distorted particle clouds \cite{Baumgarten2023}, while local stress-difference formulations address force inconsistency in prestressed states \cite{Perez2024}. Total Lagrangian MPM instead fixes the sampling geometry in a reference configuration, avoiding current-grid cell crossing at the cost of additional considerations for large rearrangement, reference updates, and contact \cite{DeVaucorbeil2020,Su2022,Bui2025}.

A third route separates material-history locations from integration locations. Weighted least squares, moving least squares, conservative Taylor least squares, continuous-gradient MLS constructions on unstructured tessellations, and cell-based integration reconstruct stresses or related quantities at mesh-associated quadrature points \cite{Wallstedt2011,SulskyGong2016,Wobbes2019,CaoUMLS2025,SongKim2024}. MPM Lite similarly transfers extensive stress to cell centers for an implicit formulation \cite{Feng2026MPMLite}. These approaches recover a mesh-associated integration rule, but local polynomial reconstruction can become sensitive to sparse or poorly distributed particles and may require enlarged support or reduced approximation order.

Data-driven quadrature suggests learning the reconstruction around the quantity that ultimately enters the governing equations. Optimized and empirical cubature make force accuracy an explicit objective, neural cubature conditions quadrature on reduced state, and recent learned finite-element rules condition integration on matrices or geometry \cite{An2008,Hernandez2017,Shen2021,Oishi2017,Yu2023,Kim2026,Daviet2025,Smaragdakis2026,Mika2026}. Within MPM, neural deformation maps have also supplied kinematic fields for Eulerian quadrature in reduced elasticity models \cite{Chen2023ROM,zong2023neural}.

For MPM, this perspective suggests learning the local reconstruction through its effect on the discrete momentum equation. Constrained learned discretizations have already shown how reconstruction coefficients can be trained through a known conservative operator \cite{BarSinai2019}. The reconstruction acts on an evolving set of particle stresses and is judged by the grid force it produces. Stress errors enter that force through quadrature weights and basis gradients; their spatial arrangement and direction therefore matter alongside their magnitude. A reconstruction that minimizes pointwise stress error need not minimize force error, while force agreement alone can leave some stress variations undetermined. The reconstruction objective must account for both the assembled response and the consistency of the recovered field.

We propose AI-assisted quadrature (AI-Q), a general framework for addressing this integration problem through constrained local stress reconstruction. A neural network uses particle and integration geometry to determine the coefficients that map particle stresses to prescribed quadrature points. For a fixed geometry, this map is linear in the supplied stresses and uses the same scalar coefficients for every tensor component. Constitutive history remains on the particles, while the reconstructed stresses enter force assembly through the prescribed quadrature weights and background-basis gradients. Training minimizes the resulting force error subject to exact constant reproduction, with additional terms promoting affine consistency and controlling stress errors weakly constrained by force agreement. Analytical stress fields provide training examples independently of a particular boundary-value solution. The same construction treats full cells and represented cut regions, linking the reconstruction to the geometry on which integration is performed. We assess AI-Q against particle quadrature and classical reconstruction for linear and B-spline bases, using two-dimensional full and cut cells, three-dimensional full cells, and boundary-value benchmarks in an explicit affine particle-in-cell (APIC) MPM solver.

Section~2 presents the formulation, consistency conditions, training procedure and solver integration. Sections~3 and~4 examine cell-level accuracy and boundary-value benchmarks, respectively. Section~5 discusses the findings and their implications, and Section~6 concludes the paper.

%% file: methodology.tex
\section{AI-assisted quadrature for the material point method}
\label{sec:method}

The material point method separates the representation of material history from the mesh on which momentum balance is solved. This separation is central to its treatment of large motion, but it also creates a numerical integration problem: the points that carry the stress move independently of the background basis against which that stress must be integrated. The proposed AI-assisted quadrature method, denoted by AI-Q, retains the material points as the carriers of constitutive history and reconstructs their stresses at mesh-associated Gauss points. The reconstruction is trained primarily through the internal force that it produces, rather than through pointwise stress agreement alone.

We develop this choice from the underlying discretization. Section~\ref{sec:foundation} identifies where conventional finite element and material point integration differ and how the difference enters the nodal force. Section~\ref{sec:objective} then compares stress-oriented and force-oriented reconstruction. The latter is realized on structured full cells in Section~\ref{sec:full}, before extending the same formulation to cut cells in Section~\ref{sec:cut}. Training data, learning objectives, and independent evaluation are presented next, followed by integration into the explicit APIC--MPM solver. Throughout the formulation, $d$ denotes the spatial dimension, bold symbols denote vectors or matrices, and superscript $\mathsf T$ denotes transpose. The Euclidean vector norm and its induced matrix norm are denoted by $\norm{\cdot}_2$. Current-volume and current-boundary measures are written as $\dd v$ and $\dd a$; in a planar calculation, volume is understood per unit out-of-plane thickness. Dimensional choices for the numerical studies are stated with the training configurations, rather than built into the reconstruction equations.

\subsection{From continuum momentum balance to particle quadrature error}
\label{sec:foundation}

\subsubsection{Continuum problem and common spatial discretization}
\label{sec:continuum}

Let $\Omega(t)\subset\R^d$ be the current material region, with Cauchy stress $\stress$, density $\rho$, velocity $\vct v$, and body force per unit mass $\vct b$. For a virtual velocity $\vct w$ vanishing on the prescribed-velocity boundary, the weak momentum balance is
\begin{equation}
 \int_{\Omega(t)}\rho\vct w\cdot\vct a\dd v
 =-\int_{\Omega(t)}\nabla\vct w:\stress\dd v
  +\int_{\Omega(t)}\rho\vct w\cdot\vct b\dd v
  +\int_{\Gamma_t(t)}\vct w\cdot\bar{\vct t}\dd a .
 \label{eq:weak}
\end{equation}
Using background basis functions $N_i$ gives the familiar semidiscrete internal force
\begin{equation}
 \vct f_i^{\mathrm{int}}
 =-\int_{\Omega(t)}\stress\nabla N_i\dd v .
 \label{eq:massforces}
\end{equation}
The integration of this quantity, rather than the continuum balance itself, is the focus of AI-Q.

\subsubsection{Element-associated and material-associated integration}
\label{sec:femmpm}

In a body-fitted finite element method, quadrature points remain attached to each element, so the integration rule is chosen for a fixed element-local basis and geometry. MPM instead stores constitutive history at moving material points and commonly uses those same points as quadrature locations \cite{Sulsky1994}. If particle $p$ has position $\vct x_p$, current volume $V_p$, and stress $\stress_p$, the standard internal force is
\begin{equation}
 \vct f_i^{\MPM}
 =-\sum_{p\in\mathcal P}V_p\stress_p\nabla N_i(\vct x_p).
 \label{eq:particleforce}
\end{equation}
The particle locations therefore evolve independently of the background basis whose gradient is being integrated. A particle may remain an accurate carrier of material history while becoming a poor quadrature location for $\stress\nabla N_i$.

Current particle volume is a physical state, not a tunable fitting weight. If a solver stores Kirchhoff stress $\mat\tau_p=J_p\stress_p$ with $V_p=J_pV_p^0$, then $V_p\stress_p=V_p^0\mat\tau_p$. AI-Q reconstructs Cauchy stress and integrates it with prescribed physical quadrature weights; normalizing network inputs never changes these stress--volume pairings.

\subsubsection{How moving quadrature enters the force error}
\label{sec:particleerror}

Write the particle stress as $\stress_p=\stress^{\mathrm{ref}}(\vct x_p)+\mat\epsilon_p$ for a smooth reference field. Subtracting the corresponding continuum force gives
\begin{equation}
 \begin{aligned}
 \vct f_i^{\MPM}-\vct f_i^{\mathrm{int,ref}}
 ={}&-\sum_pV_p\mat\epsilon_p\nabla N_i(\vct x_p)\\
 &+\left[\int_{\Omega(t)}\stress^{\mathrm{ref}}\nabla N_i\dd v
       -\sum_pV_p\stress^{\mathrm{ref}}(\vct x_p)\nabla N_i(\vct x_p)\right].
 \end{aligned}
 \label{eq:mpmerror}
\end{equation}
The first term propagates particle-state error; the bracketed term is quadrature error and remains even for exact particle stresses.

For a full cell $K_c$ with centroid $\vct x_c$, define the zeroth- and first-moment defects
\begin{equation}
 \Delta m_{0,c}=\sum_{p\in\mathcal P_c}V_p-\abs{K_c},
 \qquad
 \vct{\Delta m}_{1,c}
   =\sum_{p\in\mathcal P_c}V_p(\vct x_p-\vct x_c).
 \label{eq:particlemoments}
\end{equation}
For the local force integrand $\vct g_i=\stress\nabla N_i$, a Taylor expansion about $\vct x_c$ gives
\begin{equation}
\mathbf{E}_{ic}^{\mathrm{quad}}
=
-\mathbf{g}_i(\mathbf{x}_c)\,\Delta m_{0,c}
-\nabla\mathbf{g}_i(\mathbf{x}_c)\,
    \Delta\mathbf{m}_{1,c}
+\mathbf{r}_{ic},
\label{eq:particle_force_moment_error}
\end{equation}
where $\mathbf r_{ic}$ contains the second- and higher-order spatial variation. Thus even smooth stresses can produce force error when the moving particles no longer represent the cell volume and its first moment accurately. For an affine $\vct g_i$, matching these two moments is sufficient for exact local integration.

Cell crossing is a related but distinct mechanism. If $\nabla N_i$ is discontinuous and a particle crosses that discontinuity while $V_p\stress_p$ remains continuous, its force contribution jumps by
\begin{equation}
 \llbracket\vct f_{ip}\rrbracket
 =-V_p\stress_p\llbracket\nabla N_i\rrbracket .
 \label{eq:crossing}
\end{equation}
Smoother basis gradients can suppress this particular cell-crossing mechanism, including B-spline and compact-kernel constructions, but they do not by themselves enforce the moment conditions above \cite{Steffen2008,Bardenhagen2004,Wilson2021,CKMPM2025}. Cut cells add a third issue because the integration domain is only part of the parent cell. These observations motivate separating material-state storage from force integration while treating basis regularity, quadrature accuracy, and boundary geometry as distinct ingredients.

\subsection{From stress recovery to force-oriented reconstruction}
\label{sec:objective}

\subsubsection{Restoring mesh-associated quadrature}

The preceding discussion suggests using a quadrature rule attached to the background mesh, as in element integration, without transferring constitutive history away from the material points. Consider first a full cell $K_c$. Let $\mathcal I_c$ be the set of basis functions that are nonzero in the interior of $K_c$, let $N_{b,c}=\abs{\mathcal I_c}$ be its number of active nodes, and let $P_c=\abs{\mathcal P_c}$ be its particle count. Each full cell $K_c$ is integrated using a prescribed tensor-product
Gauss quadrature rule with $Q_{\mathrm F}$ quadrature points
$\vct{x}_{cq}$ and positive physical integration weights $W_{cq}$,
where $q=1,\ldots,Q_{\mathrm F}$. If the stress were available at these points, the cell force would be
\begin{equation}
 \vct f_{ic}^{\Gauss,\full}[\stress]
 =-\sum_{q=1}^{Q_{\full}}W_{cq}\stress(\vct x_{cq})
                                      \nabla N_i(\vct x_{cq}).
 \label{eq:gausstarget}
\end{equation}
Evaluating $\mathbf{f}_{ic}^{G,\mathrm F}[\boldsymbol{\sigma}]$
requires the values of the stress field $\boldsymbol{\sigma}$ at
the prescribed Gauss points. Since MPM carries stresses at the
moving material points, AI-Q reconstructs these Gauss-point values
from the particle stresses within $K_c$. This local reconstruction
connects the evolving material-point data to the prescribed
quadrature rule, enabling force integration on the background mesh
while constitutive history remains on the particles.

AI-Q constructs a local stress-reconstruction matrix
$\mat{R}_c^{\mathrm F}$ with $Q_{\mathrm F}$ rows and $P_c$ columns.
Let $\mathcal{N}_{\theta_{\mathrm F}}$ denote the full-cell neural
reconstruction map, with $\theta_{\mathrm F}$ collecting its trainable
network weights and biases.
The input $\mathcal{G}_c$ describes the local coordinates of the particles
and quadrature points, together with the relative particle volumes.
The map converts these geometric inputs into the reconstruction
coefficients through neural prediction and the constant-reproduction
projection introduced below.
The reconstructed stresses are then obtained as
\begin{equation}
    \mat{R}_c^{\mathrm F}
    =
    \mathcal{N}_{\theta_{\mathrm F}}(\mathcal{G}_c),
    \qquad
    \widehat{\boldsymbol{\sigma}}_{cq}^{\mathrm F}
    =
    \sum_{p\in\mathcal{P}_c}
        R_{cqp}^{\mathrm F}\boldsymbol{\sigma}_p.
    \label{eq:full_cell_stress_reconstruction}
\end{equation}
Here $R_{cqp}^{\mathrm F}$ is the coefficient associated with particle $p$
and quadrature point $q$, and
$\widehat{\boldsymbol{\sigma}}_{cq}^{\mathrm F}$ is the reconstructed
stress at $\mat{x}_{cq}$.
Each scalar coefficient acts on all components of the particle stress
tensor, making the reconstruction linear in stress for a fixed particle
configuration.
The reconstructed stresses are used for force integration, while
constitutive history and material updates remain on the particles.

We use a structured Cartesian mesh in this study because its reference geometry and basis restrictions are shared by all interior cells. The reconstruction--integration principle itself does not require a structured mesh. For an unstructured cell with an invertible reference map $\vct\varphi_c$, its Jacobian $\mat J_c$, reference query $\vct\xi_q$, reference weight $\widehat w_q$, and reference basis restriction $\widehat N_{ic}$, the corresponding ingredients are
\begin{equation}
 \begin{aligned}
 \vct x_{cq}&=\vct\varphi_c(\vct\xi_q),
 &W_{cq}&=\widehat w_q\abs{\det\mat J_c(\vct\xi_q)},\\
 \nabla N_i(\vct x_{cq})
 &=\mat J_c(\vct\xi_q)^{-\T}\nabla_{\!\xi}\widehat N_{ic}(\vct\xi_q).
 \end{aligned}
 \label{eq:unstructured}
\end{equation}
Equations~\eqref{eq:gausstarget}--\eqref{eq:full_cell_stress_reconstruction} then have the same form, with cell geometry included in $\mathcal G_c$ and an appropriate rule for each element type. This is an extension of the formulation, not an assertion that a network trained on Cartesian cells can be used unchanged on arbitrary elements. In particular, the geometric features, training distribution, and physical-coordinate consistency conditions must match the chosen mesh. The structured construction below fixes these choices for the present calculations.

\subsubsection{Why stress agreement and force agreement are different objectives}
\label{sec:forcechoice}

There are two natural ways to determine the reconstruction coefficients. A stress-oriented approach seeks agreement between reconstructed and reference stresses at the queries, and only afterwards evaluates the nodal force. A force-oriented approach instead judges the same reconstruction by the force it produces through Eq.~\eqref{eq:gausstarget}. Both approaches reconstruct the stresses at the Gauss points, with the coefficients optimized to minimize either the stress error or the error in the resulting nodal forces.

To compare these objectives, we express how the prescribed quadrature
rule maps stress reconstruction errors to nodal-force errors. Specifically, let $\vct s=\operatorname{sv}(\stress)\in\R^{n_\sigma}$ collect the independent components of a symmetric tensor, where $n_\sigma=d(d+1)/2$. The diagonal components are listed first, followed by the upper-triangular components in lexicographic order; shear components are not multiplied by two. Let $\mat S_c^{\mathrm P}\in\R^{P_c\times n_\sigma}$ and $\mat S_c^{\Gauss}\in\R^{Q_{\full}\times n_\sigma}$ collect, as rows, the stresses carried by the particles and the reference stresses evaluated at the Gauss points. Then
\begin{equation}
 \widehat{\mat S}_c^{\full}=\mat R_c^{\full}\mat S_c^{\mathrm P},
 \qquad
 \vct e_c=\vvec\!\left(\widehat{\mat S}_c^{\full}-\mat S_c^{\Gauss}\right),
 \qquad
 \widehat{\vct f}_c^{\full}-\vct f_c^{\Gauss,\full}
       =\mat B_c\vct e_c.
 \label{eq:forceoperator}
\end{equation}
Here $\vvec$ stacks matrix columns, $\widehat{\vct f}_c^{\full}$ and $\vct f_c^{\Gauss,\full}$ stack the $d$ force components at all $N_{b,c}$ active nodes in a fixed order, and $\mat B_c\in\R^{dN_{b,c}\times Q_{\full}n_\sigma}$ is the known linear assembly matrix defined by Eq.~\eqref{eq:gausstarget}. It contains the Gauss weights, background-basis gradients, tensor-component placement, and the internal-force sign. No part of $\mat B_c$ is learned.

The two objectives can be compared through quadratic measures of the
same stress reconstruction error. Let $\mat H_{\sigma,c}$ be a positive diagonal matrix that specifies a chosen query-stress metric, such as equal component weighting or volume weighting. Before normalization, the two approximation criteria have the form
\begin{equation}
 J_{\sigma,c}=\vct e_c^{\T}\mat H_{\sigma,c}\vct e_c,
 \qquad
 J_{f,c}=\norm{\mat B_c\vct e_c}_2^2
        =\vct e_c^{\T}\mat B_c^{\T}\mat B_c\vct e_c.
 \label{eq:twocriteria}
\end{equation}
They need not rank two candidate reconstructions in the same order. Stress errors are combined with different basis-gradient directions and quadrature weights before they become nodal forces; some components reinforce one another, whereas others cancel. Thus minimizing an unweighted stress error is not equivalent to minimizing the force error. A useful one-sided relation is
\begin{equation}
 \norm{\widehat{\vct f}_c^{\full}-\vct f_c^{\Gauss,\full}}_2
 \le\norm{\mat B_c\mat H_{\sigma,c}^{-1/2}}_2
               \sqrt{J_{\sigma,c}},
 \label{eq:stressforcebound}
\end{equation}
where the matrix norm is the induced Euclidean norm. Accurate stresses are sufficient to control the force for a bounded assembly operator, but the size and orientation of the stress error relative to that operator also matter.

Force agreement constrains the reconstructed stresses through their
contribution to the assembled nodal forces. If the assembly matrix
$\mathbf{B}$ has a nontrivial null space, stress errors in that space
are canceled during assembly and remain unconstrained by the force
objective. AI-Q therefore combines a primary force objective with
auxiliary stress supervision and explicit consistency conditions.
The auxiliary term penalizes discrepancies at individual Gauss points,
while the consistency conditions enforce constant reproduction and
promote affine accuracy. The normalized loss terms and their relative
weights are specified in Section~\ref{sec:losses}.

The resulting method is therefore a force-oriented stress reconstruction, not a black-box prediction of nodal forces. Substituting Eq.~\eqref{eq:full_cell_stress_reconstruction} into the quadrature gives an equivalent particle-to-force representation,
\begin{equation}
 \begin{aligned}
 \widehat{\vct f}_{ic}^{\full}
  &=-\sum_{p\in\mathcal P_c}\stress_p\vct g_{icp}^{\mathrm{eff}},\\
 \vct g_{icp}^{\mathrm{eff}}
  &=\sum_{q=1}^{Q_{\full}}W_{cq}R_{cqp}^{\full}
                                  \nabla N_i(\vct x_{cq}).
 \end{aligned}
 \label{eq:effectivegradient}
\end{equation}
The effective coupling vector $\mathbf{g}_{icp}^{\mathrm{eff}}$
describes how the stress carried by particle $p$ contributes to
node $i$ through the Gauss points of cell $c$. It combines the learned
reconstruction coefficients with the prescribed quadrature weights
and basis gradients, taking the place of
$V_p\nabla N_i(\mathbf{x}_p)$ in standard particle quadrature and
having the same units of length to the power $d-1$. This representation
identifies stress reconstruction as the learned part of the force
calculation: the network determines how particle stresses are
transferred to the Gauss points, and the prescribed quadrature rule
then assembles their nodal-force contributions. The mapping can
therefore be evaluated directly as stress reconstruction followed
by Gauss-point force assembly.

At each Gauss point, all active nodes use the same reconstructed
stress tensor, with each nodal contribution determined by its own
basis gradient. When the complete active basis satisfies partition
of unity, these gradients sum to zero at each quadrature point,
so the internal-force contributions from the cell cancel when
summed over its active nodes.

\subsection{Constrained reconstruction on structured full cells}
\label{sec:full}

\subsubsection{Reconstruction domain and prescribed quadrature}

We now specialize the general construction to a uniform Cartesian background mesh with spacing $h$ in every coordinate direction. The basis degree is $k$, with $k=1,2,3$ denoting linear, quadratic, and cubic cardinal B-splines. A reconstruction cell $K_c$ is one tensor-product polynomial knot span, not the entire support of a basis function. Accordingly, the active set in Section~\ref{sec:objective} has $N_{b,c}=(k+1)^d$ nodes on an interior span. The basis definitions and node-index conventions are collected in Supplementary Material, Section S1.

This distinction between a span and a support is essential for locality. The particle set $\mathcal P_c$ contains only particles whose current positions belong to $K_c$, even when an active basis function also extends into adjacent cells. After reconstruction, the cell force is assembled to all nodes in $\mathcal I_c$. Neighboring particles are not added to the reconstruction merely because the background basis has a wider support. A half-open ownership convention assigns a particle on a span face to one cell and prevents duplicate contributions.

\begin{figure}[!htbp]
  \centering
   \includegraphics[width=\linewidth]{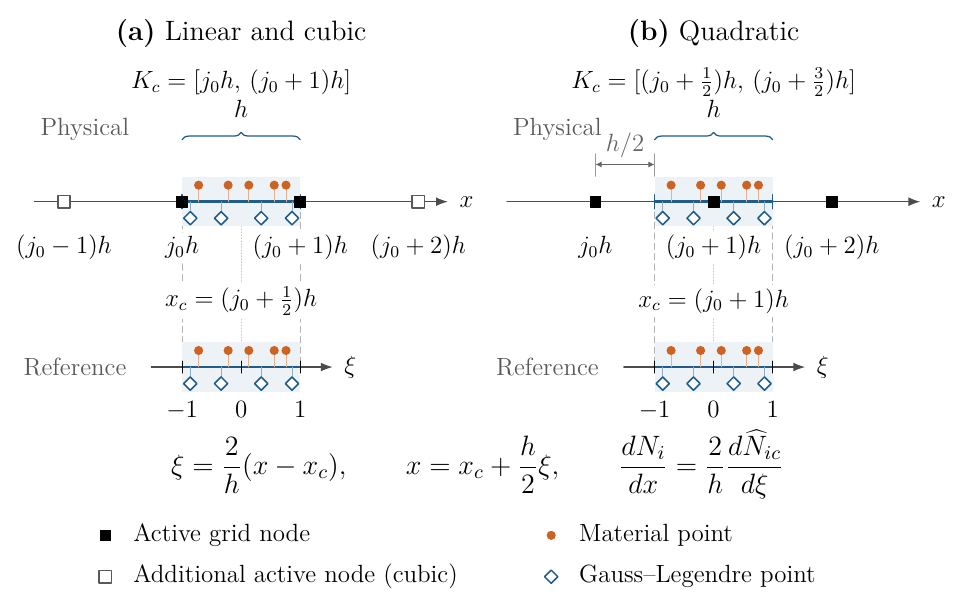}
  \caption{One-dimensional cardinal B-spline basis functions and their polynomial knot spans. AI-Q reconstructs independently on each span while assembling the resulting force to every active background node.}
  \label{fig:shape_function_span}
\end{figure}

For a span centered at $\vct x_c$, define its reference coordinates and coordinate transformation by
\begin{equation}
 K_c=\vct x_c+\frac h2[-1,1]^d,
 \qquad
 \vct\xi=\frac2h(\vct x-\vct x_c),
 \qquad
 \nabla_{\!x}N_i=\frac2h\nabla_{\!\xi}\widehat N_{ic}.
 \label{eq:localcoordinates}
\end{equation}
The closure is used in the displayed geometric definition; ownership on shared faces follows the half-open convention. For the centered quadratic cardinal basis, spans are shifted by half a grid spacing relative to the ordinary intervals between grid nodes as shown in Fig.~\ref{fig:shape_function_span}. In one coordinate, the span containing $x$ is indexed by $j_0=\lfloor x/h-1/2\rfloor$ and extends from $(j_0+1/2)h$ to $(j_0+3/2)h$, with active nodes $j_0,j_0+1,j_0+2$. This index describes the physical polynomial pieces of the basis; it does not shift the mechanical nodes or the particles.

Let $g_{\full}$ be the number of one-dimensional Gauss--Legendre points on $[-1,1]$. Denote their coordinates and weights by $\zeta_a$ and $w_a$, $a=1,\ldots,g_{\full}$. A query index $q$ enumerates the tuple $(a_1,\ldots,a_d)$ and defines
\begin{equation}
 \begin{aligned}
 Q_{\full}&=g_{\full}^{\,d},
 &\vct\xi_q&=(\zeta_{a_1},\ldots,\zeta_{a_d})^{\T},\\
 \vct x_{cq}&=\vct x_c+\frac h2\vct\xi_q,
 &W_{cq}&=(h/2)^d\prod_{b=1}^d w_{a_b}.
 \end{aligned}
 \label{eq:fullrule}
\end{equation}
Thus $\sum_qW_{cq}=h^d$, independently of the number and arrangement of particles. The rule order is fixed with the selected model. In particular, $P_c$ need not equal $Q_{\full}$ and may be either smaller or larger. A Gauss rule has polynomial exactness properties for a prescribed integrand, but increasing its order cannot by itself improve the unknown stresses supplied to it. Reconstruction and quadrature accuracy must therefore be assessed separately.

\subsubsection{Geometry-conditioned reconstruction coefficients}

The full-cell input for particle $p$ consists of its local position $\vct\xi_{cp}=2(\vct x_p-\vct x_c)/h$ and relative volume
\begin{equation}
 \nu_{cp}^{\full}=\frac{V_p}{\sum_{r\in\mathcal P_c}V_r},
 \qquad
 \vct z_{cp}^{\full}=[\vct\xi_{cp},\nu_{cp}^{\full}].
 \label{eq:fullfeatures}
\end{equation}
Brackets in a feature definition denote concatenation. In the offline full-cell bank, the particle volumes sum to the model-cell volume, so this expression coincides with normalization by that volume. Online, the adapter constructs a separate input tensor with the same normalization. It does not rescale the physical volumes, masses, or Gauss weights. The normalized volume is a descriptor of relative particle representation, not a correction of the moment defect in Eq.~\eqref{eq:particlemoments}.

A shared particle encoder $\Phi_{\theta_{\full}}$ maps this input to a vector $\vct e_{cp}^{\full}\in\R^{d_{\mathrm f}}$, where $d_{\mathrm f}$ is the feature width. The encoded particles are combined into a cell descriptor by mean pooling, sum pooling, and an explicit particle count:
\begin{equation}
 \begin{aligned}
 \vct e_{cp}^{\full}&=\Phi_{\theta_{\full}}(\vct z_{cp}^{\full}),\\
 \vct u_c^{\full}&=\left[
  \frac1{P_c}\sum_p\vct e_{cp}^{\full},\quad
  \sum_p\vct e_{cp}^{\full},\quad P_c/P_{\max}\right].
 \end{aligned}
 \label{eq:pooling}
\end{equation}
Here and below, unqualified particle sums in a cell formula run over $\mathcal P_c$, and $P_{\max}$ is the array capacity. Shared encoding and order-independent pooling follow the set-network construction of Deep Sets~\cite{Zaheer2017}. The count and pooled descriptors allow a single network to act on different particle cardinalities.

A shared particle--query decoder $\Psi_{\theta_{\full}}$ then produces one raw coefficient per pair:
\begin{equation}
 \widetilde R_{cqp}^{\full}
 =\Psi_{\theta_{\full}}\!\left(
 [\vct e_{cp}^{\full},\vct u_c^{\full},
    \vct\xi_q,\vct\xi_q-\vct\xi_{cp}]\right).
 \label{eq:decoder}
\end{equation}
Both encoder and decoder are multilayer perceptrons. The encoder input has dimension $d+1$, the pooled descriptor has dimension $2d_{\mathrm f}+1$, and the pair input has dimension $3d_{\mathrm f}+1+2d$. The output layer is linear, allowing signed coefficients. Particle stresses are applied only after these coefficients have been constructed; they do not enter the geometric feature path.

Variable-size particle sets are stored in arrays of capacity $P_{\max}$, together with a binary mask $\chi_{cp}$ that is one for a valid particle and zero for an unused slot. Invalid slots are excluded from the features, pooling, coefficient projection, and stress application. In padded-array expressions, the members of $\mathcal P_c$ are enumerated by local slots; this relabeling does not change their particle identities. Padding is a storage convention rather than an augmentation of the physical particle set. Permuting valid particles permutes the coefficient columns and the corresponding input stresses together, leaving the reconstructed stress unchanged up to floating-point summation effects.

\subsubsection{Constant reproduction and affine consistency}
\label{sec:constraints}

The reconstruction should preserve a spatially uniform stress exactly and
capture the leading spatial variation of a smooth stress field. We impose
constant reproduction through an explicit correction of the decoder output
and promote affine reproduction through a training penalty. For a full cell
$K_c$, let $\widetilde{R}^{\mathrm F}_{cqp}$ denote the raw scalar coefficient
assigned by the decoder to particle $p$ when reconstructing stress at
quadrature point $q$; the superscript $\mathrm F$ identifies the full-cell
construction. A tilde denotes a coefficient before the consistency correction.
Since the reconstructed stress is a weighted sum of the particle stresses,
exact reproduction of every constant stress field requires the coefficients
to sum to one at each quadrature point.

For a fixed cell $c$ and quadrature point $q$, collect the $P_c$ valid raw
coefficients into the column vector
$\widetilde{\vct r}_{cq}\in\mathbb R^{P_c}$, where $P_c$ is the number of
particles in the cell. The corrected coefficient vector $\vct r_{cq}$ is
the closest vector whose entries sum to one:
\begin{equation}
\vct r_{cq}
=
\underset{\vct {r}\in\mathbb R^{P_c}}{\operatorname{arg\,min}}
\frac12\left\|\vct r-\widetilde{\vct r}_{cq}\right\|_2^2
\qquad\text{subject to}\qquad
\mathbf 1^{\mathsf T}\vct r=1,
\label{eq:fullcell_constant_projection}
\end{equation}
where $\vct r$ is the candidate coefficient vector and $\mathbf 1$ is a
vector of $P_c$ ones. This Euclidean projection enforces constant reproduction
while making the smallest change to the learned coefficients.

To accommodate different particle counts, the implementation stores each cell
in $P_{\max}$ particle slots. The validity mask $\chi_{cp}$ equals one for a
slot occupied by an actual particle and zero for an unused padding slot.
Writing the projection in this padded representation gives
\begin{equation}
R^{\mathrm F}_{cqp}
=
\chi_{cp}\left[
\widetilde{R}^{\mathrm F}_{cqp}
+
\frac{1-
\displaystyle\sum_{\ell=1}^{P_{\max}}
\chi_{c\ell}\widetilde{R}^{\mathrm F}_{cq\ell}}
{P_c}
\right],
\qquad
P_c=\sum_{p=1}^{P_{\max}}\chi_{cp}>0.
\label{eq:fullcell_constant_correction}
\end{equation}
Here $R^{\mathrm F}_{cqp}$ is the corrected coefficient used in stress
reconstruction. The correction distributes the discrepancy in the coefficient
sum equally among the valid particles and keeps padded entries at zero. It is
applied whenever the coefficients are evaluated, during both training and
prediction. Consequently, if every particle carries the same stress tensor,
the reconstruction returns that tensor at every quadrature point. Using the
same scalar coefficients for all stress components also preserves
the symmetry of the particle stress tensors.

Affine reproduction additionally requires the coefficients to reproduce the
quadrature-point coordinates. Let
$\boldsymbol\xi_{cp}=2(\vct x_p-\vct x_c)/h$ be the dimensionless position
of particle $p$, where $\vct x_p$ is its physical position, $\vct x_c$
is the cell center, and $h$ is the span width. Let $\boldsymbol\xi_q$ denote
the quadrature-point position in the same reference coordinates. The
coordinate-moment defect
\begin{equation}
\boldsymbol\delta^{\mathrm F}_{cq}
:=
\sum_{p\in\mathcal P_c}
R^{\mathrm F}_{cqp}\boldsymbol\xi_{cp}
-\boldsymbol\xi_q
\label{eq:fullcell_affine_defect}
\end{equation}
measures the difference between the coefficient-weighted particle coordinates
and the target quadrature coordinate; $\mathcal P_c$ is the set of particles
in $K_c$. Together with the unit coefficient sum, a zero defect guarantees
exact reconstruction at that query for particle stresses sampled from any
affine stress field.

The connection to stress accuracy follows by considering one scalar stress
component $\sigma_a(\vct x)=\sigma_0+\vct b_\sigma\cdot
(\vct x-\vct x_c)$, where $a$ identifies the tensor component,
$\sigma_0$ is its value at the cell center, and $\mathbf b_\sigma$ is its
constant spatial gradient. Denote its reconstructed value by
$\widehat\sigma^{\mathrm F}_{a,cq}$ and the physical quadrature-point location
by $\mathbf x_{cq}$. For particle values sampled from this field, the exact constant correction gives
\begin{equation}
\widehat\sigma^{\mathrm F}_{a,cq}-\sigma_a(\vct x_{cq})
=
\frac h2\,\vct b_\sigma\cdot\boldsymbol\delta^{\mathrm F}_{cq}.
\label{eq:fullcell_affine_stress_error}
\end{equation}
Penalizing the squared magnitude of the coordinate defect therefore targets
the reconstruction error associated with linear stress variation. We impose
this condition softly during training to accommodate sparse and irregular
particle configurations. Exact affine reproduction at arbitrary query
locations requires sufficiently many affinely independent particles, and a
hard affine correction can become ill-conditioned as the particle geometry
approaches rank deficiency. The chosen formulation preserves constant stresses
for every nonempty cell while promoting affine accuracy through the training
objective described in Section~2.6.

\Needspace{13\baselineskip}
\subsubsection{Internal-force reconstruction and error decomposition}
\label{sec:fullerror}

With the query rule and projected coefficients now specified, the full-cell reconstruction is completed by applying the operator to the particle stresses and integrating the result against the background-basis gradients:
\begin{equation}
 \begin{aligned}
 \widehat\stress_{cq}^{\full}
  &=\sum_{p\in\mathcal P_c}R_{cqp}^{\full}\stress_p,\\
 \widehat{\vct f}_{ic}^{\full}
  &=-\sum_{q=1}^{Q_{\full}}W_{cq}\widehat\stress_{cq}^{\full}
                                      \nabla N_i(\vct x_{cq}).
 \end{aligned}
 \label{eq:fullforce}
\end{equation}
Here $\mat R_c^{\full}$ has $Q_{\full}$ rows and $P_c$ valid columns, and the second line is evaluated for every $i\in\mathcal I_c$. The same reconstructed stress is used for all active nodes. The particle volumes influence the geometric input, whereas the physical measure of integration is supplied by $W_{cq}$; no additional factor $V_p$ multiplies the reconstructed stress. This closes the construction introduced in Section~\ref{sec:objective}: the network supplies a constrained stress-transfer operator, and the prescribed rule converts its output into a cell contribution to the internal force.

To distinguish the accuracy of these two operations, consider a common reference stress field $\stress$ at the current configuration. For any region $D\subseteq K_c$, write $\vct f_{ic}(D;\stress)=-\int_D\stress\nabla N_i\dd v$ for its exact contribution. The force $\vct f_{ic}^{\Gauss,\full}[\stress]$ in Eq.~\eqref{eq:gausstarget} instead uses exact samples of this field at the prescribed full-cell queries. Adding and subtracting that intermediate force gives
\begin{equation}
 \begin{aligned}
 \widehat{\vct f}_{ic}^{\full}-\vct f_{ic}(K_c;\stress)
 ={}&\underbrace{\bigl[\widehat{\vct f}_{ic}^{\full}
               -\vct f_{ic}^{\Gauss,\full}[\stress]\bigr]}_{\text{reconstruction error}}\\
 &+\underbrace{\bigl[\vct f_{ic}^{\Gauss,\full}[\stress]
               -\vct f_{ic}(K_c;\stress)\bigr]}_{\text{quadrature error}}.
 \end{aligned}
 \label{eq:fullerror}
\end{equation}
The first bracket measures the effect of replacing exact query stresses by reconstructed ones while keeping the rule fixed. It includes both the approximation made by the reconstruction and any error already present in the particle stresses. The second bracket measures the error of that rule when supplied with exact stresses, and may remain nonzero even when reconstruction is perfect. There is no geometric term for a genuinely full cell because its material region is exactly $K_c$. In particular, the assumption here is full material occupancy, not merely the absence of a boundary label in a numerical classification.

The affine defect explains part of the first bracket. Define $\Lambda_c^{\full}=\max_q\sum_p\abs{R_{cqp}^{\full}}$ and $\delta_c^{\full}=\max_q\norm{\vct\delta_{cq}^{\full}}_2$. Suppose a reference stress field is twice continuously differentiable on the segments joining queries to particles, with first- and second-derivative bounds $M_{\sigma,1,c}$ and $M_{\sigma,2,c}$ in induced tensor norms. Using the tensor Frobenius norm $\norm{\mat A}_{\Frob}^2=\sum_{a,b}A_{ab}^2$, suppose that particle-stress perturbations have norm at most $\eta_c$. Taylor expansion then gives
\begin{equation}
 \max_q\norm{\widehat\stress_{cq}^{\full}-\stress(\vct x_{cq})}_{\Frob}
 \le \Lambda_c^{\full}\eta_c
   +\frac h2M_{\sigma,1,c}\delta_c^{\full}
   +\frac{d h^2}{2}M_{\sigma,2,c}\Lambda_c^{\full}.
 \label{eq:reconstructionbound}
\end{equation}
\ref{app:error} derives this estimate and the weighted force bound in Eq.~\eqref{eq:weightedforcebound}, which transfers the stress error to the first bracket of Eq.~\eqref{eq:fullerror}. The three contributions in Eq.~\eqref{eq:reconstructionbound} are amplified particle error, the affine-moment defect, and unresolved stress curvature. A unit row sum alone supplies no upper bound on $\Lambda_c^{\full}$, and the soft penalty does not by itself establish a convergence rate. These are analytical measures of the reconstructed operator, rather than quantities used by an online error controller. They also leave the quadrature bracket unchanged: increasing reconstruction accuracy and increasing quadrature order address different parts of the total force error.

The full-cell construction thus separates an assumed exact integration region, a prescribed quadrature rule, and a learned stress-transfer operator. A cut cell retains this separation, but the integration region must first be represented inside the parent span. The next section follows the same construction and shows where that additional geometric approximation enters the force error.

\subsection{Extension to cut cells in a dimension-independent form}
\label{sec:cut}

\subsubsection{Reconstruction domain and prescribed quadrature}

The full-cell construction assumes that the integration region is all of $K_c$. When a material boundary intersects that cell, the correct domain is instead $D_c=\Omega(t)\cap K_c$. Extending the full-cell rule into the empty part of the cell would introduce a geometric integration error, irrespective of the accuracy of the reconstructed stress. The cut-cell extension therefore retains the parent-cell basis and particle set, but places its quadrature points inside an approximation of the material region.

\begin{figure}[!htbp]
  \centering
   \includegraphics[width=0.8\linewidth]{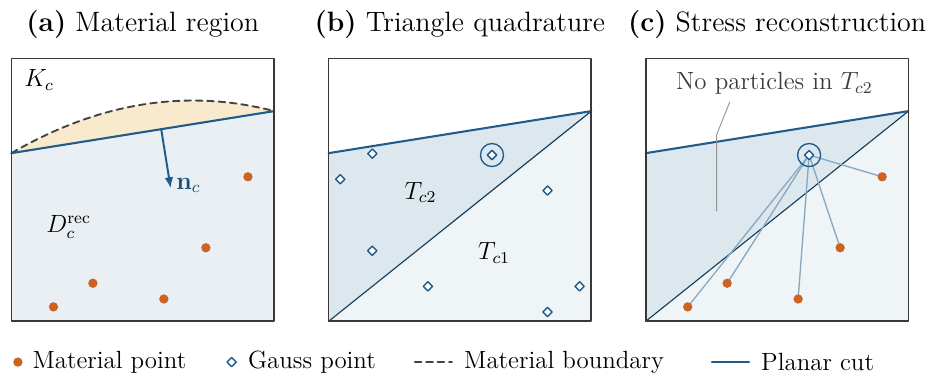}
  \caption{Cut-cell construction. A locally represented material region inside the parent span is partitioned into simplices, and prescribed quadrature points in those children are reconstructed from the complete parent particle set.}
  \label{fig:cut_cell_scheme}
\end{figure}

A locally planar boundary, as illustrated in Fig.~\ref{fig:cut_cell_scheme}, is represented by a unit normal $\vct n_c$ directed into the material and a signed offset $\kappa_c$ from the parent center. The reconstructed material region is
\begin{equation}
 D_c^{\mathrm{rec}}
 =K_c\cap\{\vct x:\vct n_c\cdot(\vct x-\vct x_c)-\kappa_c\ge0\},
 \qquad
 \alpha_c=\frac{\abs{D_c^{\mathrm{rec}}}}{\abs{K_c}}.
 \label{eq:cutgeometry}
\end{equation}
Here $\alpha_c\in(0,1)$ is the material fraction. The superscript $\mathrm{rec}$ distinguishes the represented geometry from the exact intersection $D_c$; they coincide only when the local boundary model is exact. Particle positions and supplied boundary labels determine whether such a local representation is admissible in the solver. The reconstruction formulation uses the resulting region and does not require a particular line- or plane-fitting algorithm.

Partition the represented region into $N_{\simp,c}$ nonoverlapping $d$-simplices $T_{c\ell}$, with child index $\ell=1,\ldots,N_{\simp,c}$:
\begin{equation}
 \overline{D_c^{\mathrm{rec}}}
 =\bigcup_{\ell=1}^{N_{\simp,c}}\overline{T_{c\ell}},
 \qquad
 \abs{T_{c\ell}\cap T_{cm}}=0\quad(\ell\ne m).
 \label{eq:simplexpartition}
\end{equation}
Let $\widehat T=\{\vct\eta\in\R^d:\eta_a\ge0,\ \sum_a\eta_a\le1\}$ be the reference simplex. For the vertices $\vct x_{c\ell}^{(0)},\ldots,\vct x_{c\ell}^{(d)}$, define the edge matrix $\mat A_{c\ell}=[\vct x_{c\ell}^{(1)}-\vct x_{c\ell}^{(0)},\ldots,\vct x_{c\ell}^{(d)}-\vct x_{c\ell}^{(0)}]$. A rule with $Q_{\simp}$ reference points $\vct\eta_q$ and positive reference weights $\widehat\omega_q$ gives
\begin{equation}
 \begin{aligned}
 \vct x_{c\ell q}&=\vct x_{c\ell}^{(0)}+\mat A_{c\ell}\vct\eta_q,\\
 W_{c\ell q}&=\abs{\det\mat A_{c\ell}}\widehat\omega_q,
 &\vct\xi_{c\ell q}&=\frac2h(\vct x_{c\ell q}-\vct x_c).
 \end{aligned}
 \label{eq:simplexrule}
\end{equation}
The rule is chosen to integrate constants exactly, so $\sum_qW_{c\ell q}=\abs{T_{c\ell}}$. The complete parent has $Q_{\cut}(c)=N_{\simp,c}Q_{\simp}$ query points and $\sum_{\ell,q}W_{c\ell q}=\abs{D_c^{\mathrm{rec}}}$. The two indices $\ell,q$ are retained to distinguish a simplex-local rule from its parent-cell assembly. A tensor-product Duffy construction~\cite{Duffy1982}, including its dimension-dependent Jacobian, is given in Supplementary Material, Section S2.

Although a simplex map is used to generate the integration points, the neural coordinates $\vct\xi_{c\ell q}$ remain coordinates in the parent span. Background gradients are likewise evaluated from the parent basis, not from simplex finite element functions. The simplices partition the integration region; they do not create new mechanical degrees of freedom or replace the background interpolation.

\subsubsection{Geometry-conditioned reconstruction coefficients}

For every simplex, the network uses the complete parent set $\mathcal P_c$, rather than only particles lying inside $T_{c\ell}$. Each child operator $\mat R_{c\ell}^{\cut}$ therefore has $Q_{\simp}$ query rows and $P_c$ valid particle columns. As in the full-cell module, its coefficients are constructed from geometry before being applied to stress. The additional child index identifies the subregion containing the queries; it does not change which particles supply the material state.

The cut volume feature is formed from the model input volume $V_{cp}^{\mathrm{in}}$ and the associated material-volume scale $V_c^{\mathrm{scale}}$. Together with the parent-local particle position, it defines
\begin{equation}
 \nu_{cp}^{\cut}=V_{cp}^{\mathrm{in}}/V_c^{\mathrm{scale}},
 \qquad
 \vct z_{cp}^{\cut}=[\vct\xi_{cp},\nu_{cp}^{\cut}].
 \label{eq:cutvolfeature}
\end{equation}
In cut-cell training, these are the particle volume and the complete represented material volume, respectively, not the individual simplex volume. An online adapter must express both in the coordinate and normalization convention of its model. As for the full-cell input, this operation does not modify physical particle volumes or integration weights.

The shared particle encoder $\Phi_{\theta_{\cut}}$ and pooled parent descriptor use the same construction as in Eq.~\eqref{eq:pooling}, with a separately trained parameter set $\theta_{\cut}$:
\begin{equation}
 \begin{aligned}
 \vct e_{cp}^{\cut}&=\Phi_{\theta_{\cut}}(\vct z_{cp}^{\cut}),\\
 \vct u_c^{\cut}&=\left[
  \frac1{P_c}\sum_p\vct e_{cp}^{\cut},\quad
  \sum_p\vct e_{cp}^{\cut},\quad P_c/P_{\max}\right].
 \end{aligned}
 \label{eq:cutpooling}
\end{equation}
These features describe the whole parent and are shared by its children. To identify the queried subregion, a context encoder $\mathcal U_{\theta_{\cut}}$ embeds a geometric descriptor $\vct c_{c\ell}$ into a feature $\vct e_{c\ell}^{\mathrm{ctx}}$. The descriptor contains dimensionless measures and centroids of the simplex and material region, the oriented boundary plane, and the fixed child-order information used by the model. The raw coefficients are
\begin{equation}
 \begin{aligned}
 \vct e_{c\ell}^{\mathrm{ctx}}&=\mathcal U_{\theta_{\cut}}(\vct c_{c\ell}),\\
 \widetilde R_{c\ell qp}^{\cut}
  &=\Psi_{\theta_{\cut}}\!\left(
  [\vct e_{cp}^{\cut},\vct u_c^{\cut},\vct e_{c\ell}^{\mathrm{ctx}},
   \vct\xi_{c\ell q},\vct\xi_{c\ell q}-\vct\xi_{cp}]\right).
 \end{aligned}
 \label{eq:cutdecoder}
\end{equation}
Supplementary Material, Section S3 gives an explicit descriptor and its dimensional specialization. Including child-order information makes a consistent decomposition and ordering part of the model interface; particle-permutation invariance should not be confused with invariance to an arbitrary retriangulation or retessellation.

\subsubsection{Constant projection and affine consistency}

The minimum-correction problem in Eq.~\eqref{eq:fullcell_constant_projection} is applied independently to every child-query row. With the parent mask and particle count, it gives
\begin{equation}
 \begin{aligned}
 R_{c\ell qp}^{\cut}
 &=\chi_{cp}\left[\widetilde R_{c\ell qp}^{\cut}
 +\frac{1-\sum_{r=1}^{P_{\max}}\chi_{cr}\widetilde R_{c\ell qr}^{\cut}}{P_c}\right],\\
 \vct\delta_{c\ell q}^{\cut}
 &=\sum_{p\in\mathcal P_c}R_{c\ell qp}^{\cut}\vct\xi_{cp}
                                      -\vct\xi_{c\ell q}.
 \end{aligned}
 \label{eq:cutconstraints}
\end{equation}
The row sum is one at every query, so constant stress is reproduced on every simplex. Exact affine reproduction would additionally require $\vct\delta_{c\ell q}^{\cut}=\vct0$, which is again treated as a soft condition. Because particles and queries use the same parent coordinates, the affine scalar field and gradient $\vct b_\sigma$ introduced in Section~\ref{sec:constraints} satisfy the direct counterpart of Eq.~\eqref{eq:fullcell_affine_stress_error}:
\begin{equation}
 \widehat\sigma_{\mathrm a,c\ell q}^{\cut}
      -\sigma_{\mathrm a}(\vct x_{c\ell q})
 =\frac h2\vct b_\sigma\cdot\vct\delta_{c\ell q}^{\cut}.
 \label{eq:cutaffineidentity}
\end{equation}
The soft residual consequently has the same physical meaning in both modules. For $P_c=1$, the valid coefficient must be one at every query, so the only possible prediction is the single particle stress. This limiting case makes the information content of the method explicit and explains why sparse cut cells require separate evaluation.

\Needspace{13\baselineskip}
\subsubsection{Internal-force reconstruction and error decomposition}
\label{sec:cuterror}

Applying the projected child operators to the particle stresses completes the cut-cell reconstruction. The resulting stresses are integrated first over each simplex and then summed over the parent:
\begin{equation}
 \begin{aligned}
 \widehat\stress_{c\ell q}^{\cut}
  &=\sum_{p\in\mathcal P_c}R_{c\ell qp}^{\cut}\stress_p,\\
 \widehat{\vct f}_{ic}^{\cut}
  &=-\sum_{\ell=1}^{N_{\simp,c}}\sum_{q=1}^{Q_{\simp}}
          W_{c\ell q}\widehat\stress_{c\ell q}^{\cut}
                              \nabla N_i(\vct x_{c\ell q}).
 \end{aligned}
 \label{eq:cutforce}
\end{equation}
The two lines have the same roles as in Eq.~\eqref{eq:fullforce}; only the query indexing and the integration region change. Every child uses the parent basis $N_i$ and contributes to the same active-node set $\mathcal I_c$. A simplex with no particle in its own interior can still be reconstructed from the parent data. Reusing those data does not duplicate material volume, because the physical weights belong to disjoint integration regions and each child is included exactly once.

As for the full cell, introduce the intermediate force obtained by supplying exact reference stresses to the same quadrature rule:
\begin{equation}
 \vct f_{ic}^{\Gauss,\cut}[\stress]
 =-\sum_{\ell=1}^{N_{\simp,c}}\sum_{q=1}^{Q_{\simp}}
  W_{c\ell q}\stress(\vct x_{c\ell q})\nabla N_i(\vct x_{c\ell q}).
 \label{eq:cutgausstarget}
\end{equation}
This is the counterpart of Eq.~\eqref{eq:gausstarget}, now on $D_c^{\mathrm{rec}}$. Using the exact-region force $\vct f_{ic}(D;\stress)$ defined in Section~\ref{sec:fullerror}, add and subtract both this quadrature force and the exact integral over the represented region. The full-cell decomposition then extends to
\begin{equation}
 \begin{aligned}
 \widehat{\vct f}_{ic}^{\cut}-\vct f_{ic}(D_c;\stress)
 ={}&\underbrace{\bigl[\widehat{\vct f}_{ic}^{\cut}
               -\vct f_{ic}^{\Gauss,\cut}[\stress]\bigr]}_{\text{reconstruction error}}\\
 &+\underbrace{\bigl[\vct f_{ic}^{\Gauss,\cut}[\stress]
               -\vct f_{ic}(D_c^{\mathrm{rec}};\stress)\bigr]}_{\text{quadrature error}}\\
 &+\underbrace{\bigl[\vct f_{ic}(D_c^{\mathrm{rec}};\stress)
               -\vct f_{ic}(D_c;\stress)\bigr]}_{\text{geometry error}}.
 \end{aligned}
 \label{eq:cuterror}
\end{equation}
The first two brackets have precisely the meanings established in Eq.~\eqref{eq:fullerror}: stress reconstruction is assessed with the geometry and rule held fixed, and quadrature is assessed with exact query stresses on that same geometry. The third bracket is the additional effect of representing the material boundary. It vanishes when $D_c^{\mathrm{rec}}=D_c$, even though the parent may still be only partially occupied. Taking the represented and exact regions to be $K_c$ recovers the two-term full-cell error structure. This reduction concerns the error structure; each module retains its own prescribed rule and trained operator.

The reconstruction term is controlled by the same local estimate as in the full-cell case. Define $\Lambda_c^{\cut}=\max_{\ell,q}\sum_p\abs{R_{c\ell qp}^{\cut}}$ and $\delta_c^{\cut}=\max_{\ell,q}\norm{\vct\delta_{c\ell q}^{\cut}}_2$. Under the same smoothness and particle-error assumptions as in Eq.~\eqref{eq:reconstructionbound}, with derivative bounds taken over all relevant particle--query segments, one obtains
\begin{equation}
 \max_{\ell,q}\norm{\widehat\stress_{c\ell q}^{\cut}-\stress(\vct x_{c\ell q})}_{\Frob}
 \le\Lambda_c^{\cut}\eta_c
  +\frac h2M_{\sigma,1,c}\delta_c^{\cut}
  +\frac{d h^2}{2}M_{\sigma,2,c}\Lambda_c^{\cut}.
 \label{eq:cutreconstructionbound}
\end{equation}
The length scale remains the parent spacing $h$, rather than an individual simplex diameter, because every child uses particles from throughout the parent. The weighted force bound in \ref{app:error} likewise applies with the complete $\ell,q$ sum. These estimates concern the first bracket only; quadrature order controls a different approximation, and the geometric discrepancy is bounded separately by Eq.~\eqref{eq:geometrybound}. The parallel decompositions therefore provide a common basis for evaluating both modules, while identifying boundary representation as the additional source of error in a cut cell.

\subsection{Analytical training fields and data construction}
\label{sec:aiq-training-data}

\subsubsection{Particle geometries and stress realizations}
\label{sec:aiq-training-realizations}

The offline training data pair each particle geometry with several
independently generated analytical stress fields. Evaluating each field
at the particles supplies the reconstruction input, while evaluation at
the prescribed quadrature points provides exact stress targets. Applying
the selected quadrature rule to these targets gives the reference cell
contributions to the nodal forces, allowing the reconstruction to be
trained against its effect on force integration.

Let $c$ index a sampled geometry and $s=1,\ldots,S$ its $S$ stress
realizations. The superscript $*$ identifies an analytically prescribed
field, while a hat identifies its reconstruction. For a full cell, the
particle set is $\mathcal P_c$, particle $p$ is located at $\vct x_p$,
and the $Q_{\mathrm F}$ prescribed quadrature points have positions
$\vct x_{cq}$ and physical weights $W_{cq}$. Using the full-cell
reconstruction coefficients $R^{\mathrm F}_{cqp}$, the stress samples,
reconstructed stresses, and reference force contribution to background
node $i$ are
\begin{equation}
\begin{aligned}
\boldsymbol{\sigma}_{csp}^{*}
  &=\boldsymbol{\sigma}_{cs}^{*}(\vct x_p),
\qquad
\boldsymbol{\sigma}_{csq}^{*}
  =\boldsymbol{\sigma}_{cs}^{*}(\vct x_{cq}),
\qquad
\widehat{\boldsymbol{\sigma}}_{csq}^{\mathrm F}
  =\sum_{p\in\mathcal P_c}
    R^{\mathrm F}_{cqp}\,\boldsymbol{\sigma}_{csp}^{*},
\\[4pt]
\mathbf f_{ics}^{\mathrm G,\mathrm F}
  &=-\sum_{q=1}^{Q_{\mathrm F}}
    W_{cq}\,\boldsymbol{\sigma}_{csq}^{*}
    \nabla N_i(\vct x_{cq}).
\end{aligned}
\label{eq:aiq-training-targets}
\end{equation}
Here $N_i$ is the background basis function associated with node $i$,
$\nabla N_i$ is its gradient in physical coordinates, and the superscript
$\mathrm G$ denotes evaluation with the prescribed quadrature rule. The
predicted force $\widehat{\mathbf f}_{ics}^{\mathrm F}$ uses the same
weights and basis gradients with the reconstructed stresses
$\widehat{\boldsymbol{\sigma}}_{csq}^{\mathrm F}$ in place of the
analytical samples $\boldsymbol{\sigma}_{csq}^{*}$.

For a cut cell, the index $\ell$ identifies a child simplex and $q$ a
quadrature point within that simplex. Its analytical stress sample is
$\boldsymbol{\sigma}_{cs\ell q}^{*}
=\boldsymbol{\sigma}_{cs}^{*}(\mathbf x_{c\ell q})$.
The reference and predicted cell contributions are both assembled over
all children before entering the objective for the parent cell. In both
cell types, all $S$ stress realizations share the reconstruction
coefficients associated with their geometry. Training therefore
evaluates the same operator across different stress patterns on a fixed
particle arrangement.

\subsubsection{Analytical field families and amplitude normalization}
\label{sec:aiq-training-fields}

\begin{figure}[!htbp]
  \centering
   \includegraphics[width=0.5\linewidth]{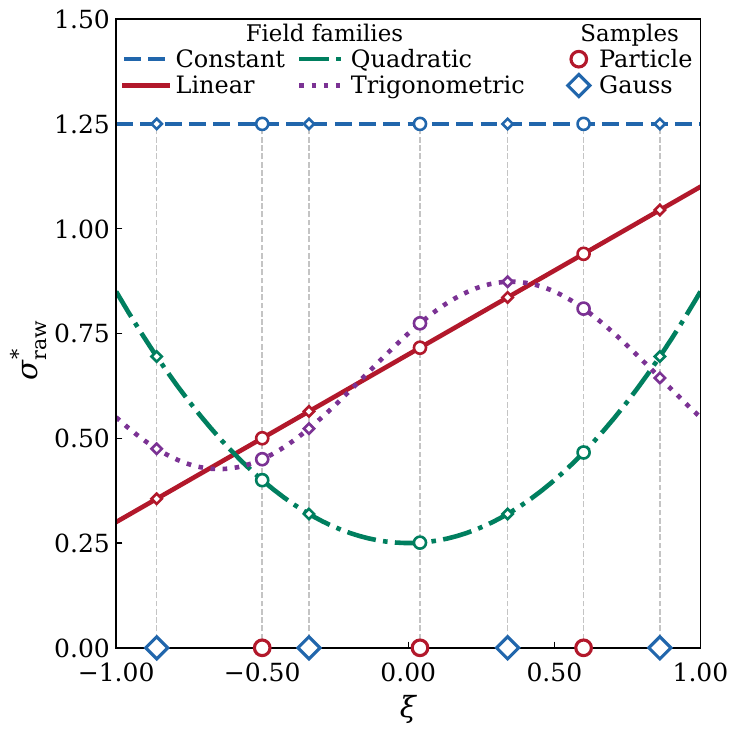}
  \caption{One-dimensional illustrations of the analytical stress families used for training: constant, affine, quadratic, and trigonometric variation. Multidimensional fields are formed componentwise in reference-cell coordinates.}
  \label{fig:analytical_scheme}
\end{figure}

Four field families provide complementary tests of the reconstruction.
Constant fields test, as shown in Fig.~\ref{fig:analytical_scheme}, the imposed reproduction condition, linear fields
probe coordinate-moment consistency, quadratic fields introduce
curvature, and trigonometric fields provide smooth nonpolynomial
variation. Their analytical definitions allow each realization to be
evaluated at arbitrary particle and quadrature locations.

We write the fields in the parent coordinates
$\boldsymbol{\xi}=(\xi_1,\ldots,\xi_d)$, where $d$ is the spatial
dimension. The notation $\boldsymbol{\sigma}_{\mathrm{raw}}^{*}$ denotes
an analytical field before amplitude normalization; the geometry and
realization indices are temporarily omitted. For the polynomial
families, $\mat A_0$, $\mat A_a$, and $\mat A_{ab}$ are
independently sampled symmetric $d\times d$ coefficient tensors. The
indices $a$ and $b$ identify the coordinate factors that each tensor
multiplies. Depending on the selected family, the field takes one of
the forms
\begin{equation}
\boldsymbol{\sigma}_{\mathrm{raw}}^{*}(\boldsymbol{\xi})
=
\begin{cases}
\mat A_0,
& \text{constant}, \\[6pt]
\displaystyle
\mat A_0+\sum_{a=1}^{d}\mat A_a\,\xi_a,
& \text{linear}, \\[6pt]
\displaystyle
\mat A_0+\sum_{a=1}^{d}\mat A_a\,\xi_a
+\sum_{1\leq a\leq b\leq d}\mat A_{ab}\,\xi_a\xi_b,
& \text{quadratic}.
\end{cases}
\label{eq:aiq-polynomial-fields}
\end{equation}
The linear family includes a constant term and hence spans affine
fields. The quadratic family contains all monomials of total degree
at most two, with each mixed product $\xi_a\xi_b$ included once.
Each coefficient tensor specifies the amplitudes of all stress
components for its spatial mode. Sampling its diagonal and
upper-triangular entries independently and reflecting the latter
across the diagonal gives independent raw coefficients for the normal
and shear components while preserving stress symmetry.

For the trigonometric family, the symmetric coefficient tensors
$\mat U_a$, $\mat V_a$, and $\mat Z_{ab}$ specify the
amplitudes of sine, cosine, and mixed modes, respectively. Let
$\mathcal E_d$ be the set of ordered coordinate pairs selected for
the mixed modes. The field is
\begin{equation}
\begin{aligned}
\boldsymbol{\sigma}_{\mathrm{raw}}^{*}(\boldsymbol{\xi})
  ={}&\mat A_0
  +\sum_{a=1}^{d}\left[
    \mat U_a\sin(\pi\xi_a)
    +\mat V_a\cos(\pi\xi_a)\right]
  \\[4pt]
  &+\sum_{(a,b)\in\mathcal E_d}
    \mat Z_{ab}\sin(\pi\xi_a)\cos(\pi\xi_b).
\end{aligned}
\label{eq:aiq-trigonometric-fields}
\end{equation}
These tensors follow the same independent sampling and symmetry
construction as the polynomial coefficients. In the two-dimensional
dataset, $\mathcal E_2=\{(1,2)\}$, giving six scalar modes:
$1$, $\sin(\pi\xi_1)$, $\cos(\pi\xi_1)$,
$\sin(\pi\xi_2)$, $\cos(\pi\xi_2)$, and
$\sin(\pi\xi_1)\cos(\pi\xi_2)$.
The final mode introduces joint variation in the two coordinates.
Specifying $\mathcal E_d$ completes the family definition in each
dimension.

For the two-dimensional full-cell dataset, the independent entries
of each coefficient tensor are drawn from a standard normal
distribution. A common scalar then sets the amplitude of the
complete field, preserving the relative amplitudes of its normal
and shear components and its spatial modes. Specifically, we define
$\mathcal X_{\mathrm{norm}}$ to be a set of uniform sampling grids on
$[-1,1]^2$, and let $\sigma_{\mathrm{amp}}>0$ be the target amplitude.
Restoring the indices $c$ and $s$, the scaling factor and the
normalized analytical field are
\begin{equation}
\begin{aligned}
\gamma_{cs} =\frac{\sigma_{\mathrm{amp}}}
  {\displaystyle\max_{\boldsymbol{\xi}\in\mathcal X_{\mathrm{norm}}}
    \left\|\boldsymbol{\sigma}_{\mathrm{raw},cs}^{*}
      (\boldsymbol{\xi})\right\|_{\mathrm{Frob}}},
\\
\boldsymbol{\sigma}_{cs}^{*}
  \left(\mathbf x_c+\frac{h}{2}\boldsymbol{\xi}\right)
  =\gamma_{cs}\,
    \boldsymbol{\sigma}_{\mathrm{raw},cs}^{*}(\boldsymbol{\xi}),
\end{aligned}
\label{eq:aiq-field-normalization}
\end{equation}
where $\vct x_c$ is the parent-cell center and $h$ its width.
The Frobenius norm sums the squares of all tensor entries before
taking the square root, so each independent off-diagonal component
contributes twice to the squared norm. The maximum field norm on
$\mathcal X_{\mathrm{norm}}$ is therefore $\sigma_{\mathrm{amp}}$
after scaling. Realizations with a numerically vanishing sampled
norm are resampled.

The sampled coefficient tensors and normalization factor are retained
for each realization, together with the dataset's mixed-mode
specification. This permits evaluation of the same analytical field
at additional quadrature points and supports comparisons between
quadrature orders with a fixed stress reference. The amplitude
normalization is part of offline data generation; during an MPM
calculation, the reconstruction acts directly on the physical
stresses carried by the particles.

\subsection{Learning objectives, optimization, and evaluation}
\label{sec:losses}

\subsubsection{Full-cell objective}
\label{sec:aiq-full-cell-objective}

The full-cell objective follows directly from the structure of the discretization. The assembled internal force is the quantity entering momentum balance and is therefore the primary target. Force matching alone, however, has a nontrivial nullspace: distinct query stresses can assemble to the same nodal force. We therefore add pointwise stress supervision to select physically meaningful reconstructions within that nullspace. Finally, exact constant reproduction is imposed analytically, while affine consistency is penalized because it controls the leading reconstruction error of a smooth stress field without requiring a potentially ill-conditioned hard affine solve on sparse particle sets. For cell geometry $c$ and stress realization
$s$, the analytical and reconstructed stresses at Gauss point $q$ are
represented by
$\mathbf{s}_{csq}^{*}=\operatorname{sv}(\boldsymbol{\sigma}_{csq}^{*})$
and
$\widehat{\mathbf{s}}_{csq}^{\mathrm F}
=\operatorname{sv}(\widehat{\boldsymbol{\sigma}}_{csq}^{\mathrm F})$.
The operator $\operatorname{sv}$ collects the $n_\sigma=d(d+1)/2$
independent components of a symmetric stress tensor in $d$ spatial
dimensions, counting each shear component once. Applying the prescribed full-cell quadrature to
these stresses gives the reference and predicted force vectors
$\mathbf{f}_{cs}^{G,\mathrm F}$ and
$\widehat{\mathbf{f}}_{cs}^{\mathrm F}$, respectively. Both stack the
cell's $d$ force components at each of its $N_{b,c}$ active nodes in
the same order.

To compare errors across stress realizations, the normalization uses
the root-mean-square value of the analytical stress components over the
$Q_{\mathrm F}$ quadrature points, bounded below by a positive stress
floor $\varsigma_{\min}$:
\begin{equation}
\varsigma_{cs}^{2}
=\max\left\{
\frac{1}{n_\sigma Q_{\mathrm F}}
\sum_{q=1}^{Q_{\mathrm F}}
\left\|\mathbf{s}_{csq}^{*}\right\|_{2}^{2},
\varsigma_{\min}^{2}
\right\}.
\label{eq:aiq-loss-stress-scale}
\end{equation}
For the two-dimensional training profile, a cell width $h$ gives the
characteristic force norm
$f_{cs}^{\mathrm{char}}=\sqrt{2N_{b,c}}\,h\varsigma_{cs}$,
measured per unit thickness. The corresponding dimensional scaling in
$d$ dimensions is $\varsigma_{cs}h^{d-1}$.

Force matching uses a relative error stabilized by
$r_0 f_{cs}^{\mathrm{char}}$, where the fixed, non-negative,
dimensionless ratio $r_0$ sets the floor for small reference-force
norms. The auxiliary stress term controls reconstruction errors that
force assembly may leave unresolved. With constant reproduction enforced
by the coefficient projection, the affine term promotes the reproduction
of linear stress fields by penalizing the coordinate defect
$\boldsymbol{\delta}_{cq}^{\mathrm F}$. This defect is the difference
between the coordinate reconstructed from particle coordinates and the
prescribed Gauss-point coordinate, both expressed in the reference cell.
The three contributions are
\begin{equation}
\begin{aligned}
\mathcal{L}_{f,cs}^{\mathrm F}
&=
\frac{
\left\|\widehat{\mathbf{f}}_{cs}^{\mathrm F}
-\mathbf{f}_{cs}^{G,\mathrm F}\right\|_{2}
}{
\sqrt{
\left\|\mathbf{f}_{cs}^{G,\mathrm F}\right\|_{2}^{2}
+\left(r_0 f_{cs}^{\mathrm{char}}\right)^{2}
}
},
\\[0.7em]
\mathcal{L}_{\sigma,cs}^{\mathrm F}
&=
\frac{
\displaystyle\sum_{q=1}^{Q_{\mathrm F}}
\left\|\widehat{\mathbf{s}}_{csq}^{\mathrm F}
-\mathbf{s}_{csq}^{*}\right\|_{2}^{2}
}{
n_\sigma Q_{\mathrm F}\varsigma_{cs}^{2}
},
\\[0.7em]
\mathcal{L}_{\mathrm{lin},c}^{\mathrm F}
&=
\frac{1}{dQ_{\mathrm F}}
\sum_{q=1}^{Q_{\mathrm F}}
\left\|\boldsymbol{\delta}_{cq}^{\mathrm F}\right\|_{2}^{2}.
\end{aligned}
\label{eq:aiq-full-cell-losses}
\end{equation}
The force term uses an unsquared norm, and the stress term is a
normalized mean squared error with equal weights for all quadrature
points and independent components. The implementation also bounds the
force-loss denominator below by the smallest positive representable
value.

With $S$ stress realizations per geometry, the objective averages these
contributions over the training geometry set
$\mathcal{D}_{\mathrm F}^{\mathrm{tr}}$, which contains
$N_{\mathrm F}^{\mathrm{tr}}$ geometries:
\begin{equation}
\mathcal{L}_{\mathrm F}
=\frac{1}{N_{\mathrm F}^{\mathrm{tr}}S}
\sum_{c\in\mathcal{D}_{\mathrm F}^{\mathrm{tr}}}
\sum_{s=1}^{S}
\left[
\mathcal{L}_{f,cs}^{\mathrm F}
+0.25\mathcal{L}_{\sigma,cs}^{\mathrm F}
+0.25\mathcal{L}_{\mathrm{lin},c}^{\mathrm F}
\right].
\label{eq:aiq-full-cell-objective}
\end{equation}
The affine defect depends only on geometry, so averaging over the same
$S$ realizations preserves its weight per geometry. The two auxiliary coefficients are fixed at $0.25$ after normalization so that stress and affine consistency regularize the force-oriented fit without competing with the primary force term on equal weight. These three loss
terms are supplemented by AdamW weight decay on the network parameters,
as described in Section~2.6.3.

The normalization floors are fixed before optimization. The stress
floor $\varsigma_{\min}$ is $10^{-8}$ times the global root-mean-square
value of the training-set stress components. The default $r_0$ is the
fifth percentile of
$\|\mathbf{f}_{cs}^{G,\mathrm F}\|_2/f_{cs}^{\mathrm{char}}$
over the training subset; an explicitly prescribed value replaces this
calibration and is recorded with the run. Both floors remain fixed
during validation and testing, while $\varsigma_{cs}$ and
$f_{cs}^{\mathrm{char}}$ are evaluated for each realization.

\subsubsection{Cut-parent objective}
\label{sec:cutloss}

For a cut cell, the network is evaluated on individual simplices, but the force being approximated belongs to the complete material portion of the parent. Denote its stacked reference and predicted forces by $\vct f_{cs}^{\Gauss,\cut}$ and $\widehat{\vct f}_{cs}^{\cut}$, and its query component vectors by $\vct s_{cs\ell q}^{*}$ and $\widehat{\vct s}_{cs\ell q}^{\cut}$. The parent-aggregated loss formulation is
\begin{equation}
 \begin{aligned}
 L_{f,cs}^{\cut}
 &=\frac{\norm{\widehat{\vct f}_{cs}^{\cut}
                        -\vct f_{cs}^{\Gauss,\cut}}_2^2}
         {\norm{\vct f_{cs}^{\Gauss,\cut}}_2^2+\varepsilon_{f,\cut}},\\
 L_{\sigma,cs}^{\cut}
 &=\frac{\displaystyle\sum_{\ell=1}^{N_{\simp,c}}\sum_{q=1}^{Q_{\simp}}
       W_{c\ell q}\norm{\widehat{\vct s}_{cs\ell q}^{\cut}
                                     -\vct s_{cs\ell q}^{*}}_2^2}
       {\displaystyle\sum_{\ell=1}^{N_{\simp,c}}\sum_{q=1}^{Q_{\simp}}
       W_{c\ell q}\norm{\vct s_{cs\ell q}^{*}}_2^2+\varepsilon_{\sigma,\cut}},\\
 L_{\mathrm{lin},c}^{\cut}
 &=\frac1{dN_{\simp,c}Q_{\simp}}
       \sum_{\ell=1}^{N_{\simp,c}}\sum_{q=1}^{Q_{\simp}}
                       \norm{\vct\delta_{c\ell q}^{\cut}}_2^2.
 \end{aligned}
 \label{eq:cutlosses}
\end{equation}
The positive denominator floors $\varepsilon_{f,\cut}$ and $\varepsilon_{\sigma,\cut}$ have the units of their respective squared-force and integrated squared-stress denominators. A code using nondimensional training data may store the same numerical epsilon in both places; that does not make these two dimensional quantities identical.

The cut objective retains the same hierarchy---assembled force first, stress and affine consistency as auxiliary constraints---but adapts the normalization to a parent composed of unequal-volume simplices. The force term is squared, and the stress term is weighted by physical quadrature volume, so the objective measures the complete parent contribution rather than treating every child or query as equally important. The weights prevent a very small simplex from receiving the same stress-error influence as a large one merely because both have $Q_{\simp}$ queries. Most importantly, the simplex forces are added before computing $L_{f,cs}^{\cut}$. If $\vct e_{cs\ell}^{f}$ denotes the nodal-force error contributed by child $\ell$, then
\begin{equation}
 \norm{\sum_\ell\vct e_{cs\ell}^{f}}_2^2
 =\sum_\ell\norm{\vct e_{cs\ell}^{f}}_2^2
  +2\sum_{\ell<m}(\vct e_{cs\ell}^{f})^{\T}\vct e_{csm}^{f}.
 \label{eq:crosschild}
\end{equation}
Independently normalized child-force losses would omit these cross terms and optimize a different quantity from the parent force assembled in the solver. They would also give more training weight to parents partitioned into more children.

Let $\mathcal D_{\cut}^{\mathrm{tr}}$ contain $N_{\cut}^{\mathrm{tr}}$ cut parents. With module-specific coefficients $\lambda_{f,\cut}$, $\lambda_{\sigma,\cut}$, and $\lambda_{\mathrm{lin},\cut}$, the objective is
\begin{equation}
 \mathcal L_{\cut}
 =\frac1{N_{\cut}^{\mathrm{tr}}S}
   \sum_{c\in\mathcal D_{\cut}^{\mathrm{tr}}}\sum_{s=1}^{S}
      \left[\lambda_{f,\cut}L_{f,cs}^{\cut}
      +\lambda_{\sigma,\cut}L_{\sigma,cs}^{\cut}
      +\lambda_{\mathrm{lin},\cut}L_{\mathrm{lin},c}^{\cut}\right].
 \label{eq:cutobjective}
\end{equation}
The coefficients and denominator floors are retained as part of the cut model specification; the full-cell values in Eq.~\eqref{eq:aiq-full-cell-objective} are not substituted for them. The common methodological choice is force-oriented supervision with auxiliary stress and affine terms, not a requirement that the two modules use numerically identical objectives.

\subsubsection{Optimization and validation-based selection}
\label{sec:optimization}

The full-cell profile uses AdamW~\cite{Loshchilov2019}. In both modules, differentiation passes through the coefficient projection, stress reconstruction, and fixed force assembly, so the force objective directly updates the geometry-to-coefficient map. A batch comprises complete geometries and their attached stress realizations. For cut cells, valid simplices may be flattened into a computational batch for the neural forward pass, but they are reassembled to their parents before loss evaluation. This preserves the statistical unit of the data while permitting batched inference.

Table~\ref{tab:optimization} gives the canonical two-dimensional full-cell architecture and optimization parameters. Encoder and decoder weights are shared over particles and queries, respectively, and the SiLU activation is $\operatorname{SiLU}(z)=z/(1+\exp(-z))$. The cut model uses an additional context encoder, but its final model-specific settings are recorded separately. No inference-time fitting, optimizer step, or constitutive recalibration is performed in the MPM solver.

\begin{table}[htbp]
\centering\small
\caption{Canonical two-dimensional full-cell training and model-selection parameters. These settings describe the specified full-cell profile, not defaults presumed to apply to every cut-cell or three-dimensional checkpoint.}
\label{tab:optimization}
\begin{tabularx}{\textwidth}{@{}lY@{}}
\toprule
Parameter & Value or definition\\
\midrule
Feature width $d_{\mathrm f}$ & 64\\
Particle encoder / pair decoder & Three hidden layers each; SiLU; linear output\\
Optimizer & AdamW\\
Maximum epochs / batch size & 300 / 32 geometries\\
Initial / minimum learning rate & $10^{-3}$ / $10^{-6}$\\
Parameter weight decay & $10^{-6}$\\
Global gradient-norm clipping & 5\\
Learning-rate reduction & Factor $0.5$ after 10 plateau epochs\\
Scheduler monitor & Validation mean of $L_{f,cs}^{\full}$\\
Checkpoint and early-stopping monitor & Validation mean unsquared relative force error\\
Early-stopping patience & 30 epochs; minimum improvement $10^{-8}$\\
Force floor calibration & Training-only fifth percentile\\
Stress floor ratio & $10^{-8}$ of training-set component RMS\\
Reported relative-force epsilon & $10^{-14}$ in normalized data units\\
Precision & Float64 for data, model, and force assembly\\
\bottomrule
\end{tabularx}
\end{table}

Validation is used to select a checkpoint, not to update its parameters. The selection metric is the mean unsquared relative parent-force error on the validation set, both for full cells and for cut cells. It is distinct from the stabilized full-cell training objective and from the squared cut-cell objective. After selection, the chosen parameters are reloaded and evaluated on the untouched test set. Reporting test performance from this frozen checkpoint keeps test data out of optimization, learning-rate scheduling, early stopping, and model selection.

\subsubsection{Force-error evaluation and reference quadrature}
\label{sec:aiq-held-out-errors}

Evaluation distinguishes agreement with the training quadrature target
from the accuracy of the resulting internal force. Training and
checkpoint selection use the prescribed Gauss rule for each model.
In the reported full-cell tests in both two and three dimensions, the selected model
retains its own stress reconstruction and force assembly, while the
reference force is computed independently by evaluating the same
analytical stress field directly at the G10 quadrature points on the
same cell. The resulting test error captures the combined
effects of stress reconstruction and the quadrature used by the model.

For cell geometry $c$ and analytical stress realization $s$, the vectors
$\widehat{\vct f}_{cs}$ and $\vct f_{cs}^{\mathrm{MPM}}$ collect
the cell contributions to all active nodes obtained with AI-Q and
standard MPM, respectively. Both methods use identical particle
positions, volumes, and stress samples. Their common reference
$\vct f_{cs}^{\mathrm{ref}}$ is evaluated with the model's training
quadrature rule for validation and with the independent G10 rule for
the reported full-cell tests in both dimensions. With a small positive
reporting floor
$\varepsilon_{\mathrm{rel}}$ in force units, fixed within each
evaluation configuration and shared across methods, the relative errors
are
\begin{equation}
\begin{aligned}
e_{f,cs}^{\mathrm{AI}}
&=
\frac{
\left\|\widehat{\vct f}_{cs}
-\vct f_{cs}^{\mathrm{ref}}\right\|_2
}{
\left\|\vct f_{cs}^{\mathrm{ref}}\right\|_2
+\varepsilon_{\mathrm{rel}}
},
\qquad
e_{f,cs}^{\mathrm{MPM}}
=
\frac{
\left\|\vct f_{cs}^{\mathrm{MPM}}
-\vct f_{cs}^{\mathrm{ref}}\right\|_2
}{
\left\|\vct f_{cs}^{\mathrm{ref}}\right\|_2
+\varepsilon_{\mathrm{rel}}
}.
\end{aligned}
\label{eq:aiq-test-force-errors}
\end{equation}
The same error definition is applied to the other reconstruction
baselines.

Full-cell training and stress reconstruction employ tensor-product
Gauss--Legendre quadrature, with G4 in two dimensions and G3 in three
dimensions; additional three-dimensional tests examine a G2
configuration. Internal forces assembled with these rules are assessed
against an independent G10 reference computed directly from the same
analytical stress field. To assess the accuracy of the reference force, quadrature convergence is checked in both two and three dimensions by reevaluating the same
analytical stress fields at successively higher Gauss orders up to G10,
with the field parameters and cell geometry held fixed. The observed
stabilization of the assembled internal forces supports the use of G10
as the common reference for the full-cell comparisons in both dimensions.

For the two-dimensional cut-cell tests,
$\vct f_{cs}^{\mathrm{ref}}$ is the prescribed G3 composite-triangle
force on the reconstructed material region. Contributions from all
child triangles are assembled into the complete parent force before
evaluating the relative error. All methods use the same reference
within each comparison.

\subsection{Incorporation into the explicit APIC--MPM solver}
\label{sec:solver}

AI-Q is incorporated into the explicit APIC--MPM solver to improve
internal-force integration from the available particle stresses. At
each time step, the current particle positions determine the
reconstruction spans, and the resulting cell forces drive the grid
update. Constitutive stresses and history evolve on the material points.

The supplied boundary labels remain attached to particle identities
and determine reconstruction eligibility. Spans containing labeled
boundary particles are considered for cut-cell reconstruction; those
without them are candidates for the full-cell model. Cut quadrature is
applied when its module is enabled and the reconstructed material
geometry is admissible. Selection uses the complete parent particle
set and checks the model's supported particle-count range.
Incompatible checkpoints and nonfinite full-cell predictions are
reported as inference errors.

These choices define a complete, disjoint partition of the occupied
spans: $\mathcal C_{\mathrm F}$ contains the parents assigned to
full-cell reconstruction, $\mathcal C_{\mathrm C}$ those assigned to
cut-cell reconstruction, and $\mathcal C_{\mathrm P}$ those retaining
particle quadrature. The reconstructed full-cell and cut-parent
contributions, $\widehat{\vct f}_{ic}^{\mathrm F}$ and
$\widehat{\vct f}_{ic}^{\mathrm C}$, are assembled with the particle
forces to give the internal force at grid node $i$:
\begin{equation}
\vct f_i^{\mathrm{AI-Q}}
=
\sum_{c\in\mathcal C_{\mathrm F}}
\widehat{\vct f}_{ic}^{\mathrm F}
+
\sum_{c\in\mathcal C_{\mathrm C}}
\widehat{\vct f}_{ic}^{\mathrm C}
-
\sum_{c\in\mathcal C_{\mathrm P}}
\sum_{p\in\mathcal P_c}
V_p\boldsymbol{\sigma}_p\nabla N_i(\vct x_p).
\label{eq:aiq-assembled-internal-force}
\end{equation}
Here $\mathcal P_c$ contains all particles assigned to span $c$.
The hatted forces include the internal-force sign, and the cut-parent
contribution includes all child simplices. Applying the quadrature
choice to the complete parent ensures that each span contributes once.

All particles otherwise follow the baseline APIC transfer and explicit update \cite{Jiang2015,Jiang2017}. AI-Q changes only the internal-force integration in Eq.~\eqref{eq:aiq-assembled-internal-force}: grid mass and momentum are still transferred from particles, the assembled forces update the grid velocity, and the updated grid velocity provides particle motion, velocity gradients, deformation gradients, volumes, and constitutive updates. Reconstructed Gauss-point stresses are transient integration quantities; constitutive state remains on the material points. The full- and cut-cell network parameters are loaded once, while their geometry-conditioned reconstruction matrices are reevaluated as particles move.

For a fixed network width, decoding the full-cell particle--quadrature
pairs requires
$O(P_cQ_{\mathrm F})$ work, and applying the resulting coefficients to
the stresses requires $O(P_cQ_{\mathrm F}n_\sigma)$. Here $P_c$ is the
particle count, $Q_{\mathrm F}$ the number of full-cell quadrature
points, and $n_\sigma$ the number of independent stress components.
For cut parents, the corresponding total quadrature count is
$N_{S,c}Q_S$, with $N_{S,c}$ child simplices and $Q_S$ points per simplex.
These reconstruction costs enter the computational assessment alongside
the force accuracy achieved at the prescribed particle resolution.

All three integration modes assemble forces through the same background
basis, retaining its required support; any positive-mass screening is
specified by the solver branch. With complete basis support, partition
of unity and linear precision give zero internal resultant and torque
for symmetric stresses, as established in~\ref{app:error}. Energy behavior
also depends on the stress-power coupling and time integration, and the
admissible time step is determined for the coupled explicit scheme.

%% file: single_cell_tests.tex
% SECTION 3: SINGLE-CELL TESTS
% Insert in the existing manuscript with \input{single_cell_tests}.
% All body text and all figure interiors are intentionally blank.
% Captions follow the agreed five-figure plan; panel details are editable.
%
% EASY FIGURE SETTINGS:
% Change the heights below to resize the empty frames.
% To insert a finished figure, replace its \SCfigureplaceholder{...} line with
% the commented \includegraphics line immediately below it.
\providecommand{\SCtrainingheight}{70mm}
\providecommand{\SCperformanceheight}{65mm}
\providecommand{\SCcomparisonheight}{65mm}
\providecommand{\SCcutperformanceheight}{90mm}
\providecommand{\SCpreviewbreak}{}
\providecommand{\SCfigureplaceholder}[1]{%
  \begingroup
  \setlength{\fboxsep}{0pt}%
  \fbox{%
    \rule{0pt}{#1}%
    \rule{\dimexpr\linewidth-2\fboxrule\relax}{0pt}%
  }%
  \endgroup
}

\section{Single-cell tests}
\label{sec:single-cell-tests}

% Write the section introduction here.

\subsection{Training and convergence}
\label{sec:single-cell-training}

The numerical study evaluates two-dimensional full- and cut-cell models and three-dimensional full-cell models. Table~\ref{tab:scope} separates these implemented scopes from the dimension-independent formulation. In two dimensions the integration children are triangles. No three-dimensional cut-cell result or unstructured-mesh result is implied by the general equations in Sections~\ref{sec:objective}--\ref{sec:cut}.

\begin{table}[htbp]
\centering\small
\caption{Model families and particle-count envelopes in the study. G$g$ denotes a one-dimensional tensor-product order, not a total point count or an exactness degree. A separate model is associated with each basis--rule pair.}
\label{tab:scope}
\begin{tabularx}{\textwidth}{@{}lYYY@{}}
\toprule
Quantity & 2D full cell & 2D cut cell & 3D full cell\\
\midrule
Basis degree $k$ & $1,2,3$ & $1,2,3$ & $1,2,3$\\
Particle count $P_c$ & Every integer $4$--$36$ & Every integer $1$--$36$ per parent & Every integer $4$--$36$\\
Rule choices & G4, G6 & G2, G3 per triangle & G2, G3\\
Queries & $Q_{\full}=16,36$ & $Q_{\simp}=4,9$ per triangle & $Q_{\full}=8,27$\\
Active parent nodes & $(k+1)^2$ & $(k+1)^2$ & $(k+1)^3$\\
State component order & $11,22,12$ & $11,22,12$ & $11,22,33,$\newline $12,13,23$\\
\bottomrule
\end{tabularx}
\end{table}

The full-cell geometries are balanced by the actual number of particles, not by a particles-per-axis parameter. The two-dimensional profile has 500 geometries at each of the 33 counts from 4 to 36. It combines five layout families with the proportions in Table~\ref{tab:layouts}. Stratified samples provide broadly distributed coverage, random gaps vary local occupancy, affine distortions alter alignment, clustered samples concentrate coverage, and boundary-crossing layouts mimic changed occupancy after transport. The last name describes a particle-placement pattern in a full material cell; it is not a cut-cell geometry.

\begin{table}[htbp]
\centering\small
\caption{Two-dimensional full-cell geometry mixture. Counts are per particle cardinality in the 500-geometry profile.}
\label{tab:layouts}
\begin{tabularx}{\textwidth}{@{}lrrY@{}}
\toprule
Layout & Fraction & Count & Geometric variation\\
\midrule
Stratified & $0.45$ & 225 & Quasi-uniform, coverage-preserving placement\\
Random gap & $0.15$ & 75 & Removal from an oversampled candidate set\\
Affine deformed & $0.15$ & 75 & Bounded stretching and shear\\
Clustered & $0.15$ & 75 & Correlated contraction and displacement\\
Boundary crossing & $0.10$ & 50 & Shifted occupancy with periodic wrapping\\
\bottomrule
\end{tabularx}
\end{table}

The position jitter is 0.30 of the local sampling-stratum width, while additional volume and particle-domain descriptor jitter are both zero. Thus the full-cell training volumes satisfy $V_p=\abs{K_c}/P_c$. The network remains capable of receiving relative-volume features, but the absence of volume jitter must not be interpreted as training over arbitrary volume distributions. The bounded layout parameters are given in Supplementary Material, Section S4 so that the distribution is reproducible without interrupting the formulation.

Each full-cell geometry is paired with 12 stress realizations, three from each family. The split is made by geometry and stratified by $P_c$, giving the counts in Table~\ref{tab:datacounts}. All realizations of one geometry remain in the same subset. The same rule applies to a cut parent: all of its simplices and all attached stresses belong to one subset. Otherwise a test realization could reuse a particle cloud or a parent geometry already observed in training, despite having different stress coefficients.

\begin{table}[htbp]
\centering\small
\caption{Two-dimensional full-cell data and evaluation profile. Counts are per basis--rule dataset; the shared geometry/field bank prevents basis and order comparisons from being confounded by different samples.}
\label{tab:datacounts}
\begin{tabularx}{\textwidth}{@{}lY@{}}
\toprule
Parameter & Specification\\
\midrule
Reference parent & Unit square; neural coordinates in $[-1,1]^2$\\
Geometries & $33\times500=16{,}500$\\
Realizations per geometry & $S=12$; three per analytical family\\
Total geometry--stress pairs & $198{,}000$\\
Training / validation / test & $80\%/10\%/10\%$ by geometry\\
Geometries per count and split & $400/50/50$\\
Total geometries per split & $13{,}200/1{,}650/1{,}650$\\
Stress pairs per split & $158{,}400/19{,}800/19{,}800$\\
Amplitude / coefficient sampling & $\sigma_{\mathrm{amp}}=1$; independent normal coefficients\\
Position / volume / domain jitter & $0.30/0/0$\\
Geometry-bank seed & 2026\\
Nominal model-training seed & 2027; each run retains its own seed\\
Arithmetic & Double precision\\
\bottomrule
\end{tabularx}
\end{table}

Paired comparisons across basis families and full-cell Gauss orders use matching particle geometries, field coefficients, and splits. The cut-cell bank additionally varies its represented material fraction and interface orientation, and includes the sparse parent counts $P_c=1,2,3$ rather than excluding them through a simplex-local particle requirement. Dataset sizes and sampling ranges are module-specific: the counts in Table~\ref{tab:datacounts} specify the two-dimensional full-cell profile and are not transferred to the cut or three-dimensional banks by assumption. In all cases, the saved geometry, field coefficients, and split identities define the samples used for training and independent evaluation.

The selected two-dimensional checkpoints achieve mean validation force errors of approximately 5.0\%, 5.7\%, and 4.9\% for the linear, quadratic B-spline, and cubic B-spline full-cell models, and 11.7\%, 12.1\%, and 11.4\% for the corresponding cut-cell models. Training and validation curves, including checkpoint locations, are provided in the Supplementary Material. The larger cut-cell residuals are consistent with partial occupancy and the inclusion of very sparse parent cells.

\subsection{Full-cell performance}
\label{sec:single-cell-full}

\subsubsection{Reconstruction accuracy}
\label{sec:single-cell-full-accuracy}

AI-Q improves steadily as the particle cloud becomes better populated. Across the three basis families, the median force error decreases from approximately 14.2--18.6\% at $P_c=4$ to 0.62--0.84\% at $P_c=36$, while P95 decreases from 84.7--92.8\% to 4.3--5.0\%. Constant fields are reproduced to numerical precision by construction; errors increase with stress variation and are largest for the trigonometric family. The complete particle-count and stress-family breakdown is provided in the Supplementary Material.

\subsubsection{Comparison with classical reconstruction methods}
\label{sec:single-cell-full-comparison}

\begin{figure}[!htbp]
  \centering
  \includegraphics[width=\linewidth]{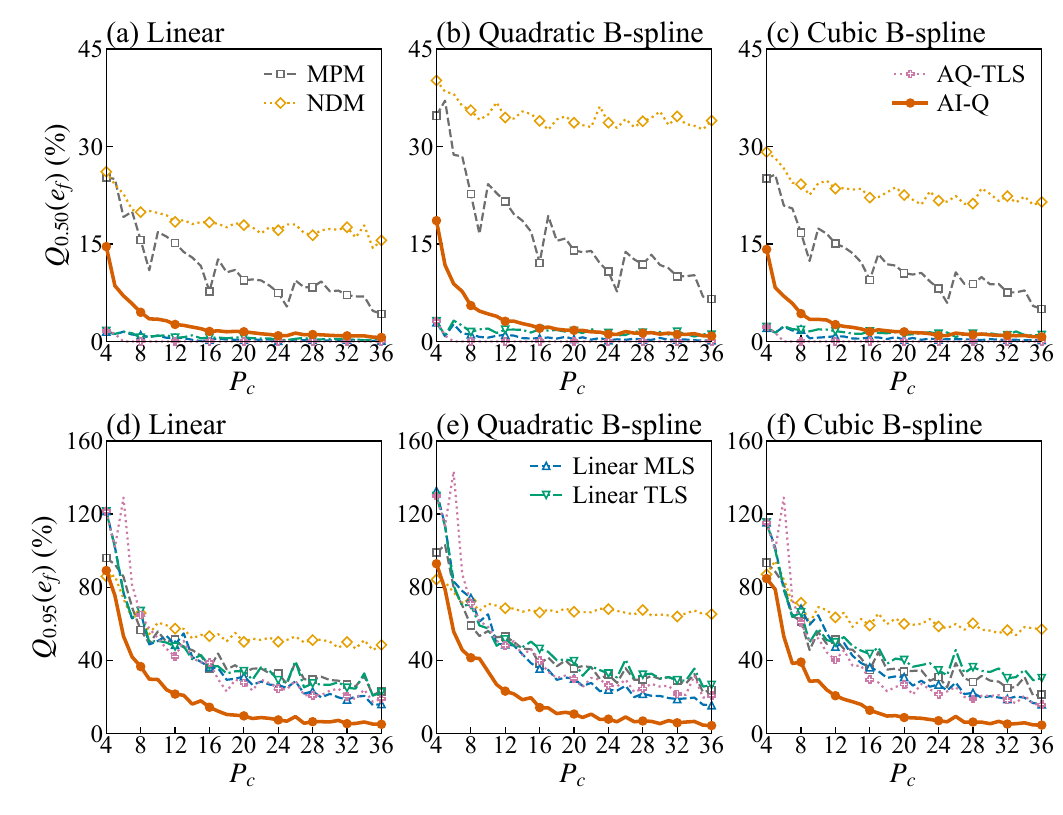}
  \caption{Comparison of full-cell internal-force reconstruction methods.
  Columns correspond, from left to right, to linear, quadratic B-spline,
  and cubic B-spline shape functions. The median (a--c) and 95th
  percentile (d--f) of the relative error are plotted against the actual
  particle count, $P_c=4,\ldots,36$. The methods are standard MPM,
  normalized double mapping (DM), linear MLS, linear TLS, adaptive
  quadratic TLS, and AI-Q; the TLS methods use the implemented
  two-dimensional extensions. All methods use the same test cases,
  pooling constant, linear, quadratic, and trigonometric stress fields
  at each particle count. Errors are expressed as percentages relative
  to the independent G10 quadrature reference.}
  \label{fig:single-cell-full-comparison}
\end{figure}

% Write the full-cell method comparison here.
% Figure reference: Fig.~\ref{fig:single-cell-full-comparison}.
Figure~\ref{fig:single-cell-full-comparison}(a--c) shows that AI-Q consistently reduces the median error relative to standard MPM and normalized double mapping. Linear MLS and adaptive quadratic TLS achieve lower median errors than AI-Q throughout the tested particle-count range, while linear TLS also generally gives low medians. These results reflect the polynomial-reproduction properties of the classical reconstruction methods. In particular, once quadratic reconstruction is activated at ($P_c\geq6$), adaptive quadratic TLS reproduces the constant, linear, and quadratic stress fields to numerical precision. Since these fields constitute the majority of the benchmark cases, its median error becomes nearly zero.

The P95 curves in Fig.~\ref{fig:single-cell-full-comparison}(d--f) provide a complementary assessment. AI-Q achieves the lowest P95 among all considered methods for every ($P_c\geq5$), across all three shape-function families. At ($P_c=4$), normalized double mapping gives slightly lower P95 values for the linear and quadratic B-spline bases, while AI-Q remains lowest for the cubic B-spline basis. By ($P_c=36$), AI-Q reduces P95 to approximately 4.3–5.0\%, compared with 15.5–30.3\% for the MLS and TLS variants. This corresponds to a reduction of approximately 69–72\% relative to the best competing method at that particle count.

The distinction between median accuracy and the upper error tail is particularly evident for adaptive quadratic TLS at ($P_c=6$). Its nearly zero median coexists with P95 values of approximately 129–143\%, whereas AI-Q gives P95 values of approximately 53–56\%. The polynomial cases benefit from exact reproduction, while the remaining nonpolynomial cases can still incur large errors under sparse sampling. Thus, the comparison identifies complementary strengths: polynomial reconstruction provides superior accuracy for stress fields represented by its basis, while AI-Q substantially reduces large internal-force errors across the mixed stress cases considered here.

\subsection{Cut-cell performance}
\label{sec:single-cell-cut}

\subsubsection{Reconstruction accuracy}
\label{sec:single-cell-cut-accuracy}

The cut-cell results expose the information limit of sparse reconstruction. At $P_c=1$, median errors are approximately 45--50\% and P95 is 186--203\%; by $P_c=36$ these decrease to 2.5--3.1\% and 13.8--16.0\%, respectively. Constant fields remain exact, while trigonometric fields are the most difficult. At fixed particle count, error also tends to increase with represented material fraction because the same observations must describe a larger occupied region. Detailed particle-count, stress-family, and material-fraction plots are given in the Supplementary Material.

\subsubsection{Comparison with classical reconstruction methods}
\label{sec:single-cell-cut-comparison}

\begin{figure}[!htbp]
  \centering
  \includegraphics[width=\linewidth]{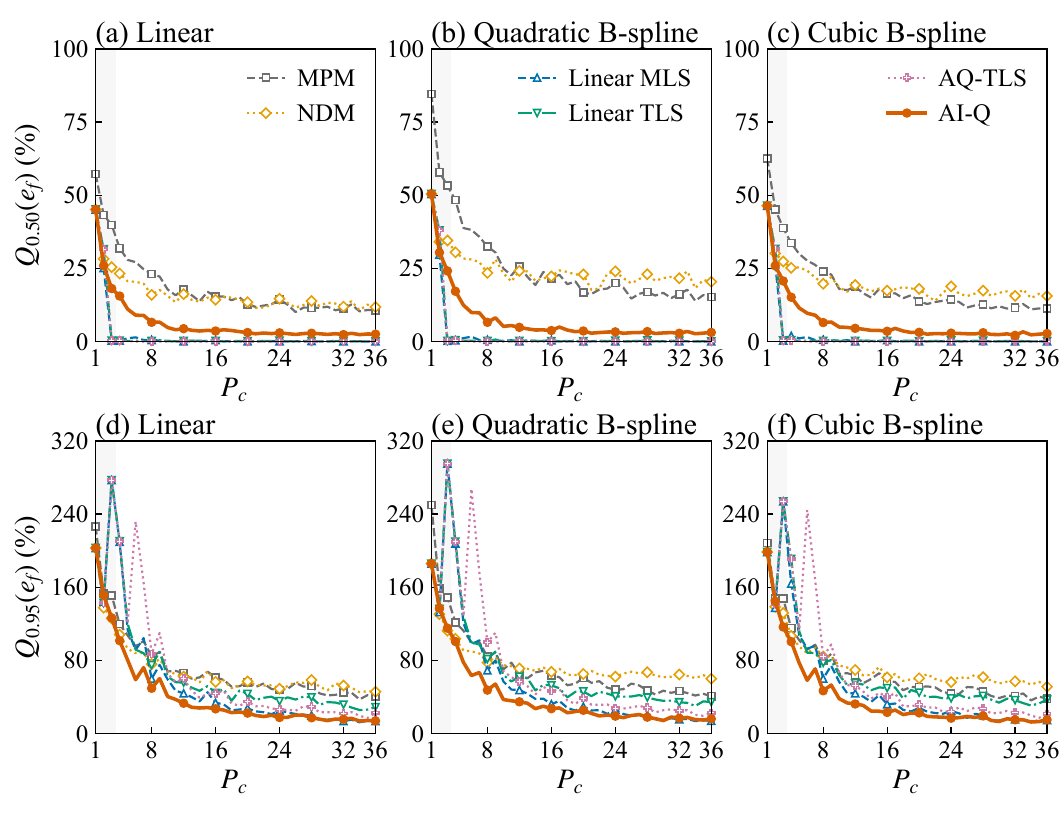}
  \caption{Comparison of cut-cell internal-force reconstruction methods.
  Columns correspond, from left to right, to linear, quadratic B-spline,
  and cubic B-spline shape functions. The median (a--c) and 95th
  percentile (d--f) of the relative error are plotted against the actual
  number of particles per parent cut cell, $P_c=1,\ldots,36$.
  The methods are standard MPM, normalized double mapping (DM), linear
  MLS, linear TLS, adaptive quadratic TLS, and AI-Q; the TLS methods use
  the implemented two-dimensional extensions. All methods use the same
  parent-cell test cases, pooling constant, linear, quadratic, and
  trigonometric stress fields at each particle count. Errors are
  expressed as percentages and evaluated after complete parent-cell
  force assembly against the G3 composite-triangle quadrature target.}
  \label{fig:single-cell-cut-comparison}
\end{figure}

% Write the cut-cell method comparison here.
% Figure reference: Fig.~\ref{fig:single-cell-cut-comparison}.
Figure~\ref{fig:single-cell-cut-comparison}(a--c) shows that AI-Q reduces the median error relative to standard MPM throughout the tested range and relative to normalized double mapping for ($P_c\geq2$). The MLS and TLS reconstructions achieve lower medians than AI-Q for ($P_c\geq3$), with adaptive quadratic TLS approaching numerical precision once quadratic reconstruction becomes available. As in the full-cell tests, these low medians reflect the polynomial-reproduction properties of the classical methods and the substantial proportion of representable polynomial fields in the benchmark.

The P95 curves in Fig.~\ref{fig:single-cell-cut-comparison}(d--f) reveal pronounced differences under sparse sampling. At ($P_c=3$), the classical reconstruction methods exhibit P95 values of approximately 254–295\%, compared with 115–126\% for AI-Q. Adaptive quadratic TLS develops a second peak at ($P_c=6$), reaching approximately 232–268\%, while AI-Q gives approximately 58–63\%. These peaks occur at the minimum particle counts associated with two-dimensional linear and quadratic fitting, respectively, and are consistent with increased sensitivity to particle configuration and stress variation under minimally sampled conditions. The coexistence of very low medians and large P95 values illustrates how accurate reproduction of polynomial cases can accompany substantial errors in the remaining cases.

Across all three shape-function families, AI-Q achieves the lowest P95 among the considered methods throughout ($4\leq P_c\leq25$). It also retains lower P95 values than MPM, normalized double mapping, and both TLS variants at every ($P_c\geq4$). Linear MLS becomes increasingly competitive at higher particle counts and gives lower P95 values at several of these counts. At ($P_c=36$), its P95 is approximately 13.5–14.6\%, compared with 13.8–16.0\% for AI-Q. The cut-cell comparison therefore shows that the relative advantage depends on particle count: AI-Q provides particularly consistent reductions in large force errors at moderate particle counts, while linear MLS becomes competitive as the particle cloud becomes denser.

\subsection{Overall assessment and discussion}

Table~\ref{tab:single_cell_overall} shows that AI-Q achieves the lowest mean and P95 errors in all six combinations of cell type and shape function. Relative to standard MPM, the mean error is reduced by approximately 68–71\% for full cells and 54–62\% for cut cells. The improvement also extends to comparisons with the classical reconstruction methods: relative to the lowest P95 among the other methods in each configuration, AI-Q reduces this statistic by approximately 41–46\% for full cells and 18–22\% for cut cells. These consistent improvements indicate that the learned reconstruction reduces both average force errors and the magnitude of higher-percentile errors across the evaluated test populations.

The classical polynomial reconstructions retain a clear advantage in median accuracy. Adaptive quadratic TLS gives median errors below the reporting threshold of ($10^{-10}\%$) in every configuration, reflecting its polynomial-reproduction properties and the substantial proportion of polynomial stress fields in the benchmark. Its larger mean and P95 values nevertheless show that this exactness coexists with substantial errors in the remaining cases. AI-Q exhibits finite median errors but lower mean and P95 values, providing a different balance between accuracy on readily representable fields and performance across the mixed stress population.

\begin{table*}[t]
\centering
\begin{threeparttable}
\caption{Overall single-cell force reconstruction accuracy.}
\label{tab:single_cell_overall}
\fontsize{9.5}{11.5}\selectfont
\setlength{\tabcolsep}{3.5pt}
\renewcommand{\arraystretch}{1.12}
\begin{tabular*}{\textwidth}{@{\extracolsep{\fill}}lrrrrrr@{}}
\toprule
& \multicolumn{3}{c}{Full cells: $P_c=4$--$36$} & \multicolumn{3}{c}{Cut cells: $P_c=1$--$36$} \\
\cmidrule(lr){2-4}\cmidrule(l){5-7}
Method & Mean (\%) & Median (\%) & P95 (\%) & Mean (\%) & Median (\%) & P95 (\%) \\
\midrule
\multicolumn{7}{@{}l}{\textit{Linear shape functions}} \\
\addlinespace[2pt]
MPM & 15.50 & 10.41 & 46.48 & 25.25 & 15.82 & 77.24 \\
Normalized DM & 20.87 & 18.18 & 55.63 & 22.24 & 14.72 & 70.72 \\
Linear MLS & 8.91 & 0.06 & 40.54 & 12.46 & 0.33 & 60.64 \\
Linear TLS & 10.00 & 0.19 & 44.48 & 15.55 & 0.85 & 68.86 \\
Adaptive quadratic TLS & 7.93 & {\boldmath$<10^{-10}$} & 41.91 & 16.09 & {\boldmath$<10^{-10}$} & 70.39 \\
AI-Q (proposed) & \textbf{5.03} & 1.49 & \textbf{21.87} & \textbf{11.57} & 3.76 & \textbf{49.91} \\
\addlinespace[4pt]
\multicolumn{7}{@{}l}{\textit{Quadratic B-spline shape functions}} \\
\addlinespace[2pt]
MPM & 19.69 & 15.20 & 49.97 & 30.72 & 22.10 & 85.55 \\
Normalized DM & 32.32 & 34.45 & 68.40 & 28.21 & 23.34 & 76.27 \\
Linear MLS & 9.38 & 0.13 & 43.67 & 12.71 & 0.38 & 64.49 \\
Linear TLS & 12.33 & 0.75 & 49.58 & 16.62 & 1.04 & 73.08 \\
Adaptive quadratic TLS & 9.09 & {\boldmath$<10^{-10}$} & 47.29 & 16.19 & {\boldmath$<10^{-10}$} & 74.99 \\
AI-Q (proposed) & \textbf{5.72} & 1.89 & \textbf{25.74} & \textbf{11.77} & 4.19 & \textbf{50.53} \\
\addlinespace[4pt]
\multicolumn{7}{@{}l}{\textit{Cubic B-spline shape functions}} \\
\addlinespace[2pt]
MPM & 16.08 & 11.35 & 45.63 & 25.69 & 16.83 & 75.25 \\
Normalized DM & 24.75 & 22.85 & 63.12 & 25.07 & 18.38 & 76.21 \\
Linear MLS & 9.14 & 0.17 & 41.55 & 12.02 & 0.38 & 59.74 \\
Linear TLS & 11.76 & 0.68 & 47.43 & 15.85 & 1.07 & 68.67 \\
Adaptive quadratic TLS & 7.45 & {\boldmath$<10^{-10}$} & 38.89 & 15.11 & {\boldmath$<10^{-10}$} & 68.09 \\
AI-Q (proposed) & \textbf{4.91} & 1.57 & \textbf{20.90} & \textbf{11.26} & 3.81 & \textbf{47.95} \\
\bottomrule
\end{tabular*}
\begin{tablenotes}[flushleft]
\fontsize{8.5}{10.3}\selectfont
\item \textit{Test populations and references.} Each method uses 19,800 full-cell cases (1,650 geometries) or 21,600 cut-cell cases (1,800 geometries). Each geometry has 12 stress cases, equally divided among constant, linear, quadratic, and trigonometric fields; particle-count groups are equally populated. Full-cell G4 reconstruction is assessed against independent G10 quadrature. Cut-cell errors use the complete parent-cell G3 composite-triangle quadrature target.
\item \textit{Accuracy.} Mean, median, and P95 (95th percentile) pool individual case errors with equal weights. All errors are percentages. Bold indicates the lowest error within each cell type and shape function. Values below $10^{-10}\%$ are displayed as $<10^{-10}$. Normalized DM denotes normalized double mapping.
\end{tablenotes}
\end{threeparttable}
\end{table*}

The advantage over the strongest classical alternative is more moderate for cut cells. Compared with linear MLS, AI-Q reduces the pooled mean error by approximately 6–7\%, while the reduction in P95 is approximately 18–22\%. The larger improvement in P95 identifies the reduction of high-percentile errors as a particularly relevant benefit in this setting. This aggregate result complements the particle-count-dependent comparisons in Section 3.3, where linear MLS becomes competitive in more densely populated cut cells.

The full-cell and cut-cell statistics should also be interpreted in the context of their respective test populations. The cut-cell set includes the additional sparse cases ($P_c=1,2,3$). Furthermore, full-cell errors are evaluated against independent G10 quadrature, whereas cut-cell errors use the parent-cell G3 composite-triangle target. The comparisons between methods are therefore made under matched conditions within each population.

These single-cell results establish the effectiveness of AI-Q in reducing average and high-percentile internal-force errors across the tested configurations, while identifying polynomial reproduction as an enduring strength of the classical methods. The subsequent MPM examples examine how these local improvements affect the accuracy of the evolving numerical solution.

\subsection{Three-dimensional full-cell performance}
\label{sec:single_cell_3d_full}

The three-dimensional full-cell tests consider linear, quadratic B-spline,
and cubic B-spline shape functions, with the actual particle count ranging
from $P_c=4$ to $36$. The G3 tensor-product rule provides $Q_F=3^3=27$
quadrature points per cell. For exact polynomial stresses of total degree
at most two, this rule is sufficient to integrate their product with the
background basis gradients for all three shape-function families. The
assessment uses the relative internal-force error $e_f$ and the same
constant, linear, quadratic, and trigonometric stress-family classification
as the two-dimensional full-cell tests.

The selected G3 checkpoints have mean validation force errors of approximately 13.1\%, 14.5\%, and 12.2\% for the linear, quadratic B-spline, and cubic B-spline models, respectively. The corresponding training and validation histories are provided in the Supplementary Material; the test results below use the independent higher-order reference defined in Section~\ref{sec:aiq-held-out-errors}.

\subsubsection{Reconstruction accuracy}
\label{sec:single_cell_3d_accuracy}

The same trend persists in three dimensions. Across the three bases, the median error decreases from approximately 29--37\% at $P_c=4$ to 4--5\% at $P_c=36$, while P95 decreases from approximately 100--106\% to 18--21\%. Constant fields are again reproduced to numerical precision, and trigonometric fields remain the most difficult. The complete AI-Q-only breakdown is reported in the Supplementary Material.

\subsubsection{Comparison with classical reconstruction methods}
\label{sec:single_cell_3d_comparison}

\begin{figure}[!htbp]
  \centering
  \includegraphics[width=\linewidth]{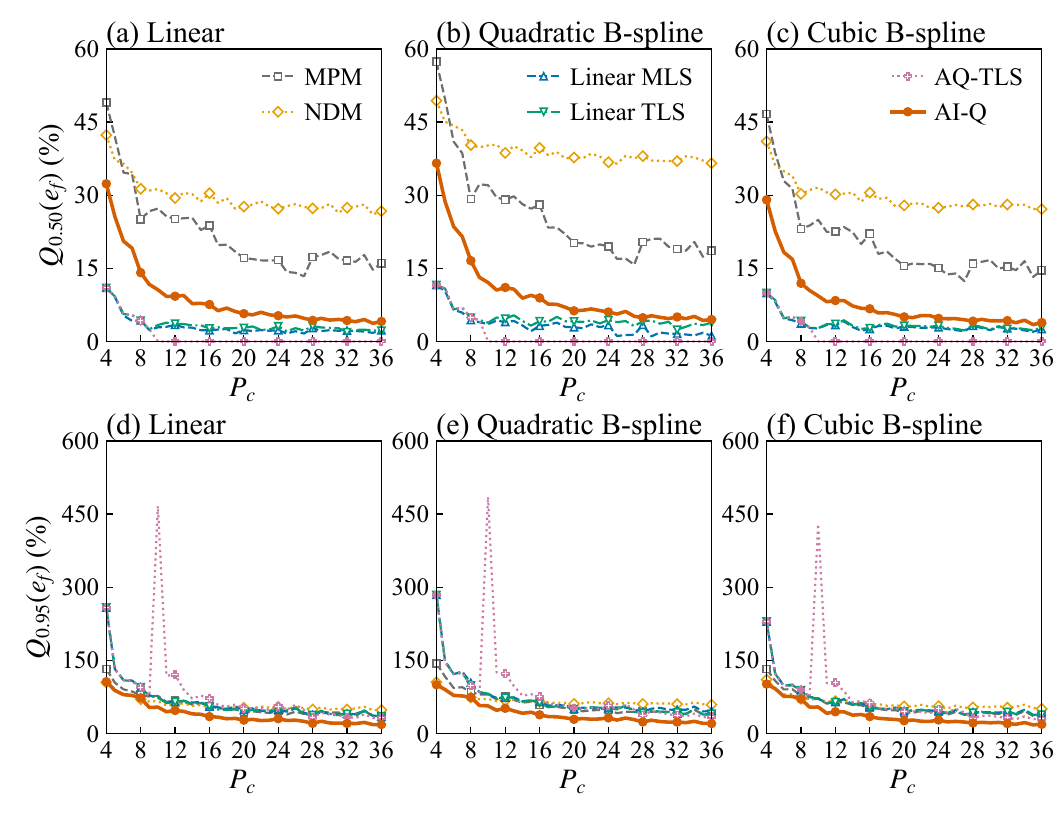}
  \caption{Comparison of three-dimensional full-cell internal-force
  reconstruction methods. AI-Q uses G3 quadrature. Columns
  correspond, from left to right, to linear, quadratic B-spline, and cubic
  B-spline shape functions. The median, $Q_{0.50}(e_f)$, in (a--c) and
  the 95th percentile, $Q_{0.95}(e_f)$, in (d--f) are plotted against the
  actual particle count, $P_c=4,\ldots,36$. The methods are standard MPM,
  normalized double mapping (NDM), linear MLS, linear TLS, adaptive
  quadratic TLS (AQ-TLS), and AI-Q. The methods are compared on the same
  test cases, pooling constant, linear, quadratic, and trigonometric
  stress fields at each particle count. Errors are expressed as
  percentages relative to the common independent higher-order
  quadrature reference.}
  \label{fig:single_cell_3d_comparison}
\end{figure}

Figure~\ref{fig:single_cell_3d_comparison}(a--c) shows that AI-Q reduces
the median error relative to standard MPM and normalized double mapping
throughout the tested particle-count range. At $P_c=36$, the median is
approximately $4$--$5\%$ for AI-Q, compared with $15$--$19\%$ for standard MPM
and $27$--$37\%$ for normalized double mapping. The MLS and TLS variants
generally achieve lower medians than AI-Q. In particular, the median of
adaptive quadratic TLS approaches zero for $P_c\gtrsim10$, consistent
with its polynomial-reproduction properties and the inclusion of constant,
linear, and quadratic stress fields in the benchmark.

The P95 curves in Fig.~\ref{fig:single_cell_3d_comparison}(d--f) provide
a complementary assessment. AI-Q gives the lowest P95 values at moderate
and large particle counts across the three shape-function families. At
$P_c=36$, its P95 is approximately $18$--$21\%$, compared with
$36$--$38\%$ for standard MPM and $30$--$35\%$ for adaptive quadratic
TLS. The reduction in higher-percentile errors therefore persists in
three dimensions, while the classical polynomial reconstructions retain
their advantage in median accuracy.

The distinction is particularly evident near $P_c=10$, where adaptive
quadratic TLS combines a nearly zero median with P95 values exceeding
$400\%$. A complete quadratic polynomial in three dimensions has ten
coefficients, placing this peak near the minimum particle count required
for such a reconstruction. This behaviour is consistent with the
sensitivity of sparse quadratic fitting when the stress field contains
variation outside the polynomial space. AI-Q shows no comparable peak.
As in the two-dimensional full-cell tests, the comparison identifies
complementary strengths: polynomial reconstruction is highly accurate
for fields represented by its basis, whereas AI-Q reduces large
internal-force errors across the mixed stress population.

% Write the overall assessment and discussion here.
% The previously prepared summary table can be inserted here when desired.

%% file: numerical_results.tex
% Numerical-results section for insertion into the main manuscript.
% Required packages: graphicx, amsmath, placeins.
% For a two-column journal layout, the figure* environments span both columns.
% Keep the supplied figures/ directory beside the manuscript, or redefine
% \resultsfigdir before \input{numerical_results}.
%
% AUTHOR CHECKS (not printed):
% 1. The uploaded maps are labelled "Absolute error [%]" but do not specify
%    their denominator. Confirm the definition in the generating scripts and
%    state it in the manuscript's error-measure subsection. No denominator
%    is inferred here from an earlier plotting recommendation.
% 2. ellipse_convergence.pdf labels its coarsest log10(h) coordinate -1.48,
%    whereas the earlier folder list contained h0.04. Check the actual h.
% 3. Confirm the beam section coordinate, the ring sampling angle/sector,
%    and the definition of L_c in the ellipse map against the final scripts.
%    The discussion does not substitute earlier default values for these.
% 4. No self-weight figure was supplied in figures.zip. A commented insertion
%    point is provided below; no self-weight results have been invented.
%
\providecommand{\resultsfigdir}{figures}

\section{Numerical results}
\label{sec:numerical-results}

The numerical assessment focuses on stress accuracy in two-dimensional
benchmarks with straight and curved boundaries. Standard MPM and AI-Q MPM
are compared through stress fields, analytical reference solutions, and
refinement studies. The cantilever example examines both particle and
Gauss-point stresses, whereas the circular and elliptical cavity examples
use particle stresses. Grid refinement and particle enrichment are
considered separately to distinguish their effects on the computed stress
distributions.

\subsection{Two-dimensional benchmarks}
\label{sec:results-2d}

The field comparisons use the two grid spacings
\(h=0.02\,\mathrm{m}\) and \(h=0.01\,\mathrm{m}\), with common colour scales
for each stress component. The accompanying error maps show the spatial
distribution of the absolute-error percentages. The refinement plots use
the section-based relative \(\ell_2\) stress error for the cantilever and
the pressure-normalised RMS stress error for the cavity problems.
Comparisons between the methods are made within each benchmark using its
prescribed sampling region.

% Optional insertion point, once the self-weight results are supplied:
% \subsubsection{Self-weight compaction}
% \label{sec:self-weight-results}
% Insert the corresponding error results and discussion here. A spatial
% stress contour is not required for this one-dimensional stress state.

\subsubsection{Self-weight compaction}
\label{sec:self-weight-results}
\begin{figure*}[!htbp]
  \centering
  \includegraphics[width=1.00\textwidth,height=1.00\textheight,keepaspectratio]
    {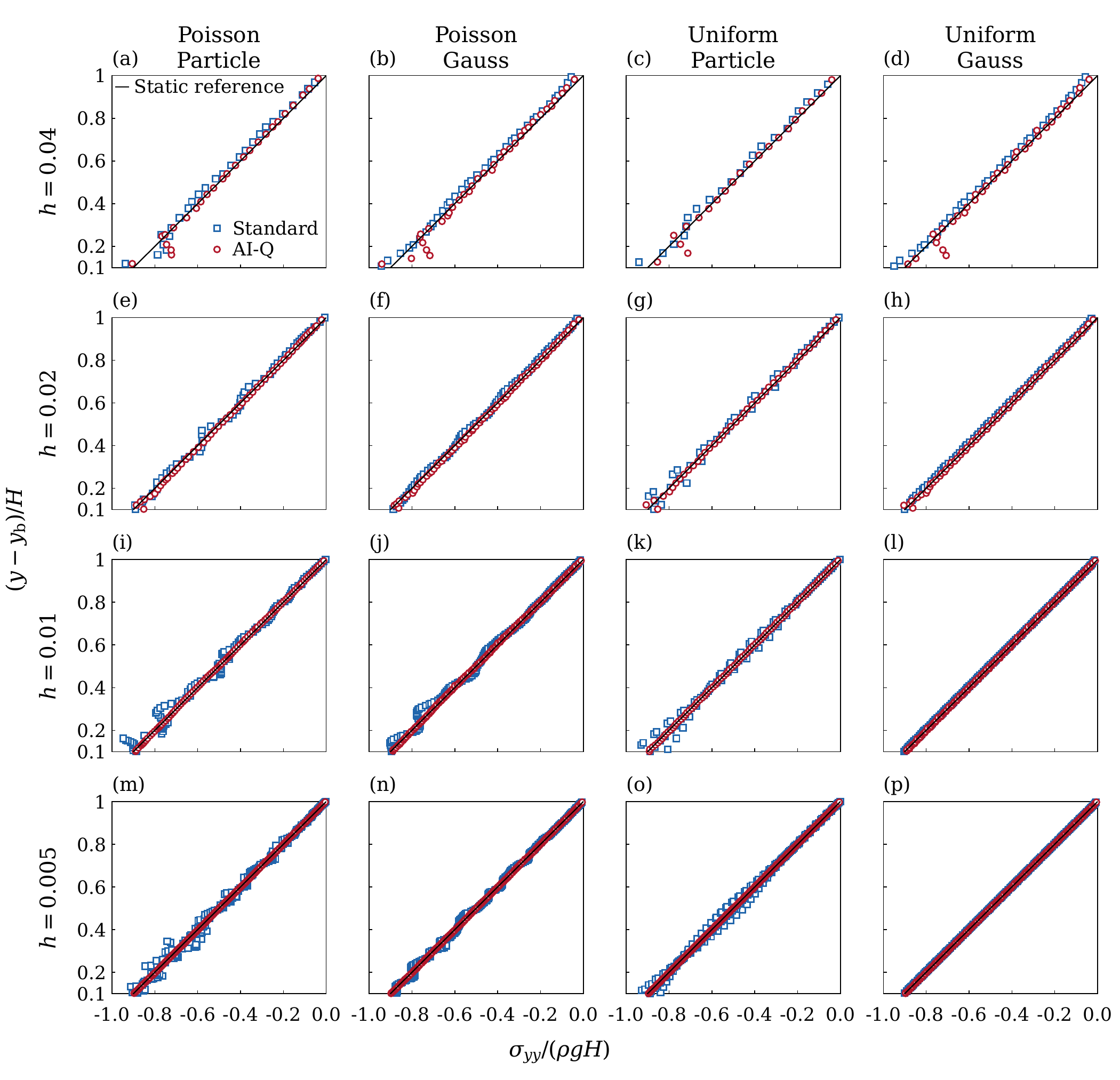}
  \caption{Self-weight compaction: normalised vertical stress profiles.
  From left to right, the columns show Poisson particle stresses,
  Poisson Gauss-point stresses, uniform particle stresses, and uniform
  Gauss-point stresses. From top to bottom, the rows correspond to
  \(h=0.04,\ 0.02,\ 0.01,\) and \(0.005\).
  The black line denotes the static reference
  \(\widehat{\sigma}_{yy}^{\mathrm{ref}}=\eta-1\).
  The displayed height interval is \(0.1\leq\eta\leq1\).}
  \label{fig:self-weight-profiles}
\end{figure*}

\providecommand{\resultsfigdir}{figures}
The self-weight compaction test examines the accuracy of a linear stress
field under gravity: a homogeneous column of height \(H\), with its
base at \(y_b\) and a traction-free upper surface. The block is modelled as a homogeneous isotropic linear elastic solid
with Young's modulus $E=1.0\times10^3\,\mathrm{kPa}$, Poisson's ratio
$\nu=0$, and reference density $\rho_0=10\,\mathrm{kg\,m^{-3}}$. Poisson and uniform particle
arrangements are compared using both particle and Gauss-point stresses.
Figure~\ref{fig:self-weight-profiles} shows the corresponding profiles
for \(h=0.04,\ 0.02,\ 0.01,\) and \(0.005\).

\begin{figure*}[!htbp]
  \centering
  \includegraphics[width=0.50\textwidth,height=0.50\textheight,keepaspectratio]    {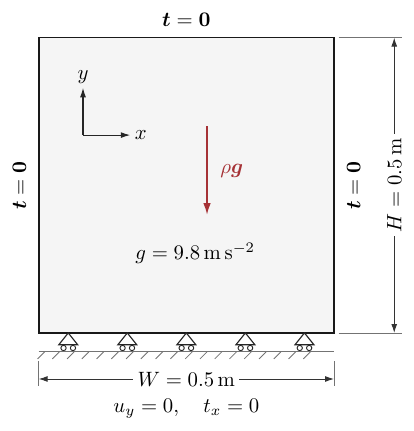}
  \caption{Self-weight compaction benchmark: geometry, boundary conditions, and loading convention used for the vertical-stress comparison.}
  \label{fig:self-weight-scheme}
\end{figure*}

Both methods reproduce the overall linear variation of compressive
stress with depth. On the coarsest grid, local deviations remain in
the lower part of the displayed profiles, including the AI-Q results.
For \(h\leq0.02\), the AI-Q profiles closely follow the static reference
for both arrangements and both stress representations. Standard MPM
retains local fluctuations in the particle stresses as the grid is
refined. Its Gauss-point profiles are smoother for the uniform
arrangement, while local variations remain visible for the Poisson
arrangement.

\begin{figure*}[!htbp]
  \centering
  \includegraphics[width=1.00\textwidth]
    {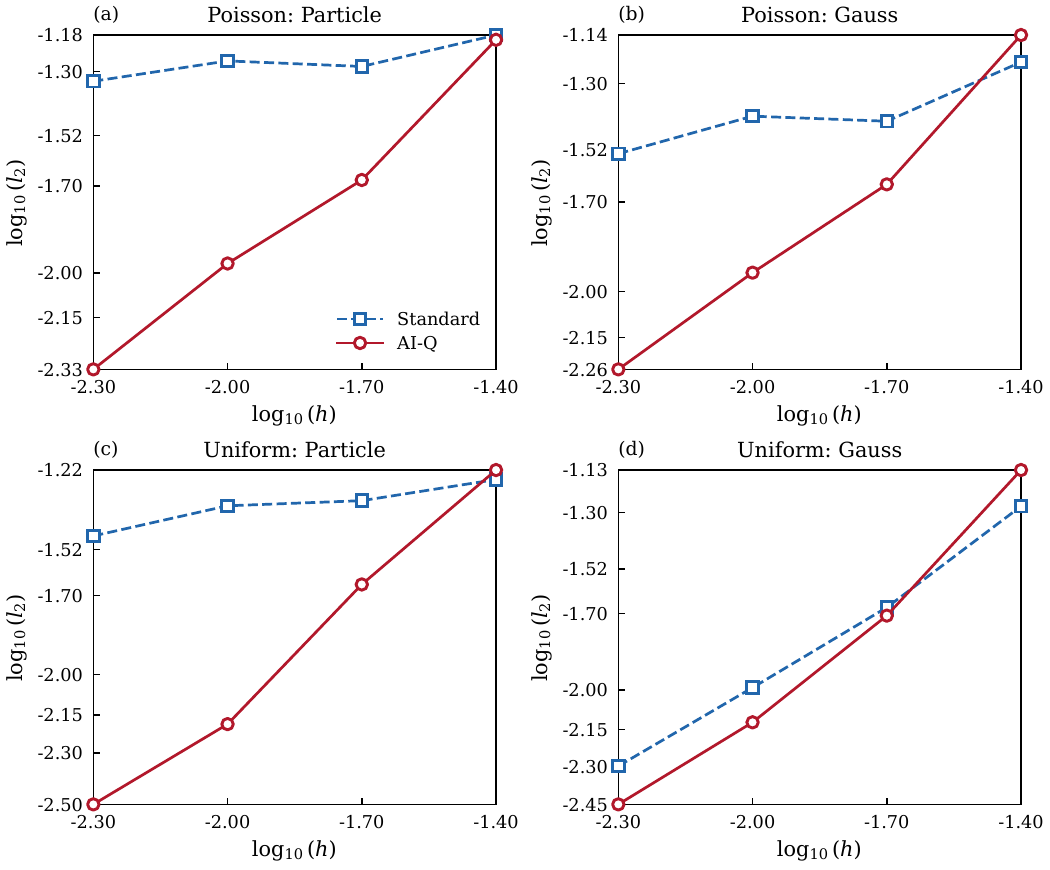}
  \caption{Self-weight compaction: grid-refinement behaviour of the
  \(\ell_2\) vertical-stress error. Panels (a,b) correspond to the
  Poisson arrangement and panels (c,d) to the uniform arrangement.
  The left and right columns report particle and Gauss-point stresses,
  respectively. Both axes are expressed on a base-10 logarithmic scale.}
  \label{fig:self-weight-convergence}
\end{figure*}

Figure~\ref{fig:self-weight-convergence} summarises the corresponding
\(\ell_2\) stress errors. AI-Q exhibits a monotonic error reduction
under grid refinement in all four comparisons and gives lower errors
than Standard MPM for \(h\leq0.02\). On the coarsest grid, its particle
errors are comparable to those of Standard MPM, whereas its Gauss-point
errors are larger. Standard MPM shows only a modest reduction in
particle-stress error, and the Poisson results vary non-monotonically
under refinement. For the uniform arrangement, its Gauss-point error
decreases substantially, so the additional improvement provided by
AI-Q is smaller in this case.

The improvement in particle stresses shows that the gain in accuracy
also carries through to the particle solution. The consistent refinement
trend of AI-Q across both arrangements suggests improved robustness to
irregular particle placement. The distinction between the particle
and Gauss-point results also illustrates the value of assessing both
stress representations when evaluating the method.

\FloatBarrier

\subsubsection{Cantilever beam}
\label{sec:cantilever-results}

\begin{figure*}[!htbp]
  \centering
  \includegraphics[width=\textwidth]{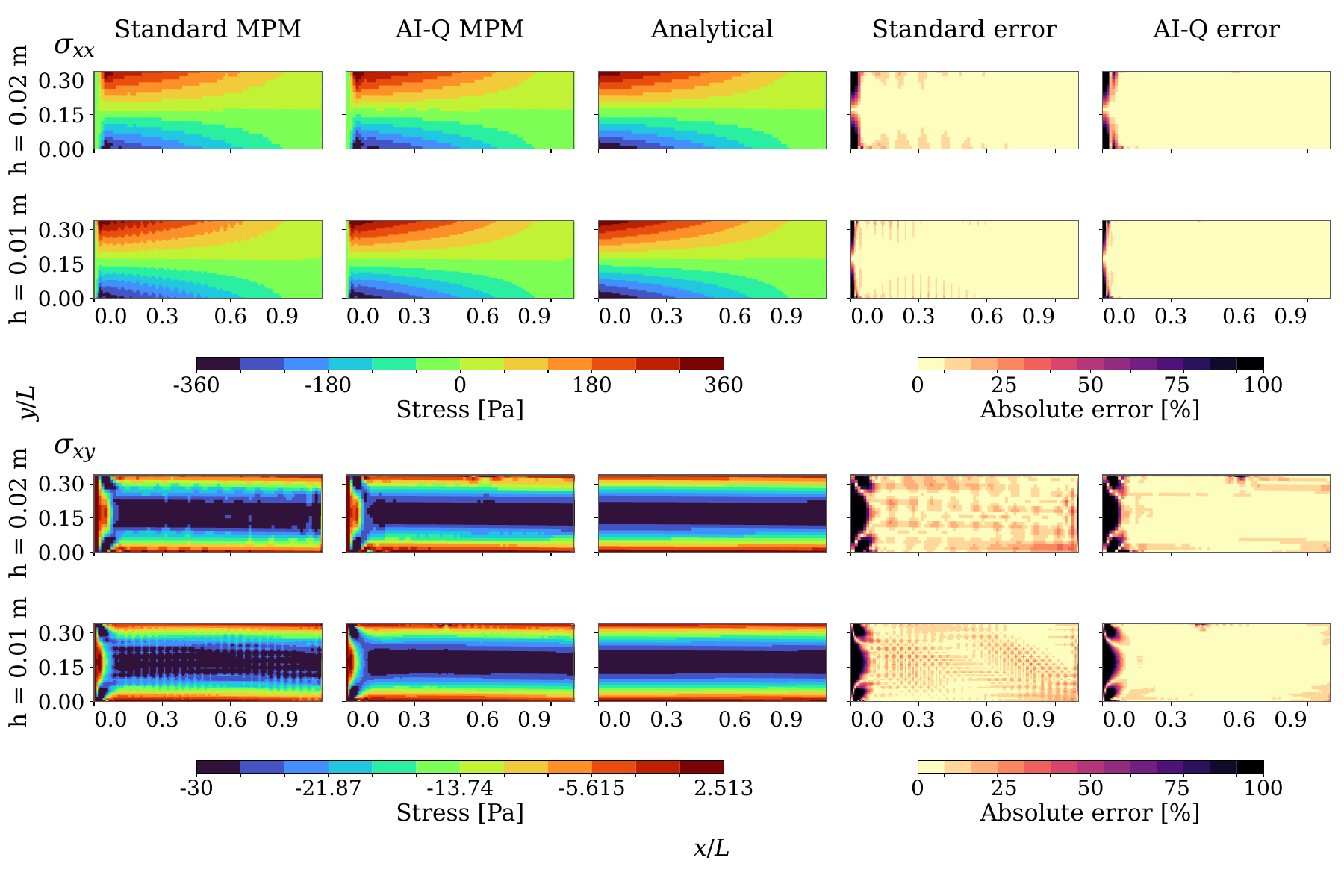}
  \caption{Cantilever beam: stress fields and absolute-error maps at
  \(h=0.02\,\mathrm{m}\) and \(h=0.01\,\mathrm{m}\).
  The upper block shows \(\sigma_{xx}\) and the lower block shows
  \(\sigma_{xy}\); within each block, the coarser grid is shown first.
  The columns contain Standard MPM, AI-Q MPM, the analytical reference,
  and the error maps of the two methods. Stresses are in Pa and the
  error maps are expressed as percentages.}
  \label{fig:beam-fields}
\end{figure*}

The cantilever tests the representation of bending and shear stresses. The beam is modelled as a homogeneous isotropic elastic solid with
Young's modulus $E=5.0\times10^1\,\mathrm{kPa}$, Poisson's ratio
$\nu=0.3$, and reference density $\rho_0=4\,\mathrm{kg\,m^{-3}}$.
The Saint-Venant analytical reference gives a linear variation of
\(\sigma_{xx}\) through the depth and a parabolic distribution of
\(\sigma_{xy}\). Figure~\ref{fig:beam-fields} compares these fields at the
two selected grid spacings. Both methods recover the overall bending
pattern. AI-Q MPM produces more regular stress bands, particularly in the
shear component, while local fluctuations remain visible in the Standard
MPM solution. The error maps show that the improvement extends across the
beam interior. The largest remaining discrepancies are concentrated near
the constrained end for both methods; the Saint-Venant field serves as
the analytical reference for the interior response.

\begin{figure*}[!htbp]
  \centering
  \includegraphics[width=0.60\textwidth,height=0.5\textheight,keepaspectratio]    {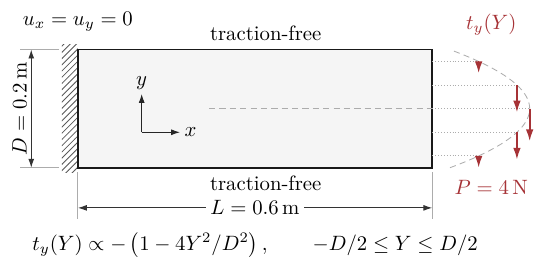}
  \caption{Cantilever benchmark: beam geometry, clamped boundary, applied end loading, and stress-section location used in the error measurements.}
  \label{fig:self-beam-scheme}
\end{figure*}

\begin{figure*}[!htbp]
  \centering
  \includegraphics[width=1.0\textwidth]{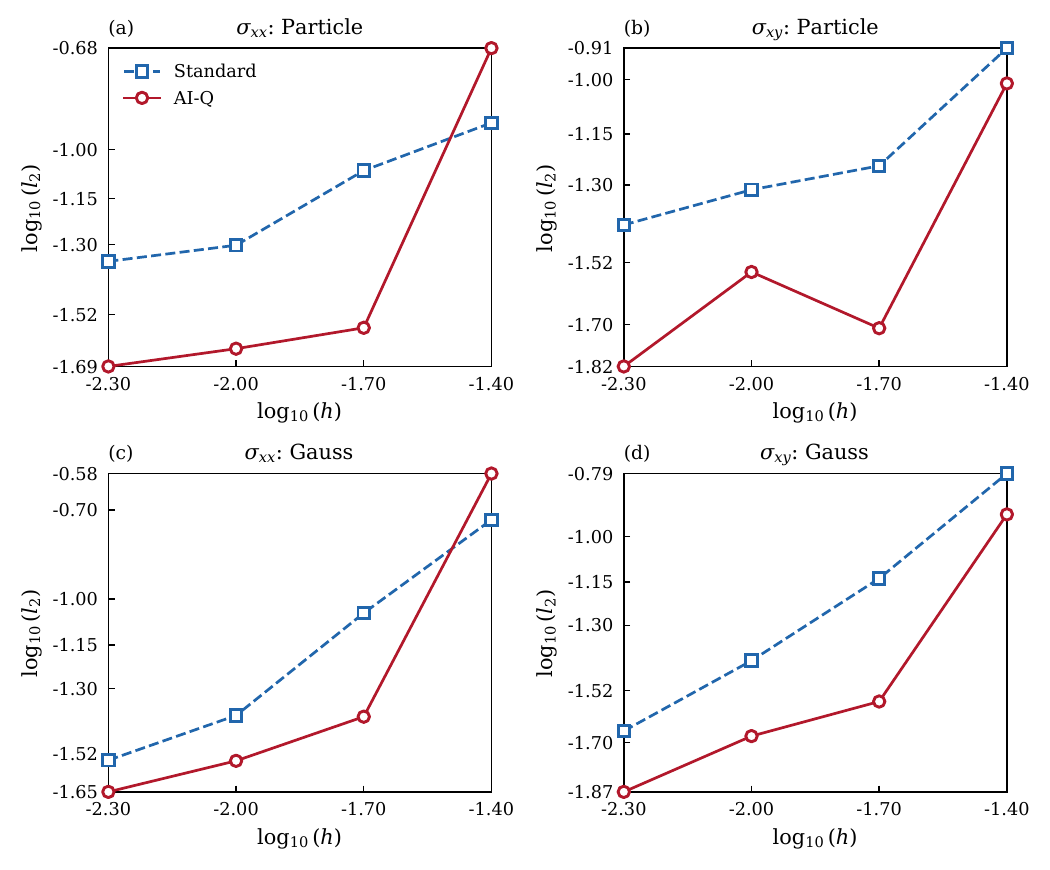}
  \caption{Cantilever beam: section-based relative \(\ell_2\) stress
  errors under grid refinement at \(\mathrm{PPC}=6\).
  Panels (a,b) report particle stresses and panels (c,d) report
  Gauss-point stresses. The columns correspond to
  \(\sigma_{xx}\) and \(\sigma_{xy}\), respectively.}
  \label{fig:beam-grid}
\end{figure*}

Figure~\ref{fig:beam-grid} shows the section-based errors for
\(h=0.04,\ 0.02,\ 0.01,\) and \(0.005\,\mathrm{m}\), with six particles
per cell (PPC). On the coarsest grid, the AI-Q normal-stress error is
larger than the Standard MPM error for both particle and Gauss-point
stresses. For \(h\leq0.02\,\mathrm{m}\), AI-Q gives lower errors for both
components and both representations. Its particle shear error increases
slightly from \(h=0.02\,\mathrm{m}\) to
\(h=0.01\,\mathrm{m}\), before decreasing on the finest grid. The
Gauss-point errors decrease more regularly. The supplementary sectional
profiles show the corresponding approach to the linear and parabolic
reference distributions, together with the remaining fluctuations in
the Standard MPM particle shear stress.

\begin{figure*}[!htbp]
  \centering
  \includegraphics[width=1.0\textwidth]{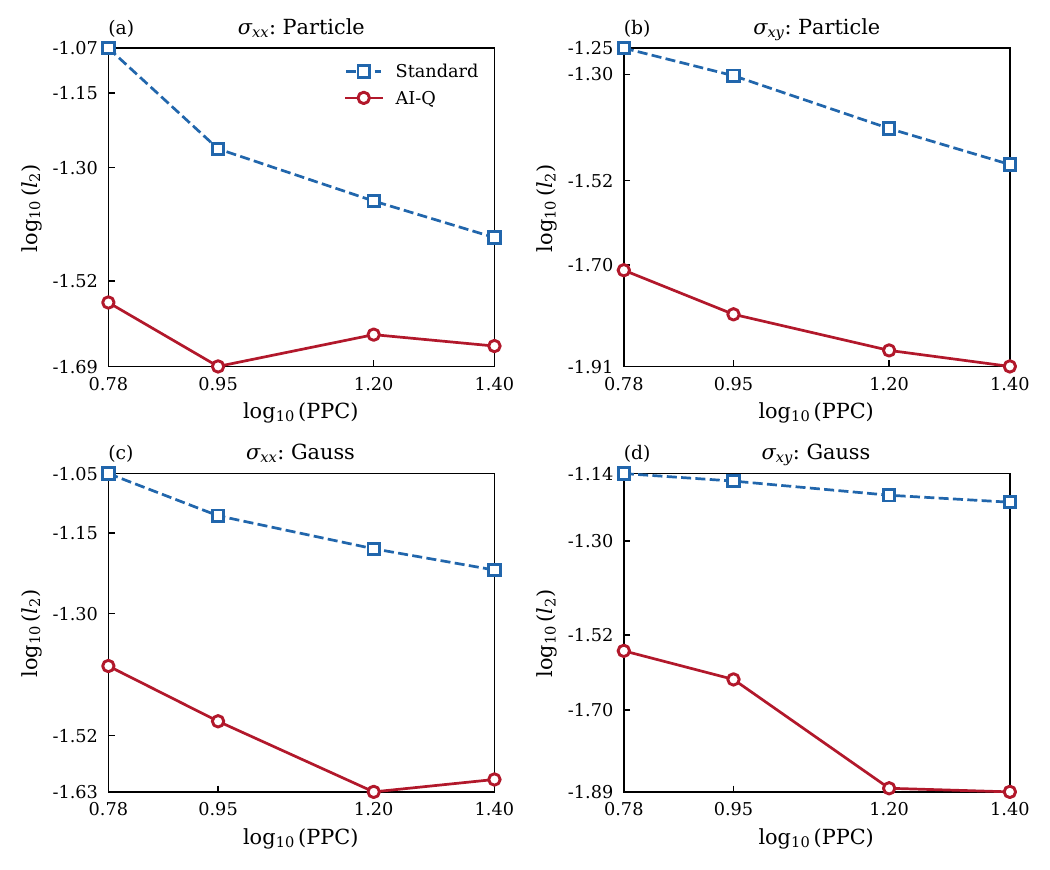}
  \caption{Cantilever beam: section-based relative \(\ell_2\) stress
  errors under particle refinement at \(h=0.02\,\mathrm{m}\).
  The tested particle counts are \(\mathrm{PPC}=6,\ 9,\ 16,\) and \(25\).
  Panels (a,b) show particle stresses and panels (c,d) show Gauss-point
  stresses; the columns correspond to \(\sigma_{xx}\) and \(\sigma_{xy}\).}
  \label{fig:beam-ppc}
\end{figure*}

The particle-refinement study in Figure~\ref{fig:beam-ppc} uses
\(h=0.02\,\mathrm{m}\) and \(\mathrm{PPC}=6,\ 9,\ 16,\) and \(25\).
Standard MPM benefits from increasing PPC in all four error measures,
although the reduction in Gauss-point shear error is modest. AI-Q
retains lower errors throughout the tested range and shows a consistent
reduction in shear-stress error. Its normal-stress errors vary
non-monotonically at the higher particle counts. Thus, particle
enrichment improves the stress approximation, but the extent of the
improvement depends on the component and stress representation.

\FloatBarrier

\subsubsection{Pressurised circular ring}
\label{sec:ring-results}

\begin{figure*}[p]
  \centering
  \includegraphics[width=\textwidth,height=0.87\textheight,keepaspectratio]{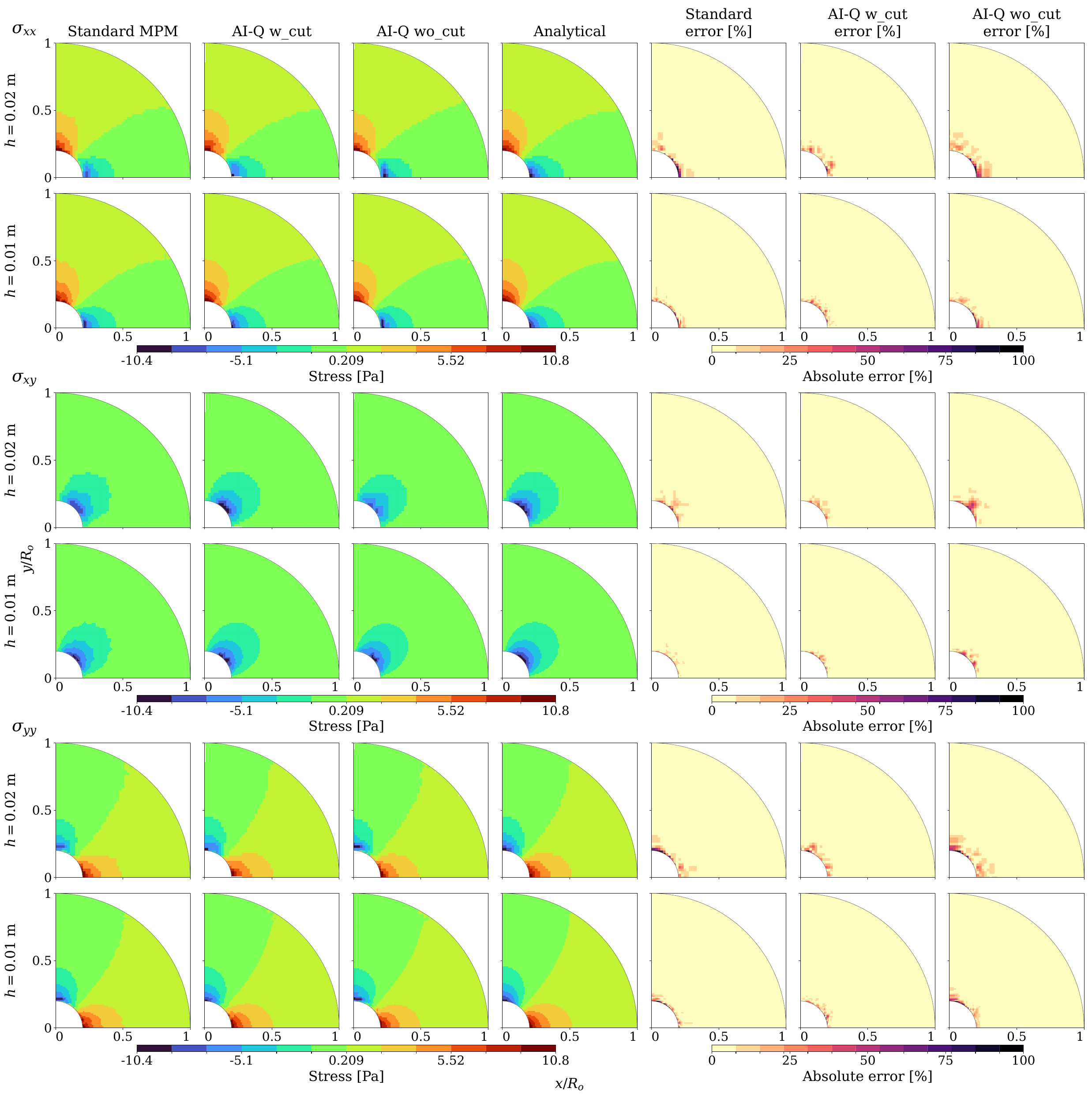}
  \caption{Pressurised circular ring: Cartesian stress fields and
  absolute-error percentage maps for the aligned particle arrangement
  with AI-Q cut-cell treatment enabled. The component blocks show
  \(\sigma_{xx}\), \(\sigma_{xy}\), and \(\sigma_{yy}\).
  Within each block, the upper and lower rows correspond to
  \(h=0.02\,\mathrm{m}\) and \(h=0.01\,\mathrm{m}\).
  The seven columns show Standard MPM, AI-Q with cut-cell treatment,
  AI-Q without cut-cell treatment, the analytical reference, and the
  corresponding three absolute-error maps. A quarter of the ring is displayed.}
  \label{fig:ring-fields}
\end{figure*}

The pressurised ring introduces a curved boundary and a strong stress
gradient near the cavity. Aligned and concentric particle arrangements
are considered, with the AI-Q cut-cell treatment either enabled or
disabled. The circular ring is modelled as a homogeneous isotropic linear elastic
solid under plane-strain conditions, with Young's modulus
$E=5.0\times10^3\,\mathrm{kPa}$ and Poisson's ratio $\nu=0.2$. Figure~\ref{fig:ring-fields} shows the Cartesian stress fields
for the aligned arrangement with cut-cell treatment. Both methods
reproduce the principal features of the analytical field. The most
visible discrepancies are concentrated near the inner boundary.
Refinement from \(h=0.02\,\mathrm{m}\) to \(h=0.01\,\mathrm{m}\)
reduces the spatial extent of this region, while local boundary
differences remain.

\begin{figure*}[!htbp]
  \centering
  \includegraphics[width=0.50\textwidth,height=0.50\textheight,keepaspectratio]    {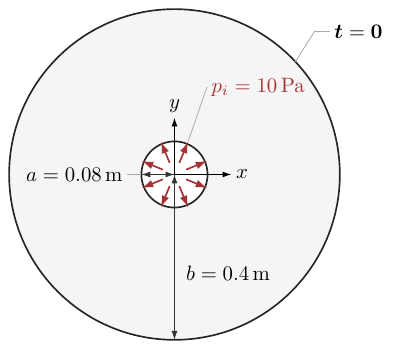}
  \caption{Pressurised circular-ring benchmark: quarter-domain geometry, inner pressure, symmetry boundaries, and radial sampling convention.}
  \label{fig:ring-pressured-scheme}
\end{figure*}

% This tall composite is deliberately placed on a full-width float page.
\FloatBarrier

\begin{figure*}[!htbp]
  \centering
  \includegraphics[width=1.0\textwidth]{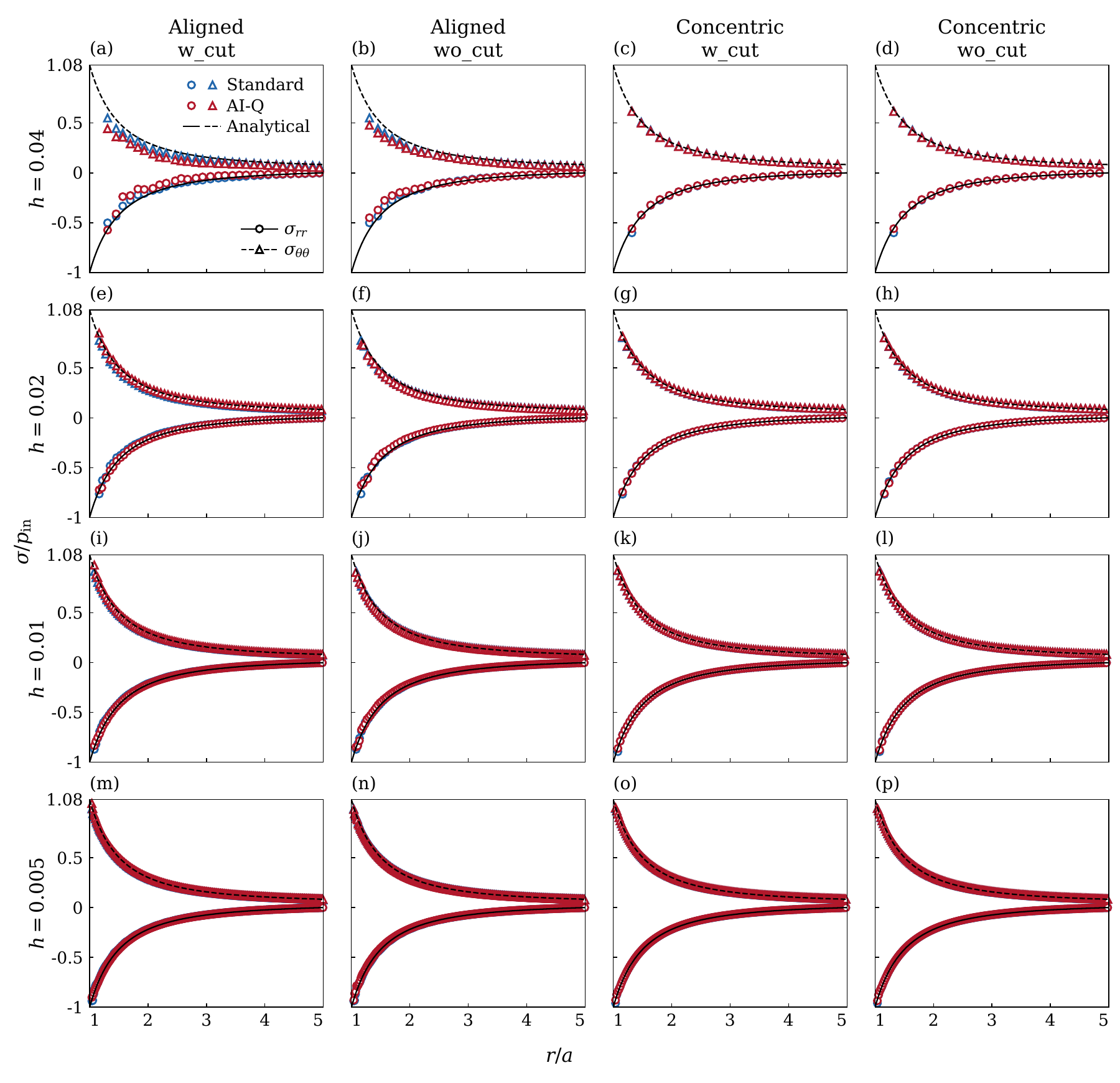}
  \caption{Pressurised circular ring: sampled radial and hoop stress
  profiles under grid refinement for aligned and concentric particle
  arrangements, with AI-Q cut-cell treatment enabled and disabled.
  The four rows correspond to \(h=0.04,\ 0.02,\ 0.01,\) and
  \(0.005\,\mathrm{m}\). Stresses are normalised by the internal
  pressure \(p_{\mathrm{in}}\), and radial positions by the inner
  radius \(a\). Analytical radial and hoop solutions are shown with
  the numerical profiles.}
  \label{fig:ring-profiles}
\end{figure*}

The sampled profiles in Figure~\ref{fig:ring-profiles} recover the
compressive radial stress and the tensile hoop stress, including their
rapid variation close to the cavity. The radial stress approaches zero
towards the outer boundary, while the hoop stress decreases with radius.
At the same grid spacing, the influence of particle arrangement and
cut-cell treatment is most visible in the inner part of the ring.

\begin{figure*}[!htbp]
  \centering
  \includegraphics[width=1.00\textwidth]{\resultsfigdir/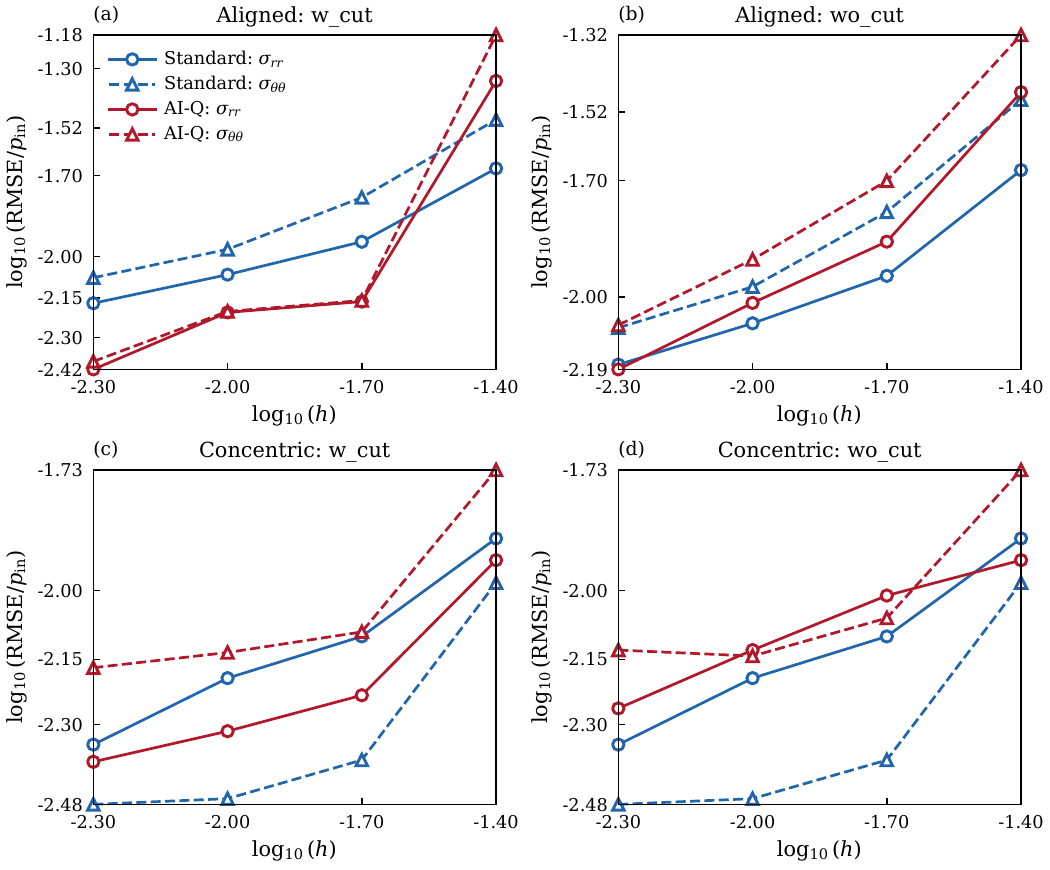}
  \caption{Pressurised circular ring: grid-refinement results for
  radial and hoop particle stresses, measured by
  \(\mathrm{RMSE}/p_{\mathrm{in}}\).
  The upper and lower rows use aligned and concentric particles;
  the left and right columns enable and disable the AI-Q cut-cell
  treatment. Standard MPM provides the same baseline in both columns
  for a given arrangement. The grid spacings are
  \(h=0.04,\ 0.02,\ 0.01,\) and \(0.005\,\mathrm{m}\).}
  \label{fig:ring-grid}
\end{figure*}

The refinement results in Figure~\ref{fig:ring-grid} show that the method
ranking depends on the particle arrangement and stress component.
For aligned particles with cut-cell treatment, AI-Q gives lower radial
and hoop errors on the three finer grids, but higher errors on the
coarsest grid. Disabling the treatment largely removes this advantage.
For concentric particles with cut-cell treatment, AI-Q improves the
radial-stress error, whereas Standard MPM retains the smaller hoop-stress
error. The AI-Q hoop error also changes only weakly on the finer grids.
The cut-cell treatment therefore benefits several configurations, but
the gain in radial accuracy does not necessarily extend to the hoop
component.

\FloatBarrier

\subsubsection{Pressurised elliptical cavity}
\label{sec:ellipse-results}

\begin{figure}
  \centering
  \includegraphics[width=\textwidth,height=0.9\textheight,keepaspectratio]{\resultsfigdir/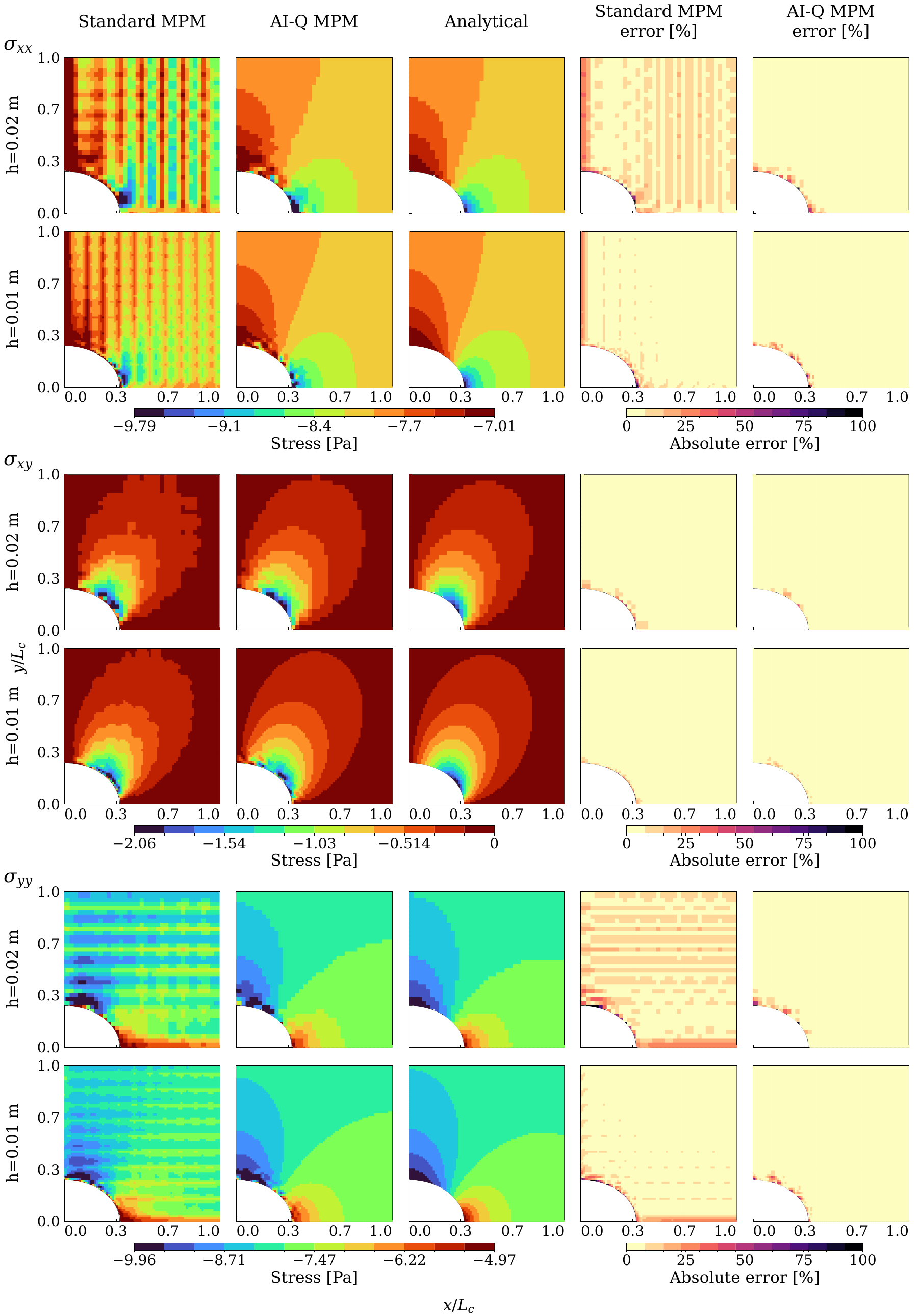}
  \caption{Pressurised elliptical cavity: stress distributions and
  absolute-error percentage maps in the selected region.
  The component blocks show \(\sigma_{xx}\), \(\sigma_{xy}\), and
  \(\sigma_{yy}\); each block contains \(h=0.02\,\mathrm{m}\) above
  \(h=0.01\,\mathrm{m}\).
  The columns show Standard MPM, AI-Q MPM, the analytical reference,
  and the error maps of the two numerical methods. Each component
  uses a common stress scale across methods and grid resolutions.}
  \label{fig:ellipse-fields}
\end{figure}

The elliptical cavity provides a non-circular boundary with a
spatially varying stress concentration. The comparison uses the
analytical reference for an elliptical cavity in an infinite plate
and the prescribed region adjacent to the cavity. The plate containing the elliptical cavity is modelled as a homogeneous
isotropic linear elastic solid under plane-strain conditions, with
Young's modulus $E=5.0\times10^3\,\mathrm{kPa}$, Poisson's ratio
$\nu=0.2$.
Figure~\ref{fig:ellipse-fields} shows pronounced spatial oscillations
in the Standard MPM normal stresses: predominantly vertical bands in
\(\sigma_{xx}\) and horizontal bands in \(\sigma_{yy}\).
These variations weaken with refinement but remain visible at
\(h=0.01\,\mathrm{m}\). AI-Q captures the main normal-stress patterns
on both grids, with the remaining discrepancies concentrated near the
cavity boundary. Both methods reproduce the principal shear-stress
distribution, for which their visual differences are smaller.

\begin{figure*}[!htbp]
  \centering
  \includegraphics[width=0.70\textwidth,height=0.70\textheight,keepaspectratio]    {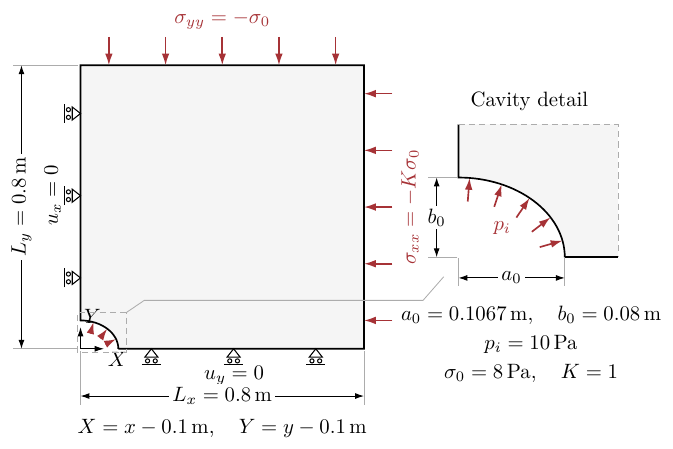}
  \caption{Pressurised elliptical-cavity benchmark: quarter-domain geometry, internal pressure, remote biaxial loading, and comparison region.}
  \label{fig:ellipse-pressured-scheme}
\end{figure*}

% Keep this tall composite full width and on its own float page.

\begin{figure*}[!htbp]
  \centering
  \includegraphics[width=\textwidth]{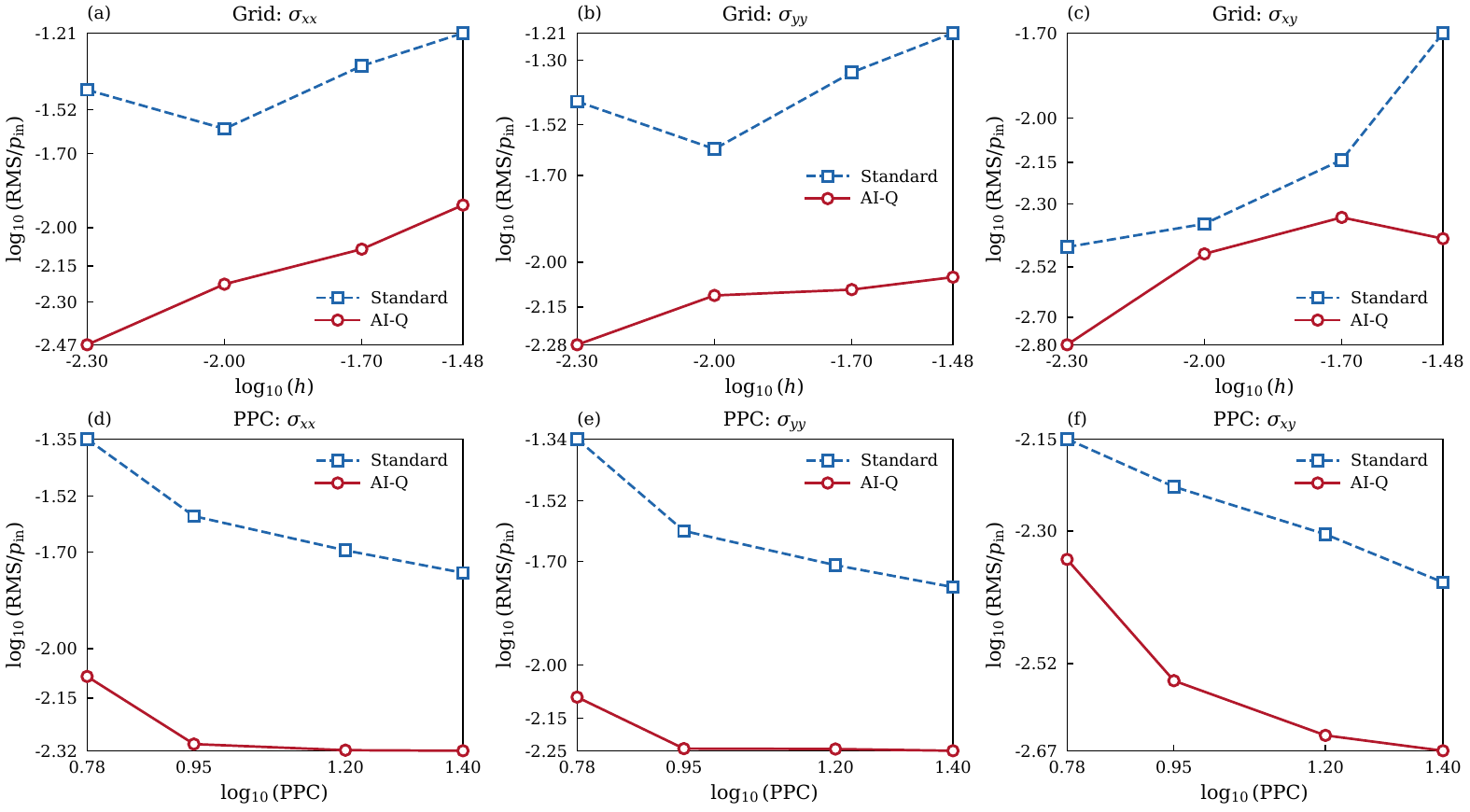}
  \caption{Pressurised elliptical cavity: pressure-normalised RMS
  particle-stress errors under (a--c) grid refinement and
  (d--f) particle refinement at \(h=0.02\,\mathrm{m}\).
  The columns correspond to \(\sigma_{xx}\), \(\sigma_{yy}\), and
  \(\sigma_{xy}\), respectively. The particle-refinement series uses
  \(\mathrm{PPC}=6,\ 9,\ 16,\) and \(25\), evaluated over the
  prescribed comparison region.}
  \label{fig:ellipse-refinement}
\end{figure*}

Figure~\ref{fig:ellipse-refinement} shows lower pressure-normalised RMS
errors for AI-Q in all three components over the tested grid and
particle resolutions. The Standard MPM normal-stress errors decrease
initially but rise on the finest grid. In comparison, the AI-Q
normal-stress errors decrease with grid refinement. Its shear error
shows a small non-monotonic change between the coarser grids before
decreasing on the finer grids.

Increasing PPC at fixed \(h=0.02\,\mathrm{m}\) reduces the Standard
MPM errors in all three components. For AI-Q, most of the improvement
in the normal stresses occurs between six and nine particles per
cell, followed by a near plateau up to 25 PPC. The shear error
continues to decrease more gradually. Once this plateau is reached,
adding particles at the same grid spacing provides little further
improvement in the normal stresses.

\FloatBarrier

\paragraph{Discussion of the two-dimensional results.}
The clearest improvements with AI-Q are the reduced stress
fluctuations in the beam interior and the smoother normal-stress
fields around the elliptical cavity. The circular-ring results
demonstrate a stronger dependence on particle arrangement and stress
component. Together, the tests show that stress accuracy responds
differently to grid refinement, particle enrichment, and cut-cell
treatment. In particular, a lower error on a resolved grid does not
imply the same ranking on the coarsest grid, and increasing PPC can
reach a level beyond which the remaining error changes little.

\subsection{Three-dimensional benchmarks}
\label{sec:results-3d}

\begin{figure}[!htbp]
  \centering
  \includegraphics[width=0.6\linewidth,height=0.60\textheight,keepaspectratio]
    {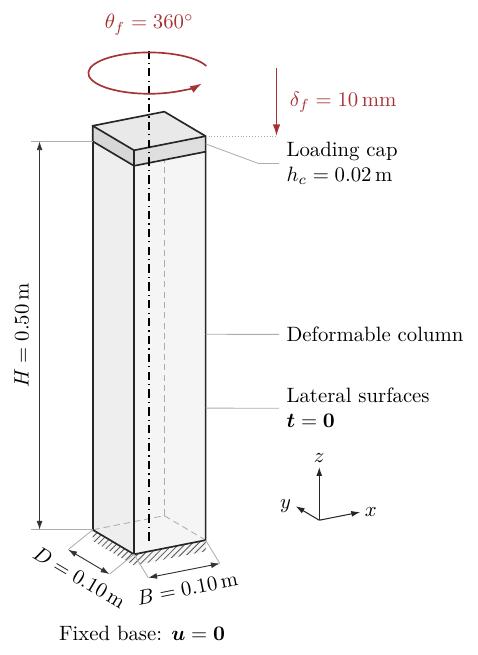}
  \caption{Three-dimensional torsion with axial settlement: initial
  geometry, fixed base, traction-free lateral surfaces, and loading-cap
  motion. The schematic indicates an end rotation of
  \(\theta_f=360^\circ\) and a downward settlement of
  \(\delta_f=10\,\mathrm{mm}\).}
  \label{fig:torsion-setup}
\end{figure}

The final benchmark examines combined torsion and axial settlement of a
three-dimensional column. The full-cell AI-Q MPM implementation is assessed
against standard APIC MPM under large deformation.
Figure~\ref{fig:torsion-setup} shows the initial geometry and end motion.
The deformable column has height \(H=0.50\,\mathrm{m}\), a square
cross-section with \(B=D=0.10\,\mathrm{m}\), and an additional loading
cap of height \(h_c=0.02\,\mathrm{m}\). The bottom face is fixed, while
the lateral surfaces are traction-free. The cap rotates about the
longitudinal \(z\)-axis and moves downwards. The end-motion values
indicated in the schematic are \(\theta_f=360^\circ\) and
\(\delta_f=10\,\mathrm{mm}\); the comparisons presented below cover the entire rotational process.
The column was modelled as a compressible neo-Hookean solid with
Young's modulus \(E=1.0\,\mathrm{MPa}\), Poisson's ratio \(\nu=0.40\),
and reference density \(\rho_0=1000\,\mathrm{kg\,m^{-3}}\).
Both simulations used explicit APIC transfers on a uniform Cartesian
grid with spacing \(h=0.01\,\mathrm{m}\), quadratic B-spline basis
functions, and a nominal initial resolution of eight material points
per grid cell. The time step was selected using a CFL number of \(0.20\)
and capped at \(\Delta t_{\max}=5.0\times10^{-5}\,\mathrm{s}\).
For internal-force assembly, standard APIC MPM used particle quadrature
throughout the domain. AI-Q used a \(2\times2\times2\) Gauss rule on
accepted full knot spans and retained particle quadrature on the
remaining spans.

\FloatBarrier

\begin{figure}[!htbp]
  \centering
  \includegraphics[width=1.0\linewidth,height=1.0\textheight,keepaspectratio]
{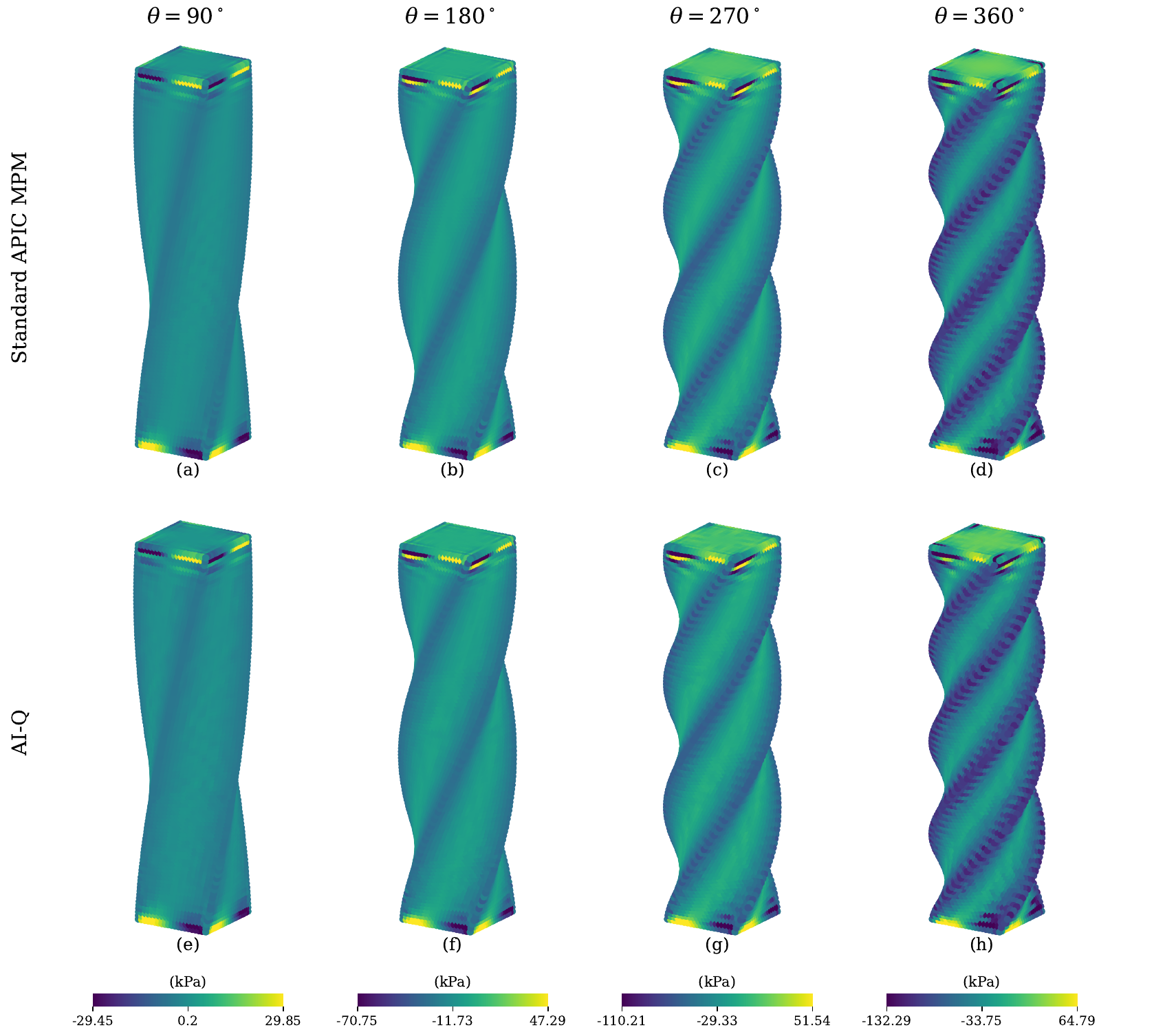}
  \caption{Deformed configurations and pressure distributions during
  three-dimensional torsion with axial settlement. Panels (a--d) show
  standard APIC MPM and panels (e--h) show AI-Q. The columns correspond
  to \(\theta=90^\circ\), \(180^\circ\), \(270^\circ\), and
  \(360^\circ\), respectively. Pressure is expressed in kPa. The two
  methods share the same colour scale within each column; the scale
  changes between rotation angles.}
  \label{fig:torsion-pressure}
\end{figure}

Figure~\ref{fig:torsion-pressure} compares the deformed configurations
and pressure distributions at \(\theta=90^\circ\), \(180^\circ\),
\(270^\circ\), and \(360^\circ\). Both methods capture the progressive
twisting of the square column and produce closely matching overall
shapes. As the rotation increases, the pressure field develops more
pronounced spatial variations near the ends and along the twisted
edges. AI-Q reproduces the main pressure patterns of the APIC solution,
with local differences visible in the end regions and surface bands. The agreement in both geometry and pressure distribution is retained through the displayed sequence, despite the substantial change in particle configuration.

\FloatBarrier

\begin{figure}[!htbp]
  \centering
  \includegraphics[width=\linewidth]
{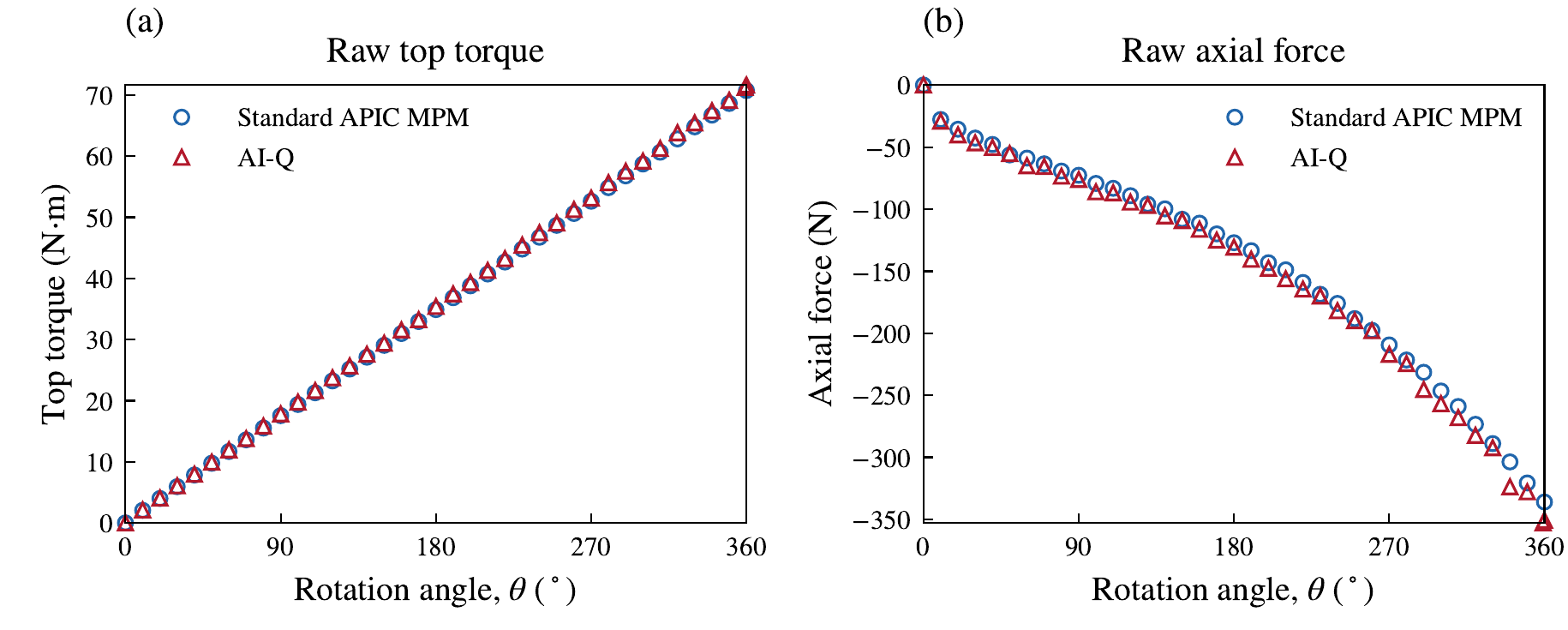}
  \caption{Global response of the three-dimensional column:
  (a) raw top torque and (b) raw axial force versus rotation angle.
  Open circles denote standard APIC MPM and open triangles denote
  AI-Q. The plotted interval is \(0\leq\theta\leq360^\circ\);
  torque and force are expressed in \(\mathrm{N\,m}\) and N,
  respectively.}
  \label{fig:torsion-responses}
\end{figure}

The corresponding torque and axial-force responses are shown in
Figure~\ref{fig:torsion-responses}. The torque increases monotonically
and approximately linearly with rotation, reaching about
\(70\,\mathrm{N\,m}\) at \(\theta=360^\circ\). The two torque curves
nearly coincide throughout the plotted interval. The axial force shows
a more nonlinear evolution: its magnitude increases more rapidly at
larger rotations. AI-Q follows the same trend and predicts a slightly
larger force magnitude towards the end of the interval. These
resultants provide complementary measures of the response: torque
characterises the torsional resistance, while axial force reflects the
coupled axial response during twisting and settlement.

\FloatBarrier

Together, the field and resultant comparisons show that the full-cell
AI-Q formulation reproduces the principal three-dimensional deformation
and load-response features of the APIC baseline over a full revolution.
Because this example has no independent analytical or converged reference,
it is used as a large-deformation compatibility and robustness test rather
than as evidence that AI-Q is more accurate than APIC. The closest agreement
is obtained in torque, with a modest difference remaining in the axial
response at the largest rotations.

% Add each 3D benchmark here after its results become available:
%   1. Geometry, loading, boundary conditions, and reference solution.
%   2. Representative stress fields and their associated error maps.
%   3. Grid/particle refinement and a short, evidence-based discussion.
% Keep any final combined 2D/3D conclusions after the completed 3D tests.

%% file: discussion_ai_q.tex
\section{Discussion}
\label{sec:aiq-discussion}

The results support AI-Q as a selective integration strategy for evolving particle distributions. Classical polynomial reconstructions retain very low median force errors when their approximation space matches the stress field, whereas AI-Q more consistently reduces the mean and upper tail of the error distribution. The boundary-value tests show the same qualified picture: the clearest gains occur for irregular sampling and curved or partially occupied cells, while the circular-ring results show that the benefit can depend on particle arrangement and stress component. This dependence is consistent with established analyses of MPM integration error \cite{Baumgarten2023}; it also motivates retaining the mixed assembly of Eq.~\eqref{eq:aiq-assembled-internal-force}, so learned and particle quadrature can coexist rather than forcing one rule onto every cell.

The trained operator is local rather than problem-specific. All boundary-value simulations in Section~\ref{sec:numerical-results} use frozen checkpoints: no benchmark trajectory, constitutive response, or stress history is used to update the network during a simulation. The single-cell test split is stricter than a random stress-realization split because complete particle geometries are held out, but it remains an in-distribution test of the synthetic geometry and analytical stress families used to construct the data. The boundary-value problems therefore provide the stronger cross-problem test: the same local operator is applied repeatedly to particle configurations and stresses generated by independent evolving simulations. A separate checkpoint is nevertheless associated with each implemented dimension, full/cut module, basis, and quadrature rule, so the present results should not be interpreted as a single universal model.

The current scope also identifies where broader validation is needed. The full-cell training profile uses equal representative particle volumes, the learned fields are smooth, and three-dimensional results use full cells only. Strong localization and sharp solution gradients remain challenging regimes for MPM and have motivated complementary spatial-resolution strategies such as mapped MPM \cite{ZhaoMappedMPM2024}; contact or fracture, highly nonuniform particle volumes, and three-dimensional cut geometry may likewise require broader training distributions or application-specific data. In implicit MPM, the reconstructed force would additionally require a consistent tangent through the geometry-dependent reconstruction \cite{SongKim2024}. These limitations concern extensions of the present operator rather than its explicit-MPM use studied here.

%% file: conclusions_ai_q.tex
% Insert after Discussion and before \appendix.
\section{Conclusions}
\label{sec:aiq-conclusions}

An AI-assisted quadrature framework has been developed for the material
point method. A local, geometry-dependent operator reconstructs stresses
at prescribed quadrature points while material history remains on the
particles. The construction combines exact constant reproduction,
force-oriented training, and known mechanical assembly, and provides
corresponding formulations for full and cut cells.

The two-dimensional single-cell studies show consistent reductions in mean and
high-percentile internal-force errors across linear and B-spline bases.
AI-Q achieves the lowest pooled mean and P95 errors among the compared
methods in all six reported two-dimensional cell-type and basis
combinations. Classical polynomial reconstructions retain lower median
errors, identifying complementary strengths in exact polynomial reproduction
and performance across mixed stress fields. The three-dimensional
full-cell tests retain the reduction in high-percentile errors at moderate
and large particle counts.

The boundary-value problems show that improved integration can carry
through to the evolving particle stresses. Self-weight compaction,
cantilever bending, and the elliptical cavity exhibit improved stress
accuracy, while the circular ring identifies particle arrangement and
cut-cell treatment as important determinants of the gain. The
three-dimensional torsion example reproduces the principal deformation and
global response of standard APIC MPM through a full revolution with axial
settlement. Together, these findings establish the value of a constrained
learned integration operator within the existing MPM structure. Future work
will address broader training distributions, three-dimensional cut cells,
runtime performance, and stress-power compatibility.

%% file: acknowledgement.tex
\section*{Acknowledgement}
Yupeng Jiang acknowledges the financial support from the International Research Training Group (IRTG) 2657, funded by the German Research Foundation (DFG) (Project-ID 433082294). Minchen Li acknowledges gift support from Genesis AI. 

%% file: theory_appendix.tex
\section{Error propagation, force balance, and stress power}
\label{app:error}

\subsection{Local reconstruction estimate}

For one full-cell query, Taylor's theorem along the segment from $\vct x_{cq}$ to $\vct x_p$ gives
\begin{equation}
 \begin{aligned}
 \stress(\vct x_p)&=\stress(\vct x_{cq})
  +D\stress(\vct x_{cq})[\vct x_p-\vct x_{cq}]
  +\mat r_{cqp}^{\sigma},\\
 \norm{\mat r_{cqp}^{\sigma}}_{\Frob}
 &\le\frac{M_{\sigma,2,c}}2\norm{\vct x_p-\vct x_{cq}}_2^2.
 \end{aligned}
 \label{eq:taylorsigma}
\end{equation}
The derivative $D\stress[\cdot]$ is a linear map from a displacement to a symmetric tensor. Multiplying by $R_{cqp}^{\full}$, summing over particles, and using the row-sum condition cancels the constant term. If $\mat\epsilon_p$ denotes the particle-stress error as in Eq.~\eqref{eq:mpmerror}, then
\begin{equation}
 \begin{aligned}
 \widehat\stress_{cq}^{\full}-\stress(\vct x_{cq})
 ={}&\sum_pR_{cqp}^{\full}\mat\epsilon_p
  +D\stress(\vct x_{cq})\!\left[\frac h2\vct\delta_{cq}^{\full}\right]
  +\sum_pR_{cqp}^{\full}\mat r_{cqp}^{\sigma}.
 \end{aligned}
 \label{eq:exactreconstructionerror}
\end{equation}
Taking norms and using $\norm{\vct x_p-\vct x_{cq}}_2\le\sqrt d\,h$ proves Eq.~\eqref{eq:reconstructionbound}. A cut query obeys the same identity with $q$ replaced by $(\ell,q)$ because its particles and query remain within the parent cell and its coordinates use the same affine map. Maximizing over all child queries proves Eq.~\eqref{eq:cutreconstructionbound} with the same parent length scale. The stress must be smooth along the connecting segments; when a stress discontinuity lies within the reconstruction region, this smooth-field estimate does not apply across that discontinuity.

For positive quadrature weights, the corresponding full-cell force error also satisfies the weighted Cauchy--Schwarz estimate
\begin{equation}
 \begin{aligned}
 \norm{\widehat{\vct f}_{ic}^{\full}
            -\vct f_{ic}^{\Gauss,\full}[\stress]}_2
 \le{}&\left[\sum_qW_{cq}
       \norm{\widehat\stress_{cq}^{\full}-\stress(\vct x_{cq})}_{\Frob}^2
       \right]^{1/2}\\
 &\times\left[\sum_qW_{cq}\norm{\nabla N_i(\vct x_{cq})}_2^2
       \right]^{1/2}.
 \end{aligned}
 \label{eq:weightedforcebound}
\end{equation}
The cut-cell form uses the complete $\ell,q$ sum. In either case, the bound controls the first bracket of the corresponding decomposition, Eq.~\eqref{eq:fullerror} or Eq.~\eqref{eq:cuterror}. The full-cell quadrature error, and the cut-cell quadrature and geometry errors, remain separate. For example, if $\stress\nabla N_i$ has a bounded extension to $D_c\cup D_c^{\mathrm{rec}}$, then
\begin{equation}
 \norm{\vct f_{ic}(D_c^{\mathrm{rec}};\stress)
           -\vct f_{ic}(D_c;\stress)}_2
 \le\abs{D_c\triangle D_c^{\mathrm{rec}}}
   \norm{\stress\nabla N_i}_{L^\infty(D_c\cup D_c^{\mathrm{rec}})}.
 \label{eq:geometrybound}
\end{equation}
The symbol $\triangle$ denotes symmetric difference of sets. In a smooth full-cell sequence with bounded coefficient amplification, a second-order stress estimate additionally requires the dimensionless affine defect to decrease at least proportionally to $h$ and particle stress errors to be $O(h^2)$. A local bound does not alone establish the convergence rate of the evolving MPM solution, which also includes mass division, transfers, temporal integration, and changing particle geometry.

\subsection{The stronger condition needed for stress-power compatibility}
\label{app:power}

Let $\vct v_{\mathrm g}=\sum_iN_i\vct v_i$ be a frozen grid velocity and let $\dot{\mat\varepsilon}_{\mathrm g}=\sym\nabla\vct v_{\mathrm g}$, where $\sym\mat A=(\mat A+\mat A^{\T})/2$. The grid internal power and particle stress power for one full cell are
\begin{equation}
 \begin{aligned}
 \mathcal P_c^{\mathrm{grid}}
  &=-\sum_qW_{cq}\widehat\stress_{cq}^{\full}
                       :\dot{\mat\varepsilon}_{\mathrm g}(\vct x_{cq}),\\
 \mathcal P_c^{\mathrm{part}}
  &=\sum_pV_p\stress_p:\dot{\mat\varepsilon}_{\mathrm g}(\vct x_p).
 \end{aligned}
 \label{eq:stresspowers}
\end{equation}
After substituting the reconstruction, their sum is
\begin{equation}
 \mathcal P_c^{\mathrm{grid}}+\mathcal P_c^{\mathrm{part}}
 =\sum_p\stress_p:\left[
 V_p\dot{\mat\varepsilon}_{\mathrm g}(\vct x_p)
 -\sum_qW_{cq}R_{cqp}^{\full}
                  \dot{\mat\varepsilon}_{\mathrm g}(\vct x_{cq})\right].
 \label{eq:powerdefect}
\end{equation}
For this quantity to vanish for arbitrary independent particle stresses and all admissible grid velocities, each bracket must vanish. This is a weighted adjoint relation between stress transfer and particle strain-rate evaluation. Neither the constant projection nor the soft coordinate penalty imposes it. The cut relation uses an additional child sum. Thus internal resultant and torque cancellation are retained, while an energy-conserving time integrator is not claimed. The distinction is especially relevant because APIC transfer, explicit stepping, damping, and boundary enforcement introduce further conditions beyond this frozen-state pairing.

\section{Analytical stress fields for the benchmark problems}
\label{app:analytical-stresses}

The static, small-deformation reference fields used in the
two-dimensional comparisons are collected below. An asterisk denotes
analytical Cauchy stress; normal stresses are positive in tension and
$\sigma_{xy}^{*}=\sigma_{yx}^{*}$. All loads are given at their final
amplitudes.

\subsection{Self-weight compaction}
\label{app:self-weight-analytical}

The column in Figure~\ref{fig:self-weight-scheme} has height $H=0.5\,\mathrm{m}$, width
$W=0.5\,\mathrm{m}$, and homogeneous mass density $\rho$.
Gravity acts downwards with magnitude $g=9.8\,\mathrm{m\,s^{-2}}$.
With $y$ measured upwards and $y_b$ denoting the base elevation, the
traction-free top gives the vertical-stress reference and its
normalisation:
\begin{equation}
  \sigma_{yy}^{*}(y)=-\rho g\,[H-(y-y_b)],
  \qquad
  \frac{\sigma_{yy}^{*}(y)}{\rho gH}
     =\frac{y-y_b}{H}-1,
  \qquad 0\le y-y_b\le H.
  \label{eq:app-self-weight-stress}
\end{equation}
This is the reference used for the vertical-stress errors in Figure~\ref{fig:self-weight-convergence}.

\subsection{Cantilever beam}
\label{app:beam-analytical}

For Figure~\ref{fig:self-beam-scheme}, $X$ measures distance from the clamped cross-section and
$Y$ measures height from the beam midline, with $0\le X\le L$ and
$-D/2\le Y\le D/2$. The length and depth are
$L=0.6\,\mathrm{m}$ and $D=0.2\,\mathrm{m}$.
The positive load magnitude $P=4\,\mathrm{N}$ acts downwards through
a parabolic end traction. With out-of-plane thickness
$t_z=1\,\mathrm{m}$, the cross-sectional second moment is
$I=t_zD^3/12$. The Saint-Venant stress field is
\begin{equation}
  \sigma_{xx}^{*}=\frac{P(L-X)Y}{I},
  \qquad
  \sigma_{yy}^{*}=0,
  \qquad
  \sigma_{xy}^{*}=-\frac{P}{2I}
     \left(\frac{D^2}{4}-Y^2\right).
  \label{eq:app-beam-stress}
\end{equation}
The vertical end traction is $t_y(Y)=\sigma_{xy}^{*}(L,Y)$, with zero
horizontal component. This field supplies the interior reference for
the fully clamped numerical beam.

\subsection{Pressurised circular ring}
\label{app:ring-analytical}

The ring in Figure~\ref{fig:ring-pressured-scheme} has inner radius $a=0.08\,\mathrm{m}$, outer
radius $b=0.4\,\mathrm{m}$, internal pressure $p_i=10\,\mathrm{Pa}$,
and a traction-free outer surface. Here $x$ and $y$ are measured from
the ring centre, $r=\sqrt{x^2+y^2}$ is the radial distance, and
$\theta$ is the polar angle measured anticlockwise from the positive
$x$-axis. For $a\le r\le b$, the Lam\'e radial, hoop, and polar shear
stresses are
\begin{equation}
  \begin{aligned}
    \sigma_{rr}^{*}
      &=\frac{p_i a^2}{b^2-a^2}
         \left(1-\frac{b^2}{r^2}\right),&
    \sigma_{\theta\theta}^{*}
      &=\frac{p_i a^2}{b^2-a^2}
         \left(1+\frac{b^2}{r^2}\right),&
    \sigma_{r\theta}^{*}&=0.
  \end{aligned}
  \label{eq:app-ring-polar}
\end{equation}
The Cartesian components used in the field comparisons follow from
\begin{equation}
  \begin{aligned}
    \sigma_{xx}^{*}
      &=\sigma_{rr}^{*}\cos^2\theta
        +\sigma_{\theta\theta}^{*}\sin^2\theta,\\
    \sigma_{yy}^{*}
      &=\sigma_{rr}^{*}\sin^2\theta
        +\sigma_{\theta\theta}^{*}\cos^2\theta,\\
    \sigma_{xy}^{*}
      &=(\sigma_{rr}^{*}-\sigma_{\theta\theta}^{*})
        \sin\theta\cos\theta.
  \end{aligned}
  \label{eq:app-ring-cartesian}
\end{equation}

\subsection{Pressurised elliptical cavity}
\label{app:ellipse-analytical}

For Figure~\ref{fig:ellipse-pressured-scheme}, the cavity-centred coordinates are
$X=x-0.1\,\mathrm{m}$ and $Y=y-0.1\,\mathrm{m}$.
The horizontal and vertical semiaxes are
$a_0=0.106666667\,\mathrm{m}$ and $b_0=0.08\,\mathrm{m}$;
the schematic rounds $a_0$ to $0.1067\,\mathrm{m}$.
The internal pressure is $p_i=10\,\mathrm{Pa}$.
The remote Cartesian stresses are $-K\sigma_0$, $-\sigma_0$, and
zero for the $xx$, $yy$, and $xy$ components, respectively, where
$\sigma_0=8\,\mathrm{Pa}$ is the vertical compressive amplitude and
$K=1$ is the horizontal-to-vertical stress ratio.
The computational lengths $L_x=L_y=0.8\,\mathrm{m}$ define the finite
quarter-domain used to approximate the infinite plate.

For this equal biaxial loading, the reference solution
has a compact representation through the
complex coordinate $z$ and dimensionless functions $F$ and $G$:
\begin{equation}
  \begin{gathered}
    z=X+\mathrm{i}Y,\qquad
    w=\sqrt{z^2-(a_0^2-b_0^2)},\\
    F=\frac{z}{w},\qquad
    G=\frac{b_0^2X+\mathrm{i}a_0^2Y}{w^3}.
  \end{gathered}
  \label{eq:app-ellipse-functions}
\end{equation}
Here $\mathrm{i}^2=-1$, and the square-root branch is chosen so that
$w/z\to1$ as $|z|\to\infty$. In the material,
$X^2/a_0^2+Y^2/b_0^2\ge1$, the reference stresses are
\begin{equation}
  \begin{aligned}
    \sigma_{xx}^{*}
      &=-p_i+(p_i-\sigma_0)\operatorname{Re}(F-G),\\
    \sigma_{yy}^{*}
      &=-p_i+(p_i-\sigma_0)\operatorname{Re}(F+G),\\
    \sigma_{xy}^{*}
      &=(p_i-\sigma_0)\operatorname{Im}G,
  \end{aligned}
  \label{eq:app-ellipse-stress}
\end{equation}
where $\operatorname{Re}$ and $\operatorname{Im}$ denote the real
and imaginary parts. These expressions are the $K=1$ specialisation
of the infinite-plate solution used in the stress comparisons.